\documentclass[twocolumn,times]{aastex62}
\usepackage{hyperref}
\usepackage{amsmath}

\shorttitle{No host halo concentration - satellite plane correlation}
\shortauthors{Pawlowski, Bullock \& Famaey}

\begin{document}

\title{\textbf{\large Do halos that form early, have high concentration, are part of a pair, or contain a central galaxy potential host more pronounced planes of satellite galaxies?}}

\author[0000-0002-9197-9300]{Marcel S. Pawlowski}
\altaffiliation{Hubble Fellow at UCI, then Schwarzschild Fellow at AIP}
\affiliation{Department of Physics and Astronomy, 
                 University of California,
                 Irvine, CA 92697, USA}
\affiliation{Leibniz-Institut f\"ur Astrophysik Potsdam (AIP),
                 An der Sternwarte 16, D-14482 Potsdam, Germany}
\email{marcel.pawlowski@uci.edu}

\author{James S. Bullock}
\affiliation{Department of Physics and Astronomy, University of California, Irvine, CA 92697, USA}

\author{Tyler Kelley}
\affiliation{Department of Physics and Astronomy, University of California, Irvine, CA 92697, USA}

\author{Benoit Famaey}
\affiliation{Universit\'e de Strasbourg, CNRS, Observatoire astronomique de Strasbourg, UMR 7550, 
 11 rue de l'Universit\'e, F-67000 Strasbourg, France}

\begin{abstract}
{
The Milky Way, the Andromeda galaxy, and Centaurus\,A host flattened distributions of satellite galaxies which exhibits coherent velocity trends indicative of rotation. Comparably extreme satellite structures are very rare in cosmological $\Lambda$CDM simulations, giving rise to the ``satellite plane problem''. As a possible explanation it has been suggested that earlier-forming, higher concentration host halos contain more flattened and kinematically coherent satellite planes. We have tested for such a proposed correlation between the satellite plane and host halo properties in the ELVIS suite of simulations. We find evidence neither for a correlation of plane flattening with halo concentration or formation time, nor for a correlation of kinematic coherence with concentration. The height of the thinnest sub-halo planes does correlate with the host virial radius and with the radial extent of the sub-halo system. This can be understood as an effect of not accounting for differences in the radial distribution of sub-halos, and selecting them from different volumes than covered by the actual observations. Being part of a halo pair like the Local Group does not result in more narrow or more correlated satellite planes either. Additionally, using the PhatELVIS simulations we show that the presence of a central galaxy potential does not favor more narrow or more correlated satellite planes, it rather leads to slightly wider planes. Such a central potential is a good approximation of the dominant effect baryonic physics in cosmological simulations has on a sub-halo population. This suggests that, in contrast to other small-scale problems, the planes of satellite galaxies issue is made worse by accounting for baryonic effects.
}
\end{abstract}


\section{Introduction}

Galaxies such as the Milky Way (MW) or Andromeda (M31) are surrounded by numerous less-luminous dwarf satellite galaxies. Within the framework of the currently favoured $\Lambda$CDM model of cosmology, which is based on the assumed existence of cold dark matter (CDM) and dark energy parametrized as a cosmological constant $\Lambda$, and implies galaxy formation to proceed hierarchically, these satellites are hosted by dark matter sub-halos and to have a variety of accretion histories, origins and infall times. Nevertheless, the structured nature of the cosmic web results in some coherences among sub-halo satellites beyond a purely isotropic distribution, with accretion preferentially happening along the directions of sheets and filaments \citep{Zentner2005, Libeskind2011, Libeskind2015}. Similarly, some sub-halos are expected to be accreted in groups, possibly resulting in coherences in the positions and orbital direction of satellite galaxies hosted by them (\citealt{LiHelmi2008}, \citealt{Wang2013}, \citealt{Wetzel2015}, \citealt{Shao2018}).
Finally, second-generation dwarf galaxies can form as highly correlated populations in the debris shed during galaxy collisions \citep{Bournaud2008, Duc2011, Hammer2013, Yang2014}. This effect, though, is purely baryonic in nature and therefore not included in dark-matter only cosmological simulations, while cosmological hydrodynamical simulations still lack the resolution to fully model this mode of galaxy formation (though see \citealt{Ploeckinger2018}).

Satellite systems are thus not expected to be completely isotropic, but the observed degree of anisotropy is of importance. The strong flattening of the MW satellite system has been discussed for decades \citep{KunkelDemers1976, LyndenBell1976, Haud1988, Majewski1994, LyndenBell1995}, but it was not before \citet{Kroupa2005} that the discrepancy between this ``plane of satellites'' and the distribution of sub-halo based satellite galaxies in cosmological simulations was measured and pointed out as a severe challenge to $\Lambda$CDM. 
Since then, this intriguing distribution of satellites around the MW, now dubbed the Vast Polar Structure (VPOS, \citealt{Pawlowski2012}), has been studied in detail, revealing alignments of additional objects such as globular cluster \citep{Forbes2009, Keller2012, Pawlowski2012}, and of later discovered satellite galaxies \citep{PawlowskiKroupa2014,Pawlowski2015a,Pawlowski2016}. 

Maybe more important than the purely spatial flattening is the strong kinematic coherence of this VPOS deduced from proper motion measurements, with not only stellar and gaseous streams preferentially aligning with it \citep{Pawlowski2012}, but also a majority of the classical MW satellite co-orbiting within the plane \citep{Metz2008,PawlowskiKroupa2013}. 
The Gaia DR2 \citep{Helmi2018} confirmed the previous finding that not all satellite galaxies share the same orbital plane \citep[e.g.][]{PawlowskiKroupa2013}, while a more detailed analysis concluded that only 6 out of 39 satellites can be claimed to have orbital poles that are conclusively misaligned with the VPOS given the current proper motion uncertainties \citep{Fritz2018}.

When \citet{Ibata2013} and \citet{Conn2013} discovered a similar satellite galaxy plane around the Andromeda Galaxy (M31), consisting of 15 out of the 27 satellites within the footprint of the PAndAS survey, this ``satellite planes problem'' was immediately regarded to be much more pressing. Like the VPOS, this Great Plane of Andromeda (GPoA) is very uncommon among simulated sub-halo systems \citep[e.g.][]{Ibata2014,Pawlowski2014}.

Similar satellite galaxy planes have been looked for in more distant host systems, and indications for planar structures have been obtained for the M81 group \citep{Chiboucas2013}, and for the non-satellite galaxies in the Local Group \citep{Pawlowski2013,PawlowskiMcGaugh2014a}. For Centaurus\,A, one or two planes of satellite galaxies have been proposed \citep{Tully2015,Mueller2016}, and recently a kinematic correlation reminicent of that of the M31 satellite plane has been discovered which places the Centaurus\,A satellite plane (CASP) in tension with simulated satellites in both dark matter only as well as hydrodynamical cosmological simulations \citep{Mueller2018}.
There is even some evidence for a statistical over-abundance of velocity-anti-correlated satellites in the SDSS that is consistent with a high abundance of satellites in co-orbiting planes (\citealt{Ibata2014b,Ibata2015}, but see \citealt{Phillips2015, Cautun2015a} for alternative views).

The resulting debate about the existence of satellite planes in the Universe and in cosmological simulations, and about the degree of the tension they constitute with $\Lambda$CDM expectations, has led to an extensive body of literature \citep[e.g.][]{Cautun2015a, Cautun2015b, BahlBaumgardt2014, Buck2015a, Buck2015b, Gillet2015, Pawlowski2014, Pawlowski2015b, Ibata2014, Ibata2015, Wang2013}. For a current review, see \citet{Pawlowski2018}. The sheer size of this body of literature, combined with the technical depth required to both discuss and explain biases and differences in the derived conclusions to untangle the controversy of the current debate, makes a detailed discussion prohibitively long for this contribution\footnote{We refer the interested reader to two of our recent contribution on this topic: \citet{Pawlowski2014} and \citet{Pawlowski2015b}.}. In the following, we thus focus on one particularly enthralling proposition as the main emphasis of this this contribution.

Satellite galaxy planes can be seen as a problem for $\Lambda$CDM cosmology due to their rarity in simulations. However, they can also be seen as a chance to potentially learn more about the Local Group host galaxies and their dark matter halos, if for example some host halo properties or special evolutionary histories strongly prefer the detection of correlated planes of satellite galaxies.
\citet{Buck2015a} propose such a solution to the satellite plane problem. 
They report the existence of thin, rotating planes of satellites that resemble the GPoA in a suite of 21 M31-like host halos. In particular, they find that the most narrow,  kinematically-coherent planes occur among the highest concentration halos in their sample and interpret this as a correlation between halo formation time and propensity for a halo to host a coherent plane of satellites.

As the physical cause of this effect they envision the following scenario. Earlier forming halos of similar mass have a higher chance to be situated at the nodes between filaments, whereas later forming halos are preferentially situated within filaments. As a consequence, halos within a filament are fed in a pattern that is closer to spherical, whereas halos in nodes of filaments are fed from only a few directions. While this appears counter-intuitive given that being situated in a node implies that a halo is fed by more filaments and thus from more directions than if it were situated within a single filament, filaments were more narrow at earlier times \citep{VeraCiro2011}. It is therefore worthwhile to test for such a correlation, though the question whether an early-forming satellite plane would survive as a coherent and narrow structure until today remains of concern \citep{Fernando2017}.

If \citet{Buck2015a} are correct then this could provide an important clue to the puzzle of satellite planes, however given that their sample included only 21 halos there is the possibility that the result was a statistical anomaly. 
Another concern is that the effect could be caused by differences in the radial distribution of sub-halo satellites. If the halos with higher concentration also contain more compact satellite systems, as seems to be the case for this sample \citep{Buck2015b}, a comparison measuring only the absolute height of the satellite planes will be biased to find more narrow planes in the higher-concentration cases. 

We set out to address these concerns. In the following, we will test whether the concentration, formation time, virial radius (or mass) of a halo, or the radial concentration of the sub-halo population, has an effect on the thickness and kinematic coherence of satellite galaxy planes. We do this using 48 MW and M31- like halos from the ELVIS simulation suite – thereby alleviating the concern of low-number statistics in the host sample. We look for a correlation between host halo properties and the occurrence and properties of satellite galaxy planes, and control for the effect of the radial distributions by exploring randomized samples with the same radial distributions drawn from isotropy. We further utilize the fact that the ELVIS suite consists of 24 isolated hosts and 24 in paired configurations as found in the Local Group to test for an effect of environment on the width or kinematic coherence of planes of satellites. In addition, we use the new PhatELVIS suite of 12 hosts simulated both as dark-matter-only simulations and with an added analytical MW-like disk potential \citep{Kelley2018}, to investigate the influence that the additional tidal disruption by a central baryonic disk galaxy has on the width and kinematic coherence of satellite galaxy planes. Since such disruption is the main effect that the inclusion of baryonic physics has on the satellite galaxy system \citep{GarrisonKimmel2017}, this comparison will provide hints on the feasability to address the planes of satellite galaxies problem with more realistic hydrodynamical cosmological simulations.
We compare all planes with the observed GPoA, which consists of 15 satellite galaxies with a root- mean-square (rms) height of $\Delta_\mathrm{rms} = 12.6 \pm 0.6$\,kpc (with a 99\% confidence upper limit of 14.1\,kpc), out of which 13 show line-of-sight velocities consistent with co-orbiting around M31\citep{Ibata2013}.

\section{Method and sub-halo selection}

\subsection{Top 30 sub-halos within $r_{\mathrm{vir}}$}
\label{subsect:top30}

\begin{figure}
   \centering
   \includegraphics[width=88mm]{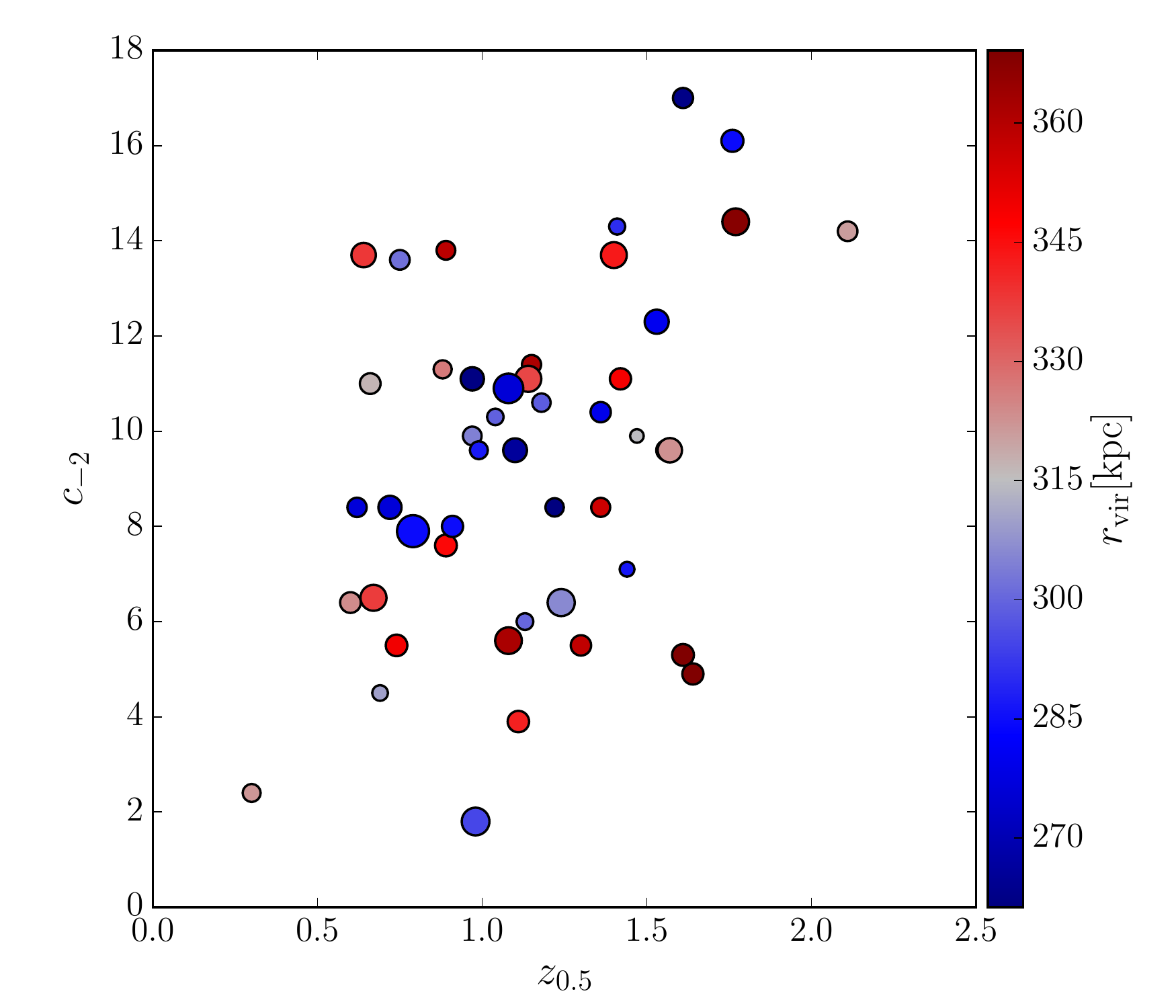}
   \includegraphics[width=88mm]{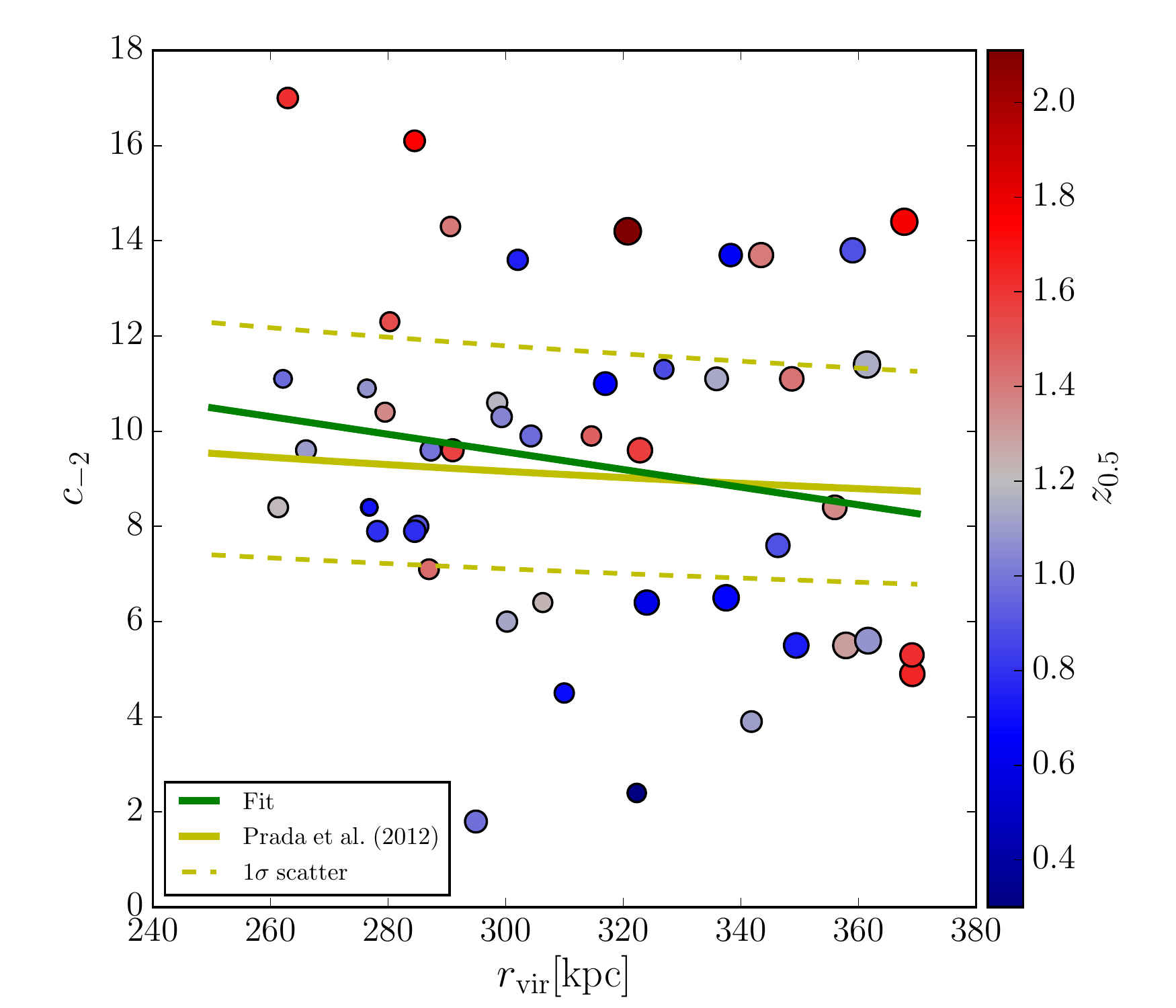}
   \caption{
Overview of the host halo properties in the ELVIS suite. The top panel plots the halo concentration parameter $c_{-2}$\ against the formation redshift $z_{0.5}$, and color-codes for the virial radius $r_\mathrm{vir}$. Larger symbol size indicate halos that host more extended sub-halo systems as measured via $\Delta_\mathrm{r}^\mathrm{subhalos}$. It is apparent that earlier-forming halos on average have higher concentration.
The bottom panel compares the virial radius $r_\mathrm{vir}$\ and concentration parameter $c_{-2}$. A linear fit (green line) indicates that more compact host halos (small $r_\mathrm{vir}$) tend to have slightly higher concentrations while larger halos have lower concentration, albeit with large scatter. This is in line with the expected mass-concentration relation, shown by converting $M_\mathrm{vir}$\ to $r_\mathrm{vir}$\ (yellow line from \citealt{Prada2012}). The dashed yellow lines indicate a $1\sigma$\ scatter of 0.11 dex around the concentration relation \citep{Maccio2008,DuttonMaccio2014}. While the ELVIS halos follow the overall concentration relation well, the nine highest-concentration halos of the 48 host halos are on average offset by $2\sigma$\ relative to the mean concentration relation.
  }
              \label{fig:haloprops}
\end{figure}

As a first step, we reproduce the earlier analysis, now using the 48 MW- and M31-like host halos of the Exploring the Local Volume in Simulations (ELVIS) project by \citet{GarrisonKimmel2014}. We use the publicly available present-day ($z = 0$) halo catalogs, and rank all sub-halos of a given host by the maximum mass $M_{\mathrm{peak}}$\ they had over their history. This is consistent with the ranking used by \citet{Buck2015a} and with prevalent abundance matching schemes. The mass range covered by the ELVIS halos ($M_{\mathrm{vir}} = 1.0~\mathrm{to}~2.8\,\times\,10^{12} \mathrm{M}_{\sun}$) is comparable with that of the halos analysed by \citet{Buck2015a} ($M_{\mathrm{200}} = 0.8~\mathrm{to}~2.1\,\times\,10^{12} \mathrm{M}_{\sun}$), especially when considering that $M_{\mathrm{200}}$\ is systematically smaller than $M_{\mathrm{vir}}$\ because the latter is calculated over a larger halo volume (encompassing an average density of 97 instead to 200 times the critical density of the universe). The resolutions of the two simulation suites are comparable, too. ELVIS uses a particle mass of $m_\mathrm{p} = 1.9 \times 10^5 \mathrm{M}_\sun$\ and a force softening of $\epsilon = 141$\,pc for the highest-resolution regions, while the simulations of \citet{Buck2015a} have a particle mass of $m_\mathrm{p} \approx 1.5 \times 10^5 \mathrm{M}_\sun$\ and force softening of between $\epsilon = 370\ \mathrm{to}\ 540$\,pc.
 
To investigate the effect of a central baryonic host galaxy, we additionally analyze the 12 hosts of the Phat ELVIS simulation suite by \citet{Kelley2018}. This extension to the ELVIS project consists of twelve dark-matter-only zoom simulations, which each were re-run with an embedded galaxy potential grown to match the observed Milky Way disk and bulge today. The host halos cover a mass range of  ($M_{\mathrm{vir}} = 0.7~\mathrm{to}~2.0\,\times\,10^{12} \mathrm{M}_{\sun}$), while the central host disk galaxy is always $6.9\,\times\,10^{10} \mathrm{M}_{\sun}$\ at $z=0$. The galaxy potential consists of a stellar and a gaseous disk as well as a bulge component, and is grown starting from redshift $z = 3$. The resolution of PhatELVIS is better than in the ELVIS suite, with a particle mass of $m_\mathrm{p} = 3 \times 10^4 \mathrm{M}_{\sun}$\ and a force softening of $\epsilon = 37$\,pc for the highest-resolution regions. Due to the differences in resolution and assumed cosmological parameters -- Planck 2015 results \citep{Planck2016} instead of WMAP7 \citep{Larson2011} as in ELVIS -- we refrain from combining the ELVIS and PhatELVIS samples despite the temptation of a larger sample size\footnote{ We checked that our main conclusions remain unchanged if combining the samples.}. Instead, the Phat ELVIS simulations will be used in Sect. \ref{sect:newtests} to determine whether the changes to the orbital properties of a satellite system caused by the presence of a baryonic central galaxy potential can have a substantial effect on the properties of planes of satellite galaxies.

We then look for correlations of the properties of the thinnest and most co-rotating satellite planes among the sub-halos and the following properties of their host halo:
\begin{itemize}
\item $r_{\mathrm{vir}}$, the virial radius measured as the radius of a sphere centered on the host halo that contains a density of 97 times the critical density $\rho_\mathrm{crit}$\ of the universe.  
Since this definition implies an unequivocal relation between $r_{\mathrm{vir}}$\ and the virial mass $M_{\mathrm{vir}}$\ we thus could also refer to the latter which might be seen as more fundamental. However, we decided to refer to the virial radius in the following because it is more directly linked to the issue of spatial distributions.
\item $c_{\mathrm{-2}}$, the halo concentration determined as $c_{\mathrm{-2}} = r_{\mathrm{vir}}/r_{\mathrm{-2}}$, where $r_{\mathrm{-2}}$ is the radius where $\rho r^{2}$ peaks. This differs slightly from the concentration parameter in \citet{Buck2015a}, who instead of $r_{\mathrm{vir}}$\ use $r_{\mathrm{200}}$, the radius at which the halo contains a density of 200 times the critical density. Since $r_{\mathrm{200}} < r_{\mathrm{vir}}$, this results in somewhat lower numerical values of concentration for a given halo.
\item $z_{\mathrm{0.5}}$, the formation redshift of the host halo, measured as that redshift at which the host's main progenitor first acquires 50 per cent of the host halo's $z = 0$\ virial mass $M_{\mathrm{vir}}$.
\item $\Delta_{\mathrm{r}}^{\mathrm{subhalos}}$, the root-mean-square (rms) of the radial distances from the center of their host of the selected sub-halos (30, or 27 in case of the PAndAS-like selection volume).
\end{itemize}

Figure \ref{fig:haloprops} summarizes these host halo properties. Additionally, the lower panel compares the ELVIS halo concentrations $c_{\mathrm{-2}}$\ to the average concentration relation of \citet{Prada2012} which uses the same cosmology (yellow line), plotted against virial radius $r_{\mathrm{vir}}$\ instead of virial mass $M_{\mathrm{vir}}$. The dashed lines indicate a $1\sigma$\ scatter of 0.11 dex (or a factor of 1.3) as reported by \citet{DuttonMaccio2014} and used in \citet{Buck2015a}. While the study of \citet{DuttonMaccio2014} is based on a slightly different cosmology, the typical scatter around the average mass-concentration is found to be virtually identical among different cosmologies \citep{Maccio2008}, and thus applicable to the ELVIS simulations, too. While the sample of host halos of \citet{Buck2015a} was selected to especially sample the wings of the concentration distribution, Fig. \ref{fig:haloprops} shows that the ELVIS sample also contains nine hosts that are offset from the mean concentration relation by $\approx 2\sigma$\ towards high concentrations.

\begin{figure}
   \centering
   \includegraphics[width=80mm]{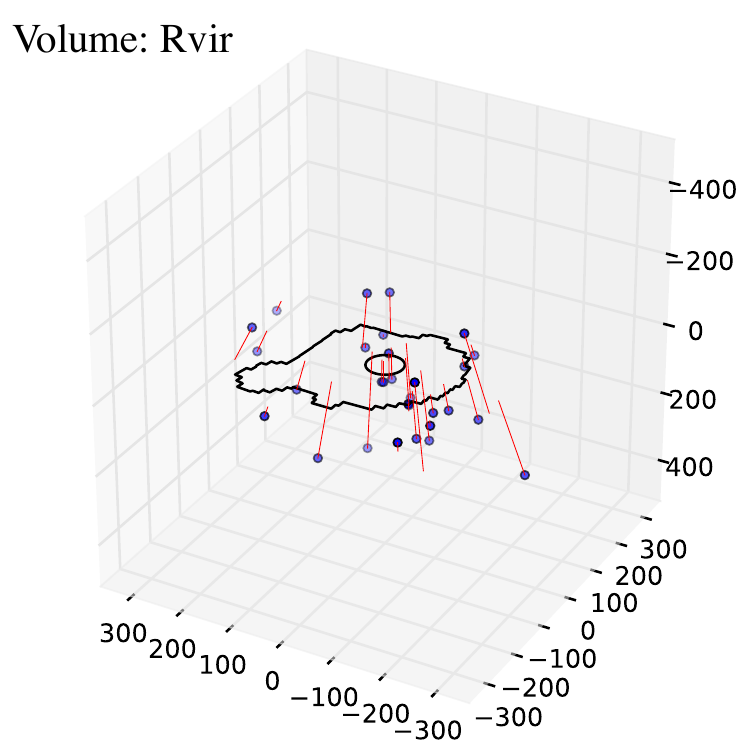}
   \includegraphics[width=80mm]{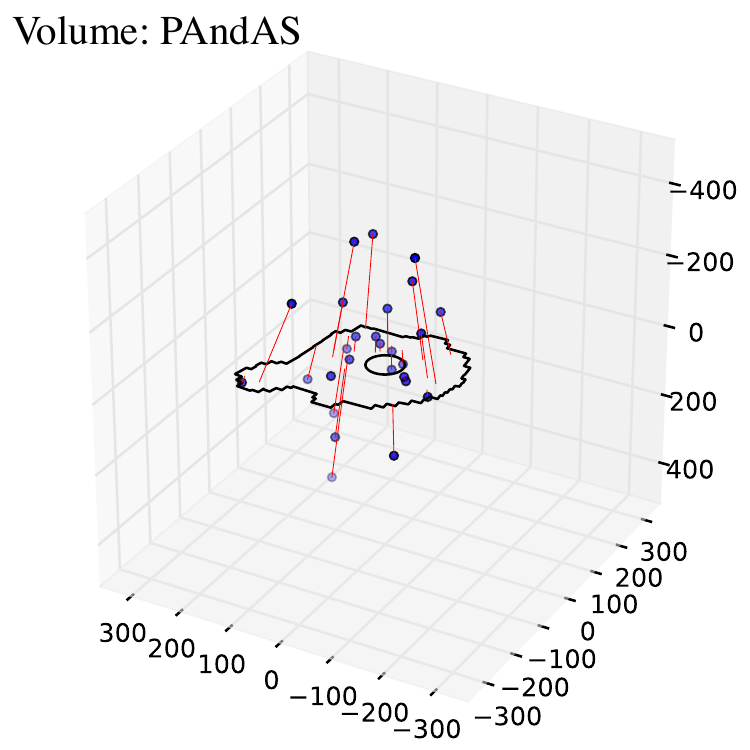}
   \caption{
Comparison of the selection volumes for the top-30 and the PAndAS-like samples. Shown in blue are the Cartesian positions (axes in kpc) of the selected sub-halos belonging to the ELVIS halo iCharybdis. The black polygon indicates the PAndAS footprint at the distance of M31, centered on the position of the host halo. The black circle indicates the inner exclusion region within $2.5^{\circ}$\ around the host. The red lines indicate the projection of the sub-halos into this plane along the line-of-sight from the virtual MW. Clearly, selecting sub-halos from the whole virial radius (upper panel, Sect. \ref{subsect:top30}) results in many objects outside of the region actually observed by PAndAS, while some close to the host fall within the inner exclusion region, too. When assessing the existence of sub-halo structures similar to the GPoA in simulation, it is therefore essential to select model satellites only from the actually observed volume (lower panel, Sect. \ref{subsect:PAndAS}).
   }
              \label{fig:selectionvolumes}
\end{figure}

Like \citet{Buck2015a}, we first select the top $N_{\mathrm{sh}} = 30$\ sub-halos within the virial radius of each of our 48 host halos, excluding all sub-halos within the innermost 30\,kpc. This is our ``simulated'' sample. We also generate a ``randomized'' sample of sub-halo systems. For this, we keep the radial distances and absolute velocities of all sub-halos, but draw random directions from an isotropic distribution for the position and velocity vectors. 
This constitutes a simple but necessary and important test: are potential differences in the properties of satellite planes due to an increase of sub-halo coherence that is linked to host-halo properties (such as concentration or formation redshift), or are such differences driven by the radial distribution of the sub-halo system itself?

To find the thinnest planes for each number of sub-halos between 3 and $N_{\mathrm{sh}}$, we proceed as follows. We generate 10,000 normal directions that are approximately evenly distributed on a hemisphere. Each of these describes a possible satellite plane centered on the host halo. For each of these planes, we rank all $N_{\mathrm{sh}}$\ sub-halos by their distance from the plane, and measure the rms height of the $N_\mathrm{inPlane} = 3, 4, 5, ... N_{\mathrm{sh}}-1, N_{\mathrm{sh}}$\ closest sub-halos. For the kinematic analysis, we also determine the number of co-orbiting satellites by counting the number of positive and negative angular momenta projected onto the normal vector. The number of co-orbiting satellites $N_{\mathrm{Coorb}}$\ is the larger of the two (and thus always $N_\mathrm{inPlane}/2 \leq N_{\mathrm{Coorb}} \leq N_\mathrm{inPlane}$). For each combination of $N_\mathrm{inPlane}$\ and $N_{\mathrm{Coorb}}$\ that occurs in a sub-halo system, we record the smallest plane height $\Delta_{\mathrm{rms}}$.

\subsection{PAndAS-like selection of sub-halos}
\label{subsect:PAndAS}

\begin{figure*}
   \centering
   \includegraphics[width=88mm]{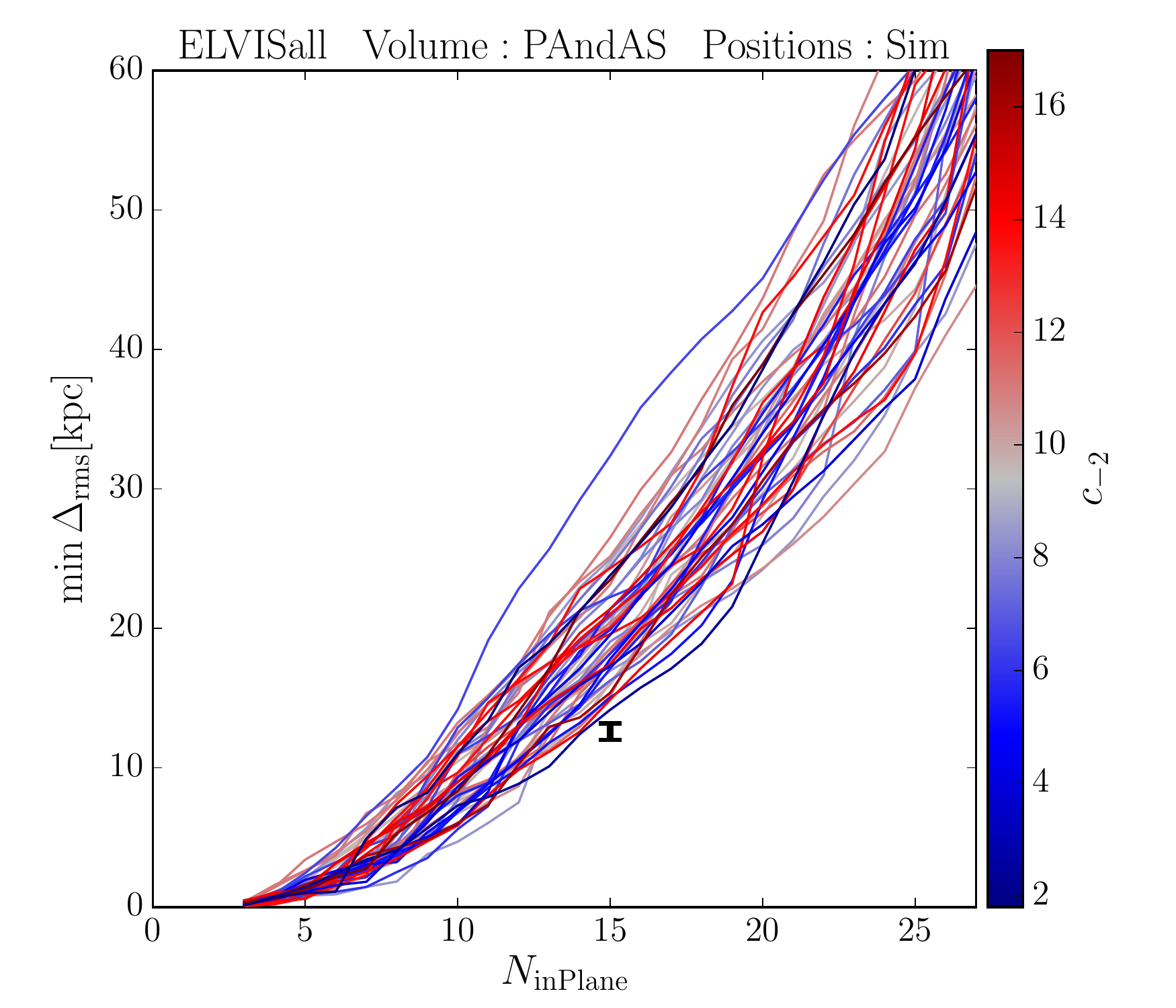}
   \includegraphics[width=88mm]{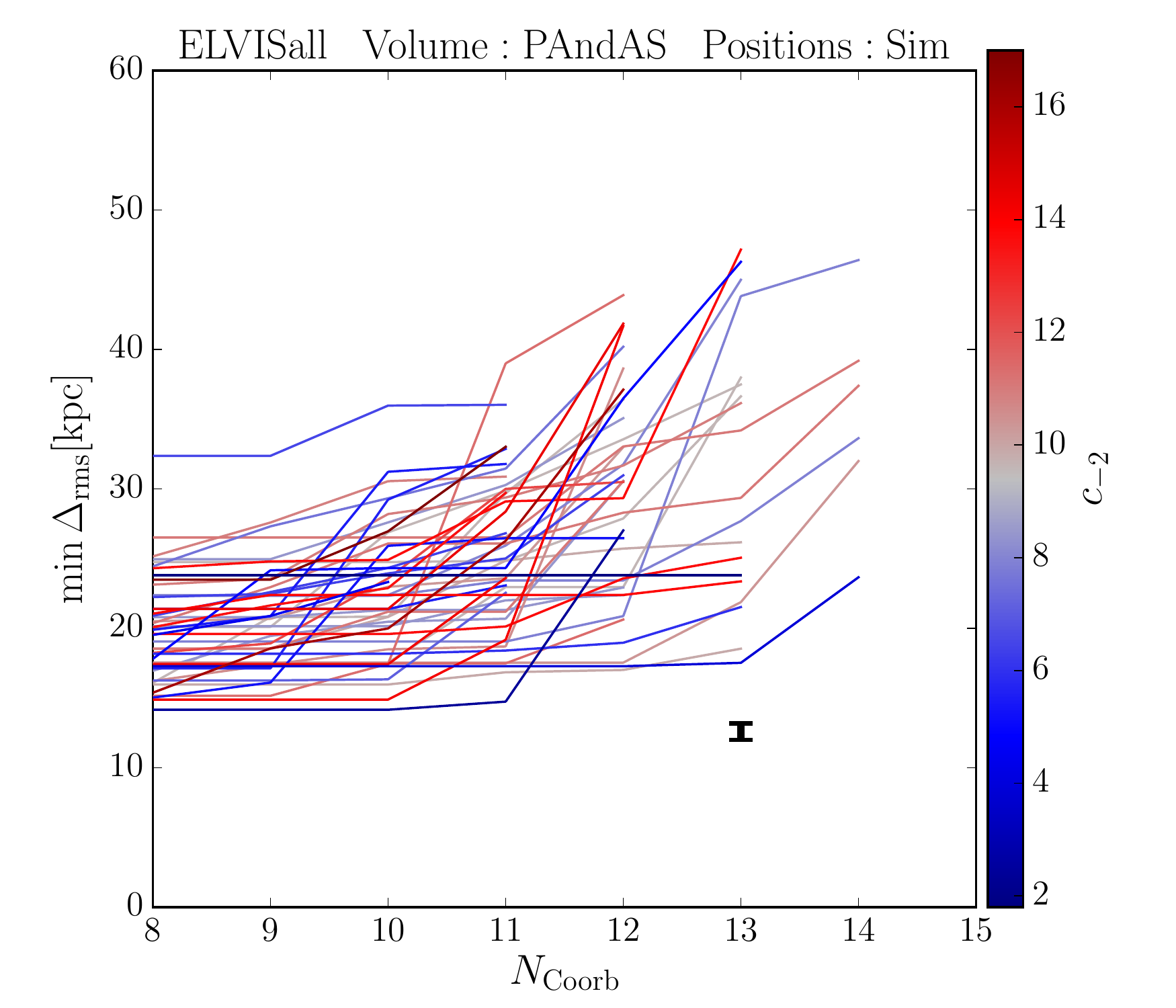}
   \caption{
Comparison of satellite plane properties of the observed GPoA with those found for satellite systems in ELVIS and selected from within the mock PAndAS volume.
The {\it left} panel plots minimum plane height against the number of satellites in the plane, irrespective of the number of co-orbiting satellites (see Fig. \ref{fig:planeheights}). The {\it right} panel plots the plane height against the number of co-orbiting satellites for planes consisting of exactly 15 satellites. The lines stop once no plane with a given number of co-orbiting satellites $N_\mathrm{Coorb}$\ is present for the corresponding halo's satellite system (see Fig. \ref{fig:planeheightscorot}).
Lines are color-coded by the concentrations $c_\mathrm{-2}$\ of their host halos. No correlation of plane heights or kinematic coherence with host halo concentration is found. The black error bar gives the plane height and degree of kinematic coherence for the GPoA around M31 of $\Delta_\mathrm{rms} = 12.6 \pm 0.6$\ \citep{Ibata2013}, which consists of 15 satellites out of which 13 show correlated line-of-sight velocities. None of the 48 simulated satellite systems contains a satellite plane with properties that are as extreme as observed.
   }
              \label{fig:summary}
\end{figure*}

As already pointed out in \citet{Pawlowski2014}, the satellite selection volume has a major effect on the distribution of sub-halos selected from a simulation and thus on the deduced properties of satellite planes (see Fig. \ref{fig:selectionvolumes}). Volumes that differ in size and shape from the one actually observed can bias both towards and away from finding narrow satellite planes, depending on their shape. Furthermore, the number of satellites used to find satellite planes has a major influence on the resulting properties: the more satellites are in the sample, the higher is the chance to find more extreme properties among sub-samples of a fixed number. This is why the selection of 30 (instead of the observed 27) satellites, from the virial radius (instead of the PAndAS footprint),  but allowing for satellites close to the host in projection (in contrast to the observational study by \citealt{Ibata2013}, which excluded the inner $2.5^{\circ}$\ around M31) makes any comparison between the observed GPoA and satellite planes in the simulation futile.

Defining the satellite selection volume by the virial radii of the individual host halos furthermore risks introducing the looked-for anti-correlation between host halo concentration and thickness of the most narrow satellite planes. If there is a relation between the overall extent of a satellite system and the height of the most narrow satellite plenes (see Sect. \ref{subsect:rmsthicknessrelation})\footnote{In its simplest form, this can be thought of as if satellite systems of hosts with different $r_\mathrm{vir}$\ can be assumed to be re-scaled versions of the same underlying distribution,  as can be expected from the self-similarity of $\Lambda$CDM halos over a large mass range.}, then linking the volume from which satellite are selected to $r_\mathrm{vir}$\ introduces a correlation between $r_\mathrm{vir}$\ and the minimum plane heights $\Delta_{\mathrm{rms}}$. The definition of $r_{\mathrm{vir}}$\ implies an unequivocal relation to the virial mass $M_{\mathrm{vir}}$\ of the form
$$M_{\mathrm{vir}} = 97 \times \rho_\mathrm{crit} \times \frac{4}{3} \pi r_{\mathrm{vir}}^{3},$$
such that a correlation with $r_{\mathrm{vir}}$\ also implies a correlation with $M_{\mathrm{vir}}$.
In $\Lambda$CDM, it is well established that dark matter halos follow a halo mass-concentration relation, in the sense that halos with larger $M_{\mathrm{vir}}$\ have lower $c_{\mathrm{-2}}$\ \citep{Bullock2001,Ludlow2014}. Thus, a correlation with $M_{\mathrm{vir}}$\ also implies an anti-correlation with the concentration parameter. Putting all this together, more concentrated host halos tend to be less massive, have smaller virial radii ( see also lower panel of Fig. \ref{fig:haloprops}), their satellite systems are thus selected from a smaller volume, which biases the satellite planes found in them to be more narrow. Thus, it is conceivable that the selection volume chosen by the previous study introduces the very correlation they reported. However, it is unclear whether this artificially introduced correlation is discernible given low number statistics, the substantial scatter in the mass-concentration relation, and the stochasticity of satellite plane heights.

Another distinction in our analysis that improves upon the initial exploration by \citet{Buck2015a} is that we restrict the analysis to using only the 1D line-of-sight velocities, in contrast to the previously described analysis (Sect. \ref{subsect:top30}) which uses the full 3D velocities of the sub-halos to determine the kinematic coherence. The latter are not available observationally (proper motions at the distance of M31 are extremely difficult to measure, \citealt{Sohn2012, vanderMarel2012, Salomon2016}).

Finally, we point out that in principle one should consider observational uncertainties since these are an additional source of dispersion in the satellite coherence (they make planes appear thicker and velocities appear less coherent). We ignore this in the rest of this work for the sake of simplicity -- and because the frequency of satellite planes comparable to the observed one turns out to be zero even without considering this effect -- but we will re-visit the point in a future publication (Pawlowski et al. in prep.).

To compare with the observed GPoA, we use the same ELVIS simulations as discussed before, but select sub-halos from a volume determined by the PAndAS footprint area.  For this, we observe each of the isolated host halos from a random direction, while for the paired hosts the direction of observation is from the partner host. To select only sub-halos that fall into the survey area covered by PAndAS, we mock-observe the sub-halos by projecting their positions into a spherical coordinate system with angles centered on the host halo. The mock-observer is situated at a distance to the host halo of 780\,kpc, consistent with the distance to M31. We then select the top 27 sub-halos ranked by their mass at infall ($m_{\mathrm{max}}$) that lie within the PAndAS survey polygon. This number is chosen to be identical to the number of satellites considered by \citet{Ibata2013} and \citet{Conn2013}. The maximum allowed distance of a sub-halo from the host halo center is 500\,kpc, which makes the selection volume consistent with the most-distant M31 satellite in the observed sample (Andromeda XXVII at a most-likely distance of $\sim 480$\,kpc from M31; \citealt{Conn2012}). Like \citet{Ibata2013}, we also exclude the region within $2.5^{\circ}$\ of the host halo in projection and exclude all sub-halos from that volume, because the high background this close to the host hampers the detection of satellite galaxies and negatively affects distance measurements. This selection results in a total volume of about 0.07\,Mpc$^{3}$, or a volume corresponding to that of a sphere with radius 257\,kpc. As before, we also generate a random sample by randomizing the sub-halo positions and velocities before applying the PAndAS selection.

For the plane-fitting analysis, we consider the full 3D position vectors of the 27 selected sub-halos for each host, as well as the line-of-sight velocity vectors as seen from the direction from which the PAndAS footprint was applied (within the host halo rest frame). The plane finding routine itself is identical to the one described before.

\section{Spatial Coherence}
\label{subsect:spatial}

In Fig. \ref{fig:summary}, the properties of the observed GPoA (black error bars) are compared to those of satellite planes identified in the simulations. The figure uses the PAndAS-like selection volume to mimic the selection biases present in the observed satellite sample. No planes with plane heights as narrow as observed are identified. While for some halos planes of 15 satellites can be defined which contain 13 kinematically correlated satellites, analogous to the observed system, none of these simulated planes are simultaneously as narrow as the observed structure. The color-coding by each halo's concentration highlights that no correlation with the concentration parameter of the host halos is present. Consequently, the presence of a narrow, kinematically highly correlated plane of satellite galaxies in the observed satellite system of M31 does not support the conclusion that M31 lives in a high-concentration halo.

\subsection{Plane thickness vs. halo properties}

\begin{figure*}
   \centering
   \includegraphics[width=88mm]{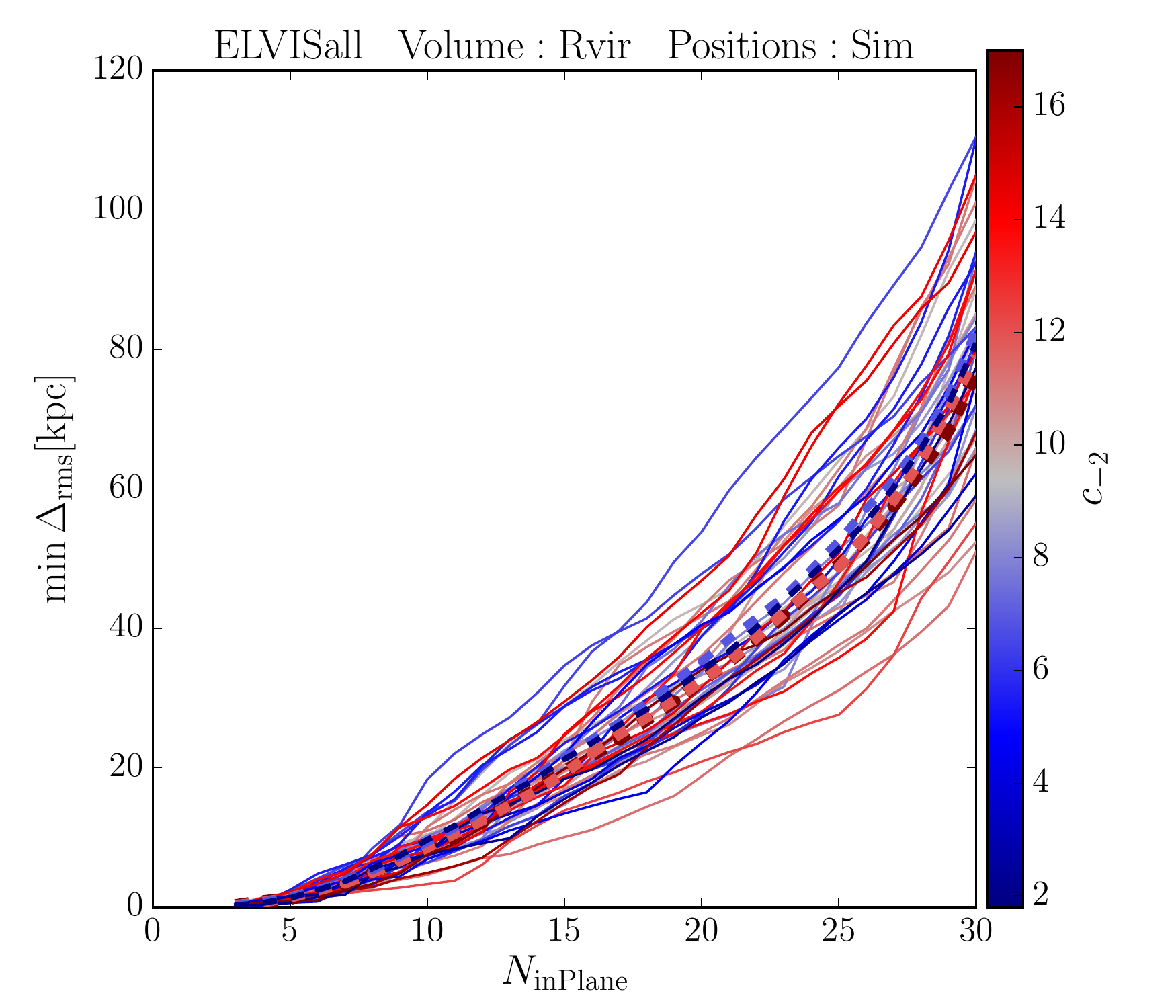}
   \includegraphics[width=88mm]{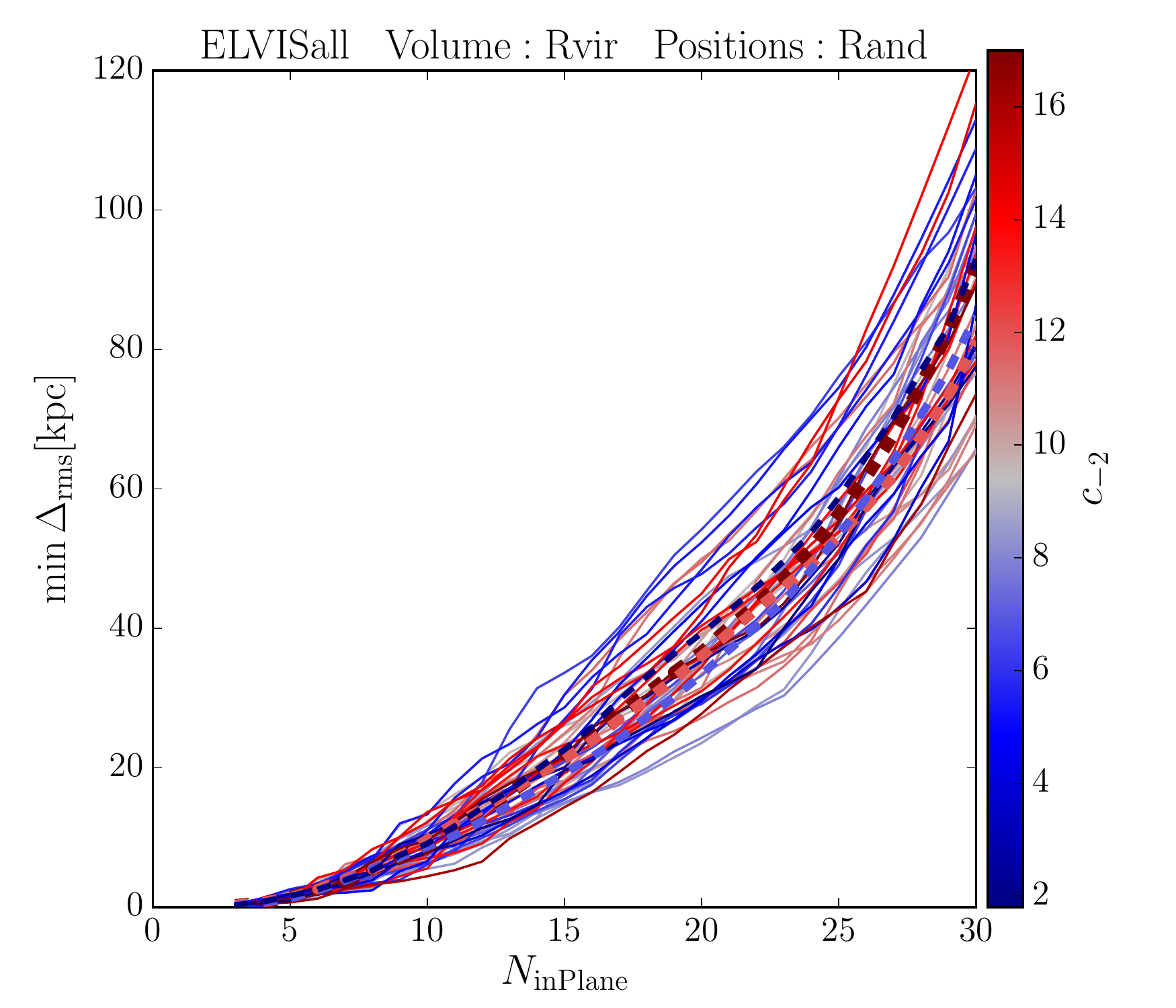}
   \includegraphics[width=88mm]{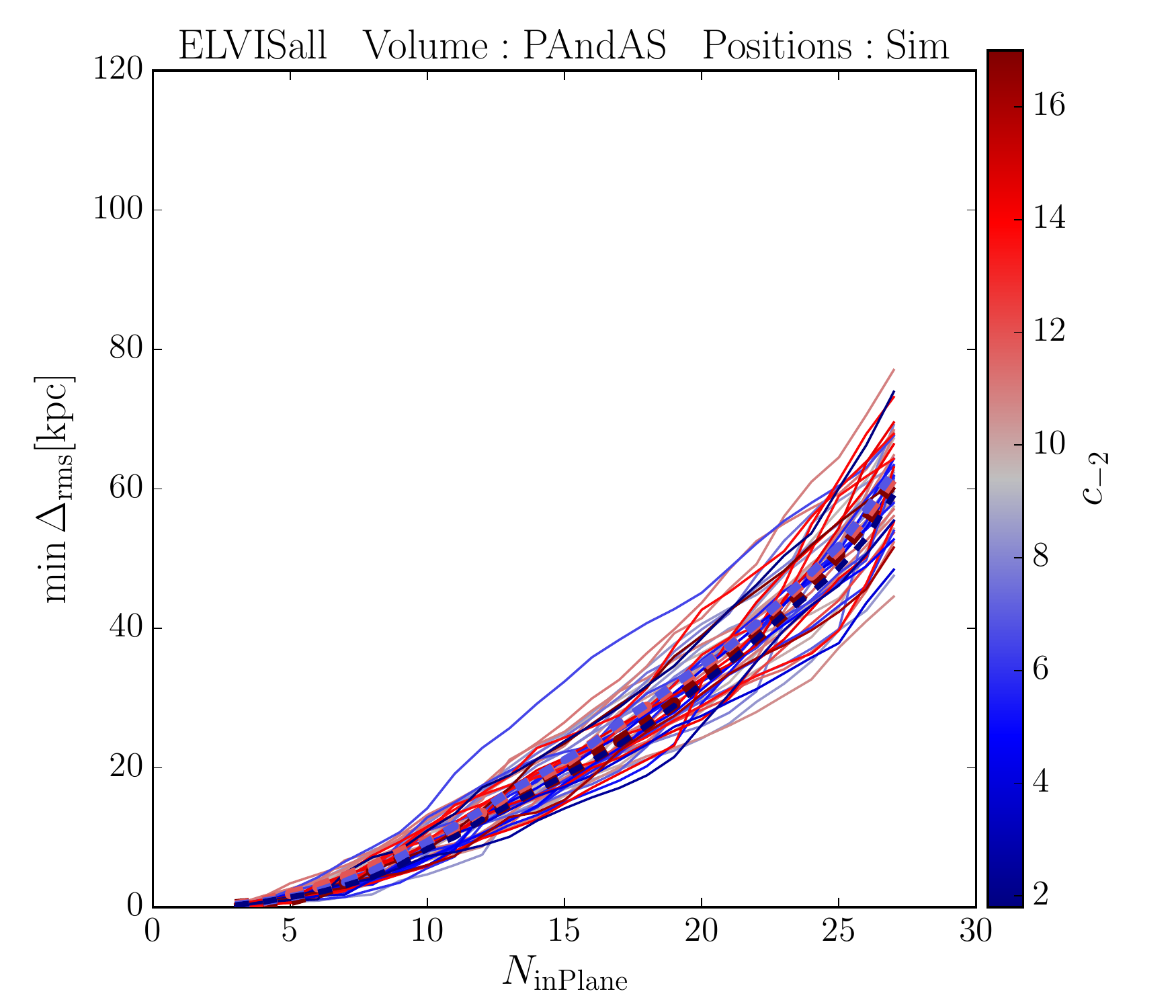}
   \includegraphics[width=88mm]{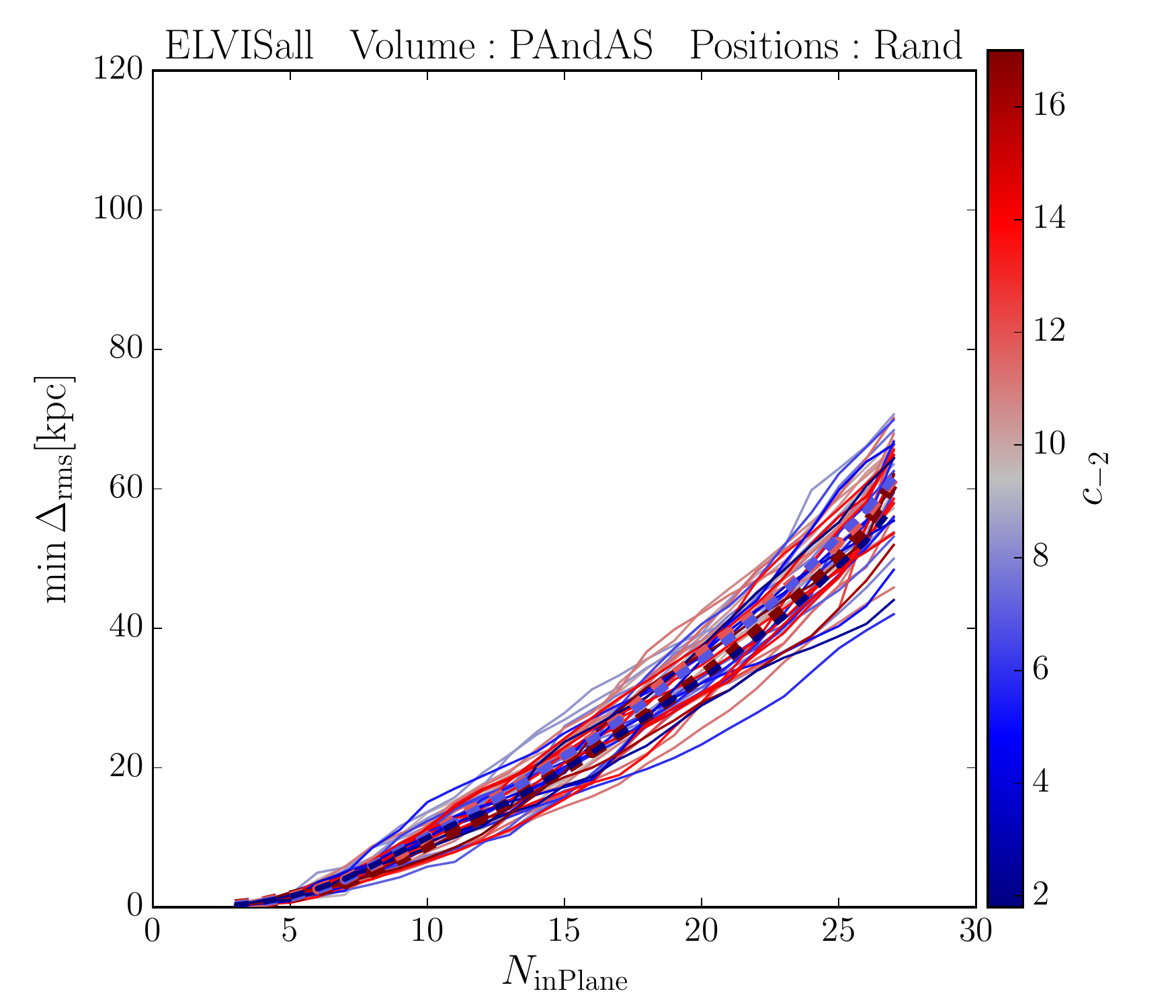}
   \caption{
Minimum plane height vs. number of satellites in the plane (without any constraint on the number of co-orbiting satellites), color-coded for host halo concentration. Shown are the results for 30 sub-halos in the virial radius (top left), 30 sub-halos with randomized positions and velocities (top right), 27 sub-halos selected from within the PAndAS volume (bottom left) and 27 sub-halos selected from the PAndAS volume after randomizing their positions and velocities (bottom right). The thick, dashed lines give the average plane heights for combinations of 12 sub-halo systems ranging from the 12 most-concentrated (red) to the 12 least concentrated (blue).
   }
              \label{fig:planeheights}
\end{figure*}

\begin{figure*}
   \centering
   \includegraphics[width=80mm]{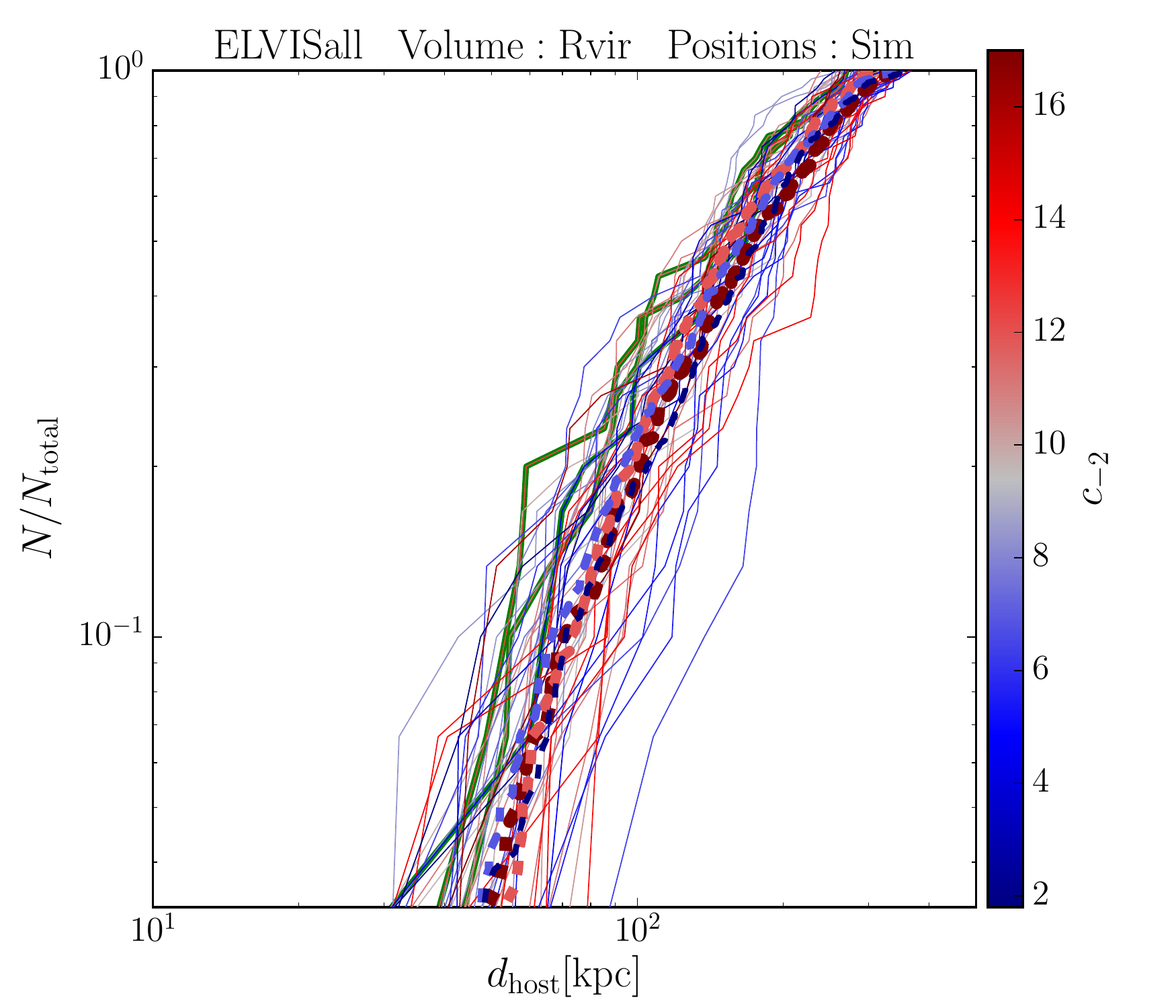}
   \includegraphics[width=80mm]{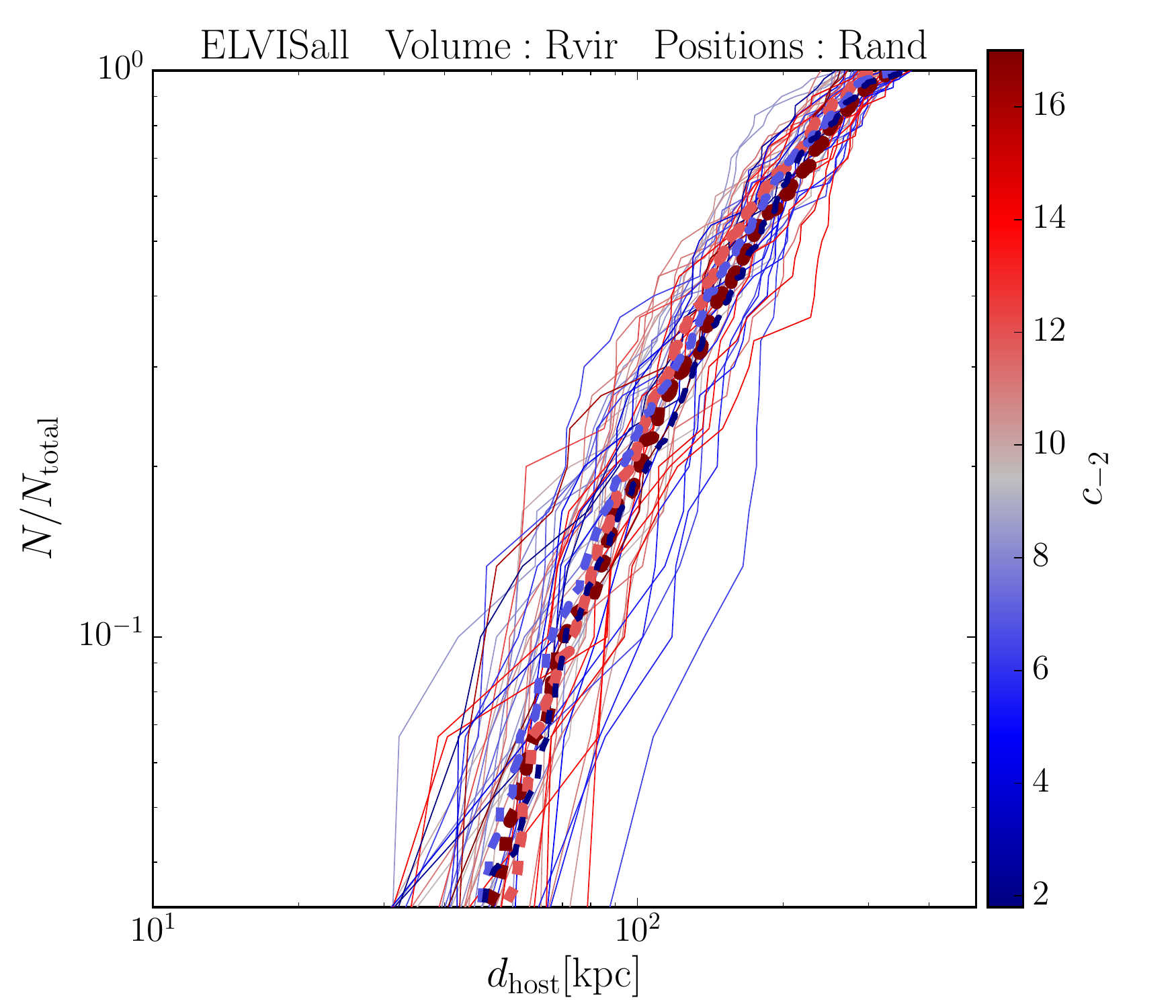}
   \includegraphics[width=80mm]{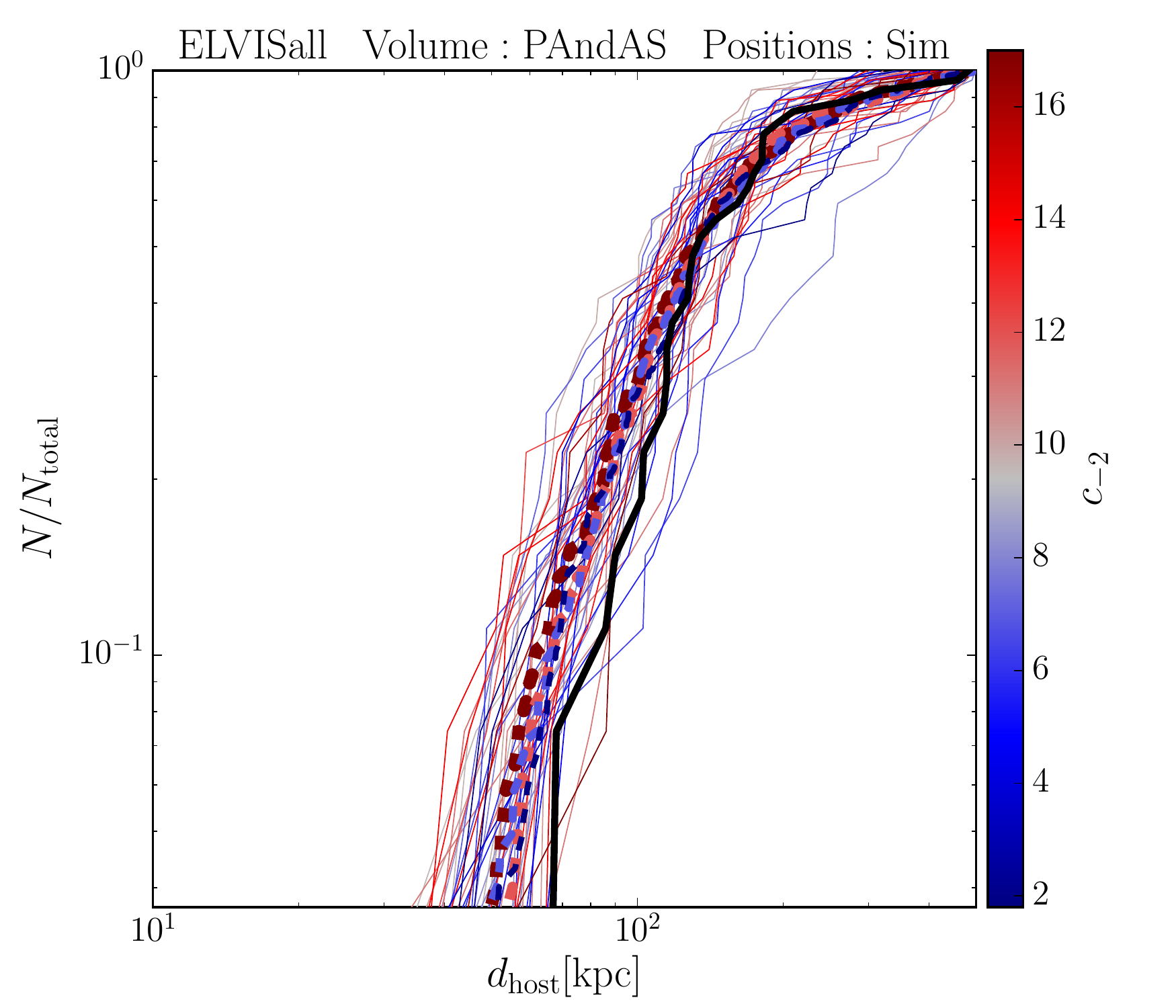}
   \includegraphics[width=80mm]{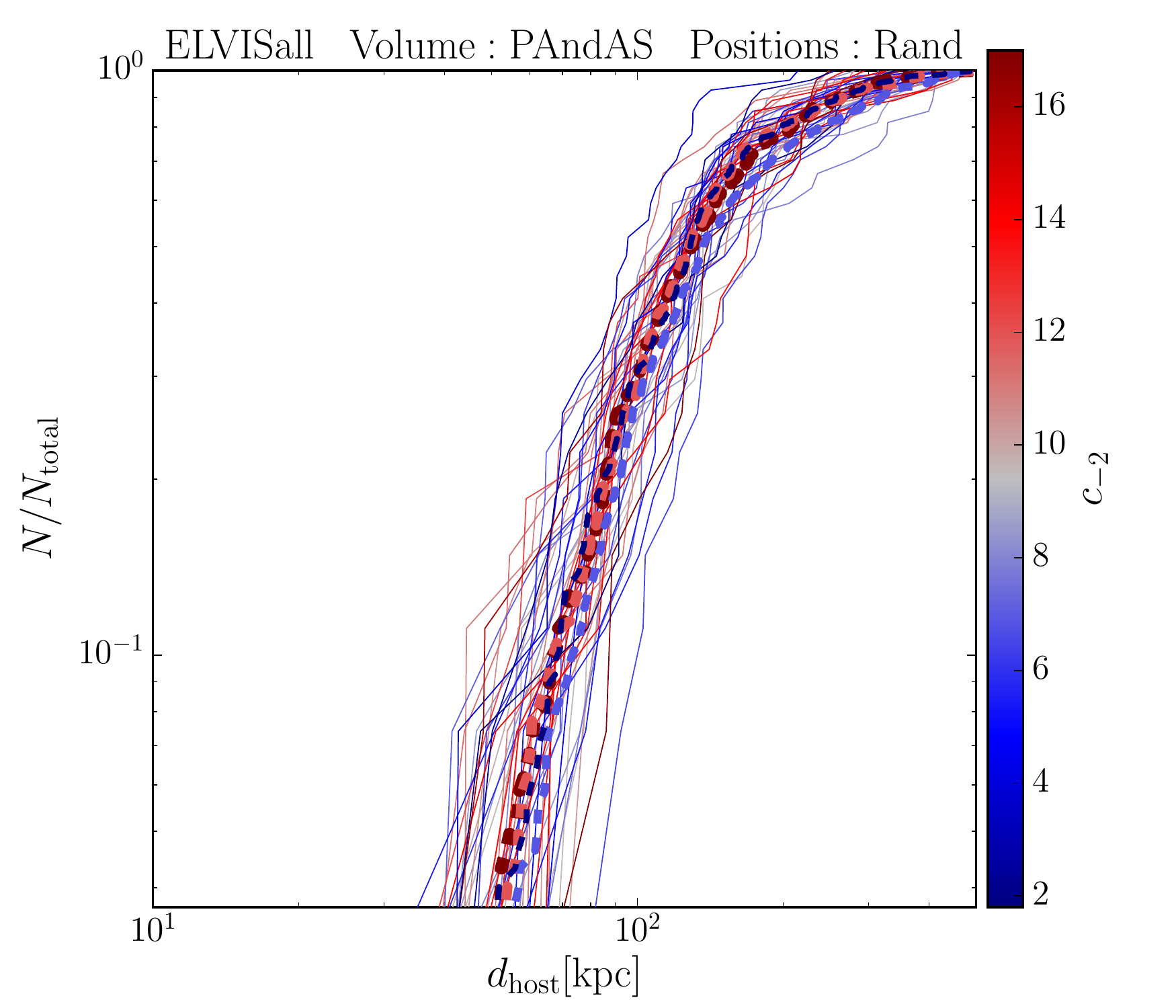}
   \includegraphics[width=80mm]{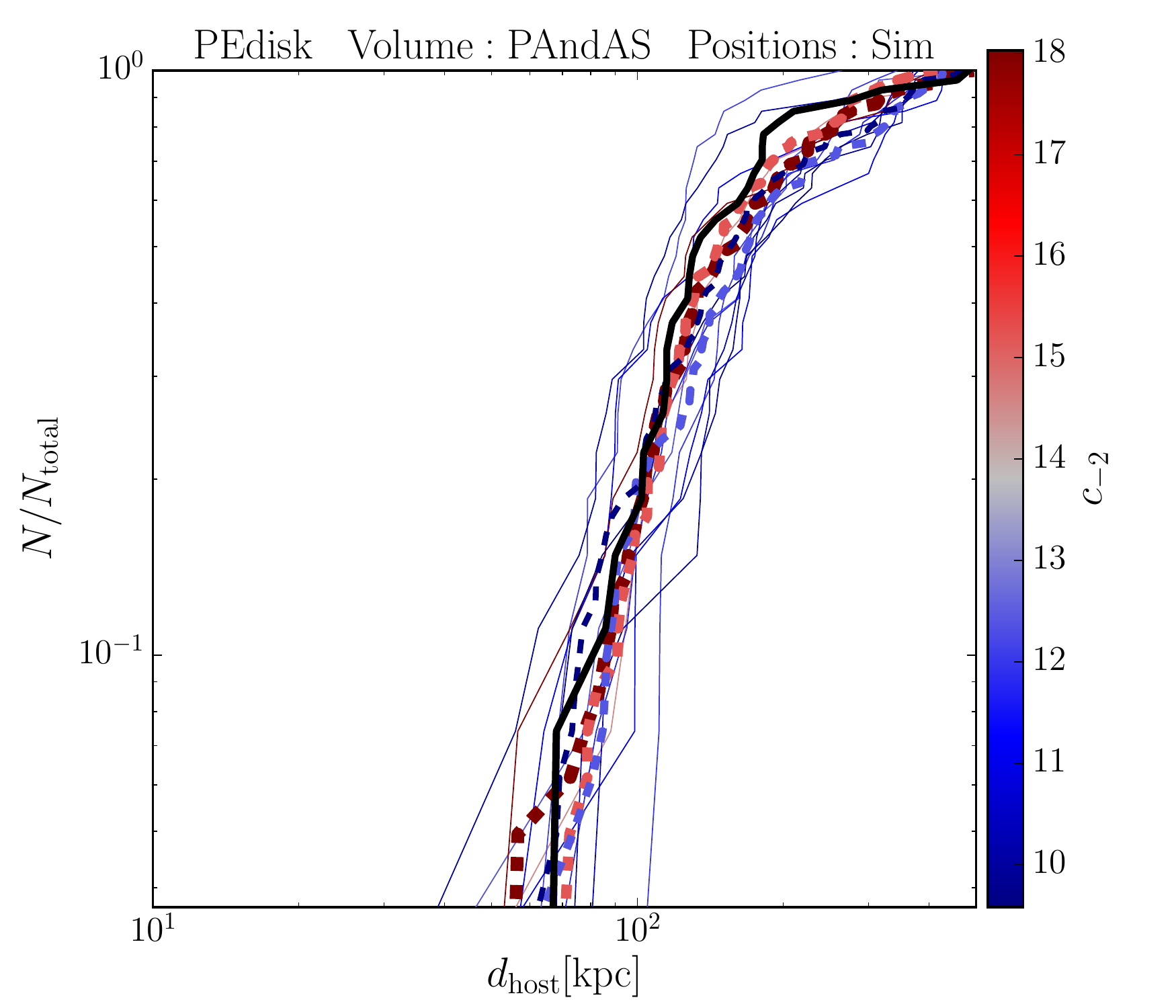}
   \caption{
Radial distribution of sub-halos in the samples color-coded by host halo concentration. The top four panels are for the same sub-halo samples as those in Fig. \ref{fig:planeheights}. The thick dashed lines again give the average radial profiles for different bins in halo concentration. The green lines correspond to those sub-halo distributions that contain planes of 15 satellites that are more narrow than the observed GPoA around M31. They are more radially concentrated than the average of the radial distributions, especially in the inner regions of the distribution. They are also considerably more radially concentrated in the inner regions than the 27 observed M31 satellites, illustrated by the black line in the lower left panel (assuming their most-likely positions from \citealt{Conn2012}). The fifth panel shows the radial distribution of the sub-halos selected from a PAndAS footprint in the Phat ELVIS simulations which include a central disk potential. The sub-halos tend to be found at larger radial distances from the host than in the dmo ELVIS simulations and their radial distribution matches better with the observed one (black line), in particular in the inner regions.
   }
              \label{fig:radialdistr}
\end{figure*}

\begin{figure*}
   \centering
   \includegraphics[width=59mm]{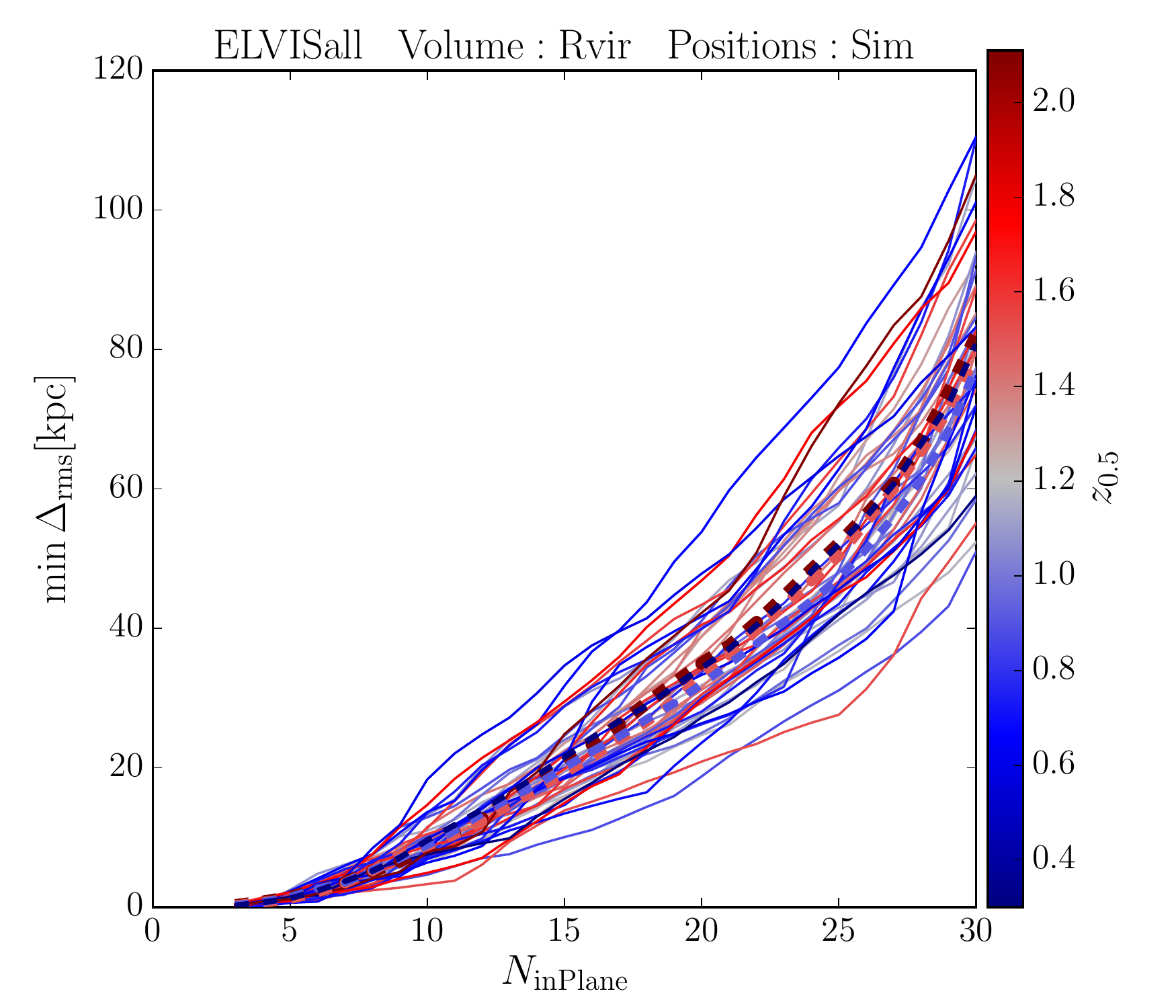}
   \includegraphics[width=59mm]{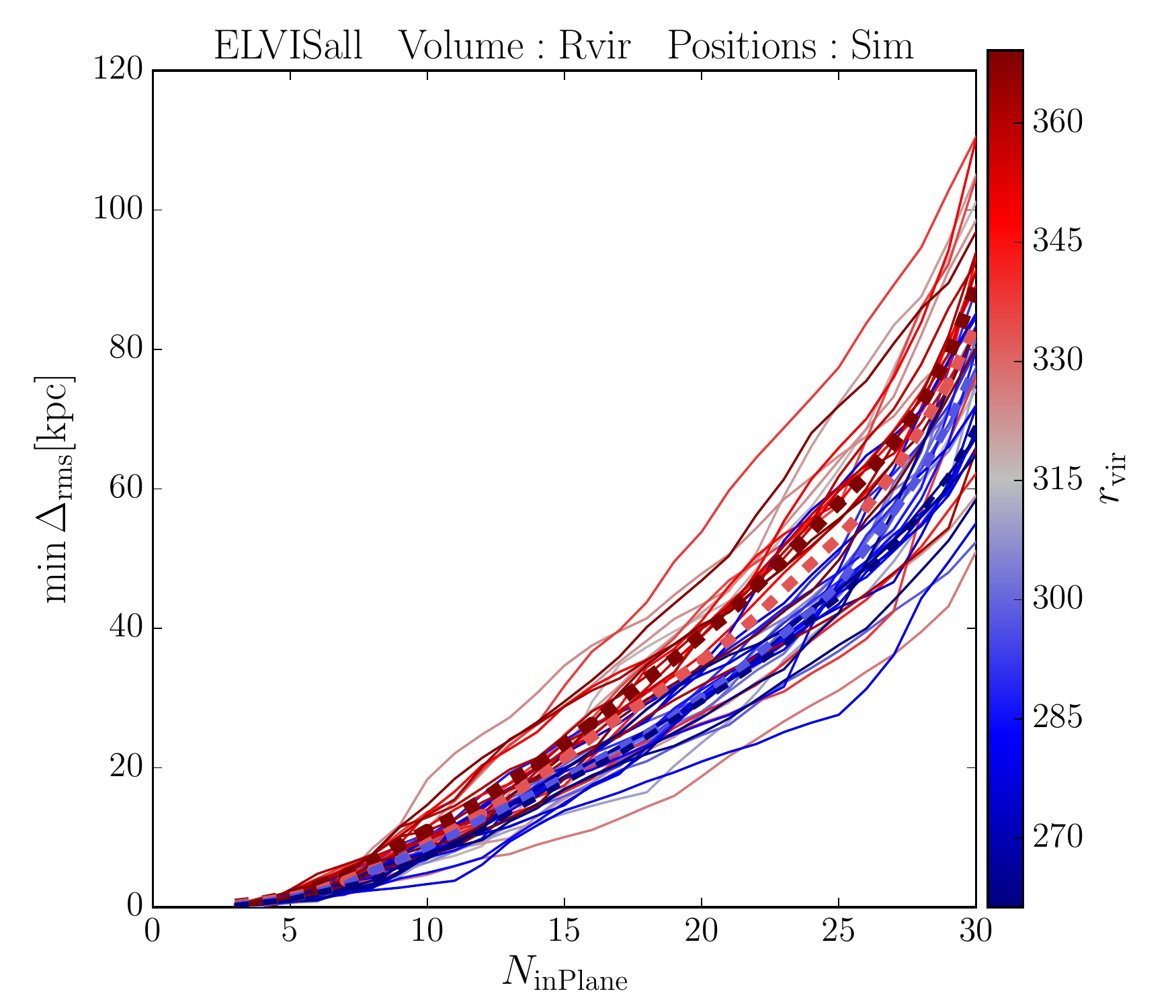}
   \includegraphics[width=59mm]{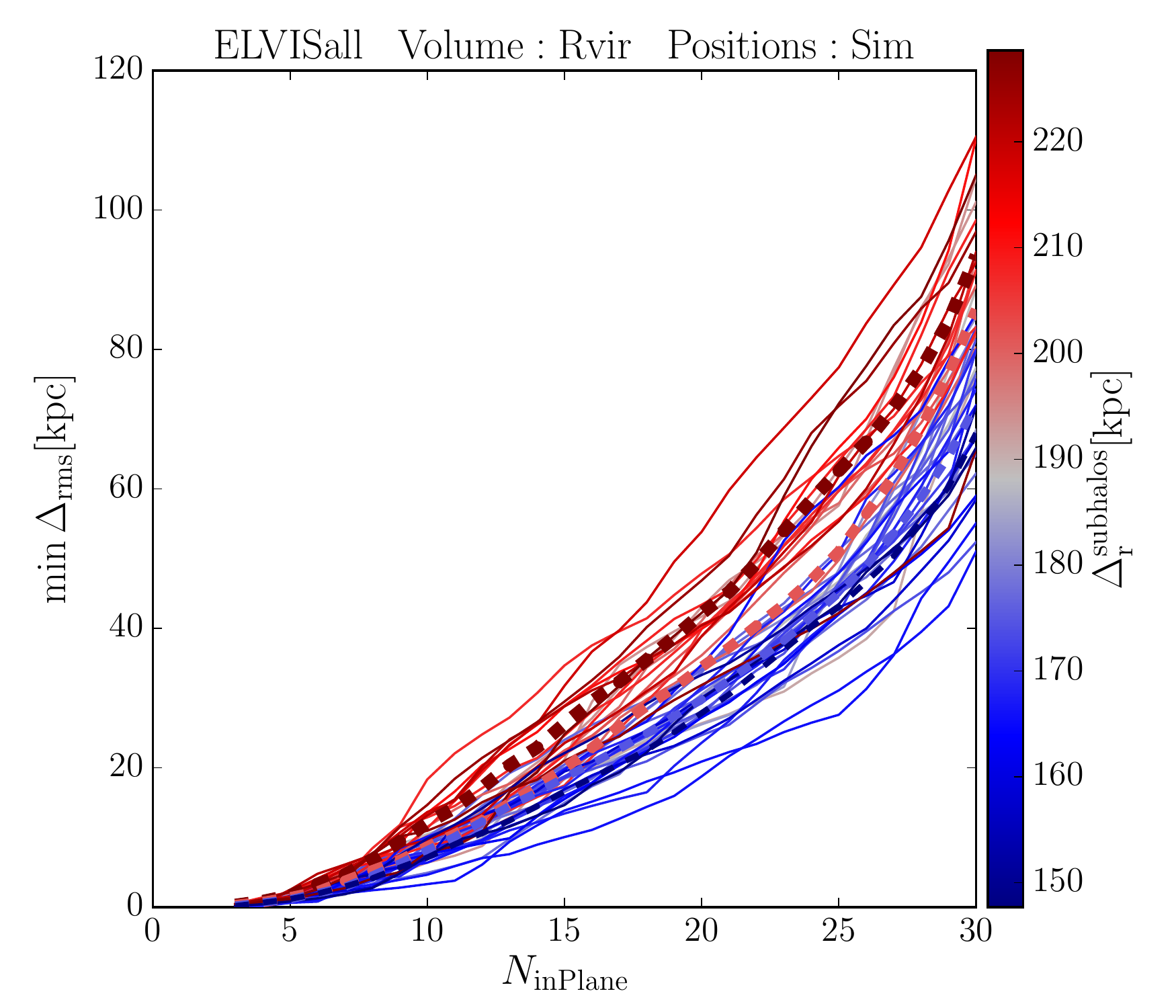}
   \includegraphics[width=59mm]{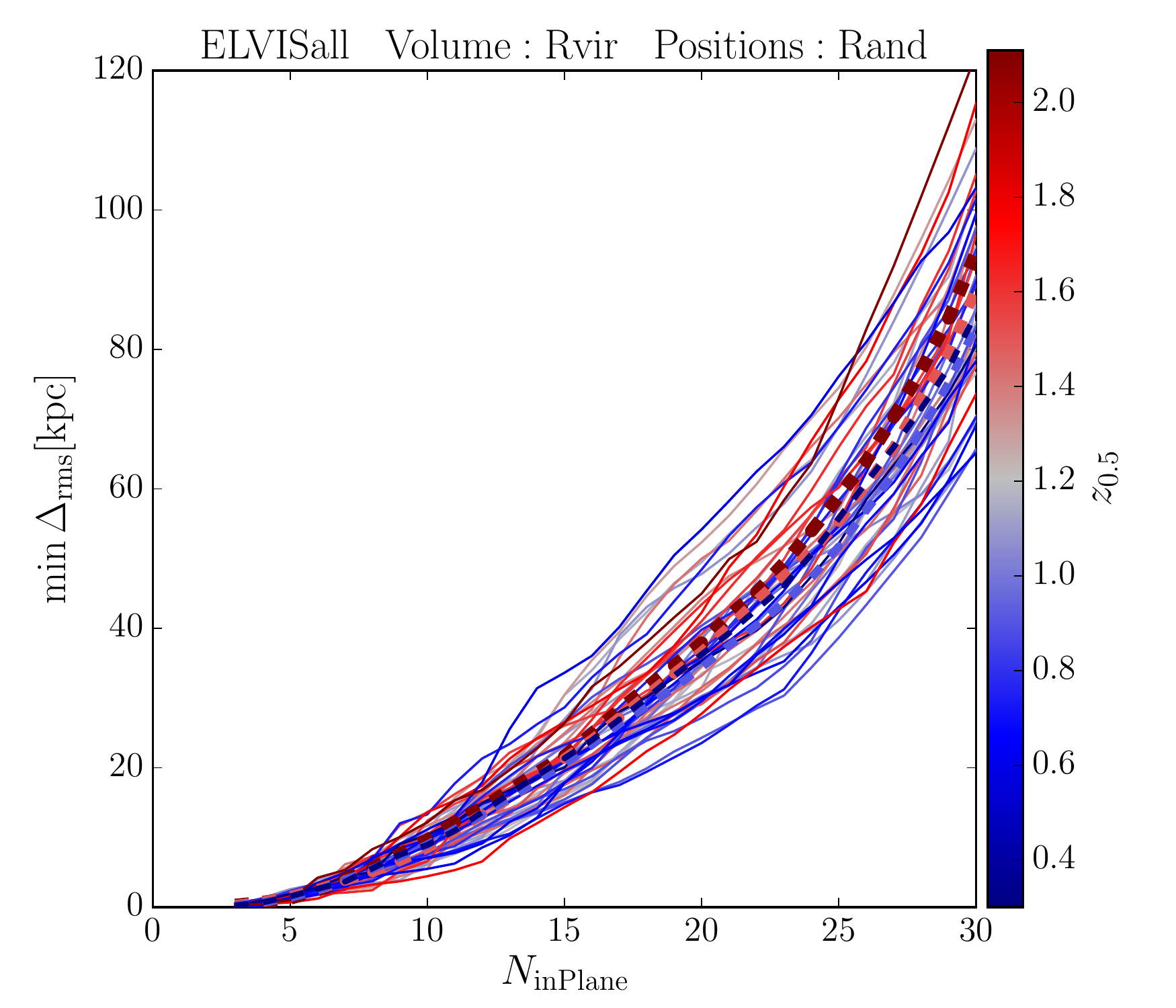}
   \includegraphics[width=59mm]{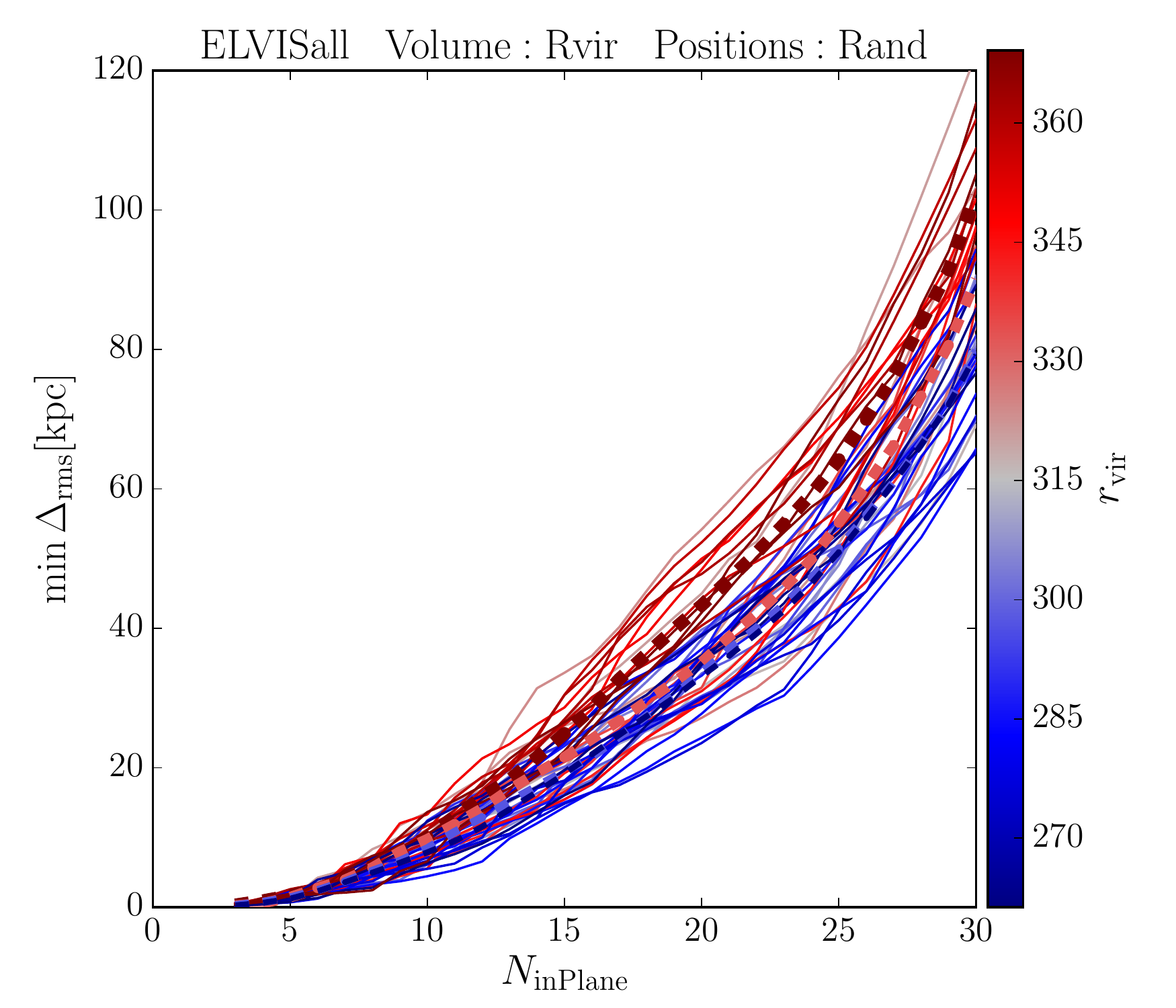}
   \includegraphics[width=59mm]{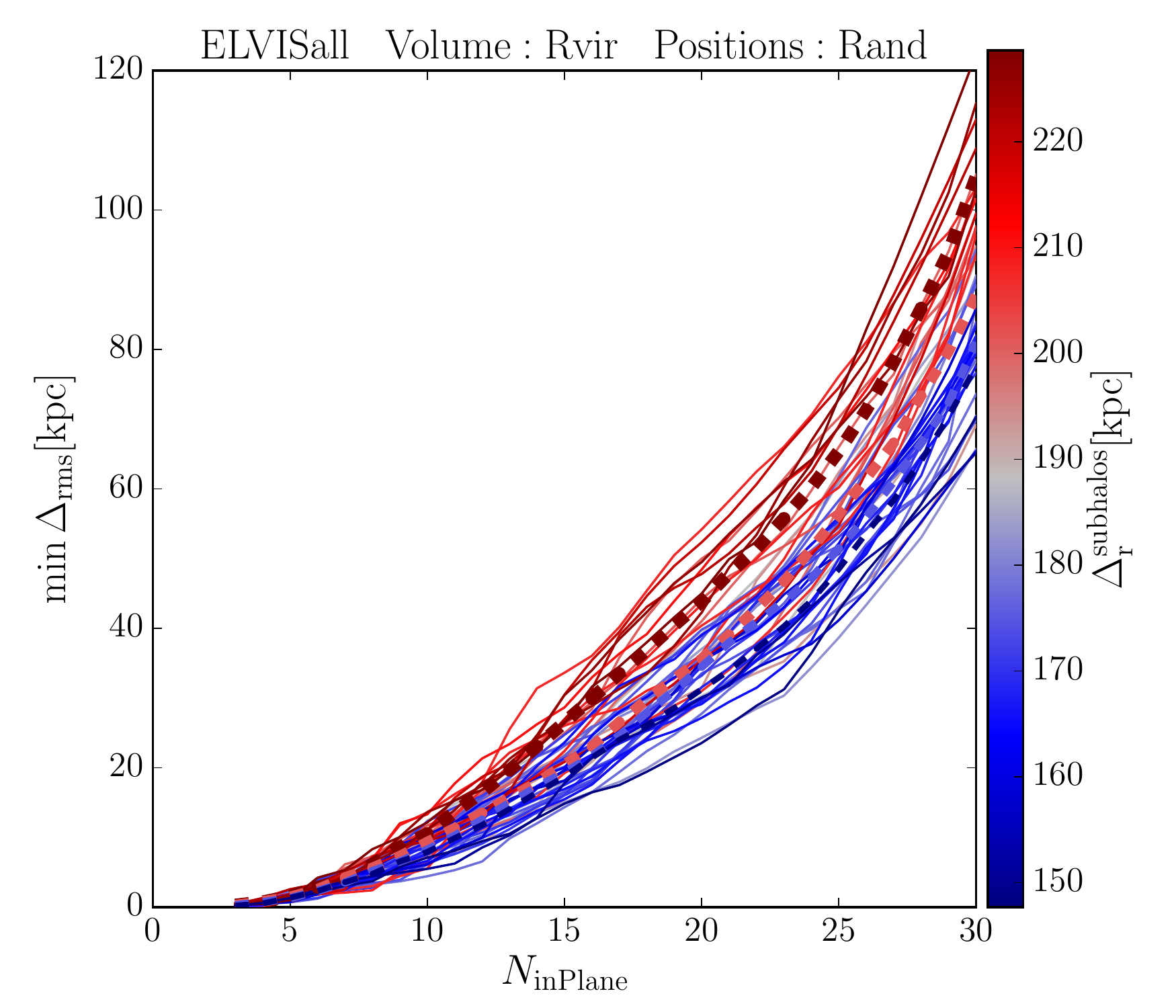}
   \caption{
Same as Fig. \ref{fig:planeheights}, but the lines are now color-coded for host halo formation redshift $z_{\mathrm{0.5}}$, virial radius $r_{\mathrm{vir}}$, and rms radius of the sub-halo system $\Delta_{\mathrm{r}}^{\mathrm{subhalos}}$. The top row gives the results for the 30 top-ranked sub-halos selected from within the respective virial radii, the bottom row gives the sample of randomized positions (but identical radial distances).
The formation redshift does not correlate with the height of the thinnest satellite planes, but there is a clear correlation with the virial radius and the radial distribution of the sub-halo system. These are also present for the samples with randomized positions, indicating that the radial distribution drives the differences in plane heights, and not an additional positional or kinematic coherence within the simulated sub-halo systems of these hosts.
   }
              \label{fig:planeheightsother}
\end{figure*}

Figure \ref{fig:planeheights} presents the minimum plane height for each possible number of satellites in a plane $N_\mathrm{inPlane}$. Each line corresponds to one host's sub-halo system, and the lines are color-coded by the concentration of the host halo. While some of the halos hosting the thinnest planes have higher concentrations, no clear trend between the halo concentration and the minimum plane thickness is apparent. We would have expected to see a gradient of higher concentrations for lines at lower $\min \Delta_{\mathrm{rms}}$\ at a given satellite number.

There are, however, a few minor indications that can be interpreted to be consistent with the findings of \citet{Buck2015a}. Three of the halos do in fact contain satellite planes of $N_\mathrm{inPlane} = 15$\ that are at least as narrow as the 99 per cent upper limit of the GPoA around M31\footnote{We would like to stress again that due to the very different selection volume, this does not warrant to claim consistency between the simulations and the observed M31 satellite system. Furthermore, to resemble the GPoA these satellite planes would simultaneously need to show the same degree of kinematic coherence, which they do not (see Sect. \ref{subsect:kinematics} below).} ($\Delta_\mathrm{rms} \leq 14.1\,\mathrm{kpc}$). Two of theses three halos seem to be of higher-than-average concentration, too, while the two halos with the largest $\min \Delta_{\mathrm{rms}}$\ at $N_\mathrm{inPlane} = 15$\ are of lower concentration. In addition, the average of the minimum plane heights for the 12 most-concentrated hosts (thick dashed red line) tends to be slightly below that of the 12 least-concentrated hosts (thick dashed blue line).

However, before we jump to this conclusion we need to take a step back and compare the findings with our randomized control sample (upper right panel in Fig. \ref{fig:planeheights}). Interestingly, the randomized sample also has a high-concentration halo at lowest $\min \Delta_{\mathrm{rms}}$. Like in the simulated sample, the two most-pronounced high-$\min \Delta_{\mathrm{rms}}$\ halos up to $N_\mathrm{inPlane} = 20$\ have low concentration, too. Furthermore, even the average $\min \Delta_{\mathrm{rms}}$\ for the 12 most- and least-concentrated hosts show the same behavior as in the simulated sample, with the more concentrated ones having slightly lower $\min \Delta_{\mathrm{rms}}$. All indications for a possible correlation between host halo concentration and the height of satellite planes previously discussed for the simulated sample are therefore also present in a sample of satellites drawn from isotropy. This randomized sample does not have any formation history that could be responsible for this. The similarities, therefore, are hints that indeed the radial distributions of the sub-halos might be most responsible for driving $\min \Delta_{\mathrm{rms}}$, since this is the only property that the simulated and randomized samples have in common.

More support for this interpretation can be found in the radial profiles of the sub-halo systems (Figure \ref{fig:radialdistr}). The systems belonging to the lowest-concentration hosts tend to be slightly less radially concentrated, especially in the inner parts (compare the dashed red and blue lines, which show the average radial profiles for the high- and low-concentration halos, respectively). More importantly, those systems that contain planes of 15 satellites which are at least as narrow as the GPoA (underlaid with a thicker green line) are more radially concentrated than the average sub-halo distributions in high-concentration halo. This confirms that the radial concentration is the factor driving the tendency to lower minimum plane heights, but that this is only due to an overall more compact scaling of the satellite system which is not accounted for when measuring plane heights in absolute distance.

The PAndAS samples do not even show a weak indication of a correlation between $\min \Delta_{\mathrm{rms}}$\ and $c_{\mathrm{-2}}$\ in Fig. \ref{fig:planeheights}, which demonstrated that the selection volume is an important aspect of any comparison between simulations and observations. None of the 48 ELVIS halos contains a sub-halo plane that is as narrow as the observed GPoA. Figure \ref{fig:radialdistr} reveals that the PAndAS volume affects the typical radial distribution of selected sub-halos compared to the spherical volumes of radius $r_{\mathrm{vir}}$. It results in a somewhat steeper inner slope within $d_\mathrm{host} \lesssim 150\,kpc$\ and then a more extended tail to larger distances, the latter because the selection function allows sub-halos to be found up to 500\,kpc from the host if they lie within the PAndAS footprint, instead of cutting off at the virial radius. Note that the radial profile of the 27 observed M31 satellites among which \citet{Ibata2013} discovered the GPoA (black line in Fig. \ref{fig:radialdistr}) is \textit{less} radially concentrated the vast majority of sub-halo systems, especially in the inner regions of the halos. This is another indication that it is the radial sub-halo distribution that drives $\min \Delta_{\mathrm{rms}}$\ to lower values, not the presence of actual satellite planes. In this regard it is important to note that the analysed simulations are collision-less, and that the host halos do not contain central galaxy disks, whose tidal effects can result in a depletion of sub-halos in the innermost regions which results in more radially extended sub-halo systems \citep{GarrisonKimmel2017, Sawala2017}. This effect on the distribution of satellite sub-halos is present in the Phat ELVIS sample that includes a central galaxy potential. As a consequence, for this set of simulations we find a better match with the observed radial distribution of M31 satellite galaxies, provided that the sub-halos are selected from the PAndAS footprint (see Sect. \ref{sect:newtests}).

\citet{Buck2015a} have concluded that the narrowness of satellite planes correlates with halo formation time, in the sense that earlier forming halos contain more narrow planes. They based this conclusion on using halo concentration as a proxy for formation time, and argued to have found an anti-correlation between plane height and halo concentration. We can not reproduce the reported correlation between host concentration and the width of satellite planes, which appears to invalidate this line of argument. Thus, in Figure \ref{fig:planeheightsother} we more directly check for a possible correlation with halo formation time, by comparing to the halo formation redshift $z_{\mathrm{0.5}}$\ (left panels). Maybe unsurprisingly, no correlation is apparent between $z_{\mathrm{0.5}}$\ and the minimum plane heights either. However, color-coding by the virial radius $r_{\mathrm{vir}}$\ of the host (middle panels) or the rms radius of the sub-halo sample $\Delta_{\mathrm{r}}^{\mathrm{subhalos}}$\ (right panels) reveals a clear gradient (and thus correlation) which confirms our preceding discussion: systems with smaller total radial extent result in apparently more narrow satellite planes if measured by absolute rms height.

\subsection{Average halo properties}

\begin{figure*}
   \centering
   \includegraphics[width=44mm]{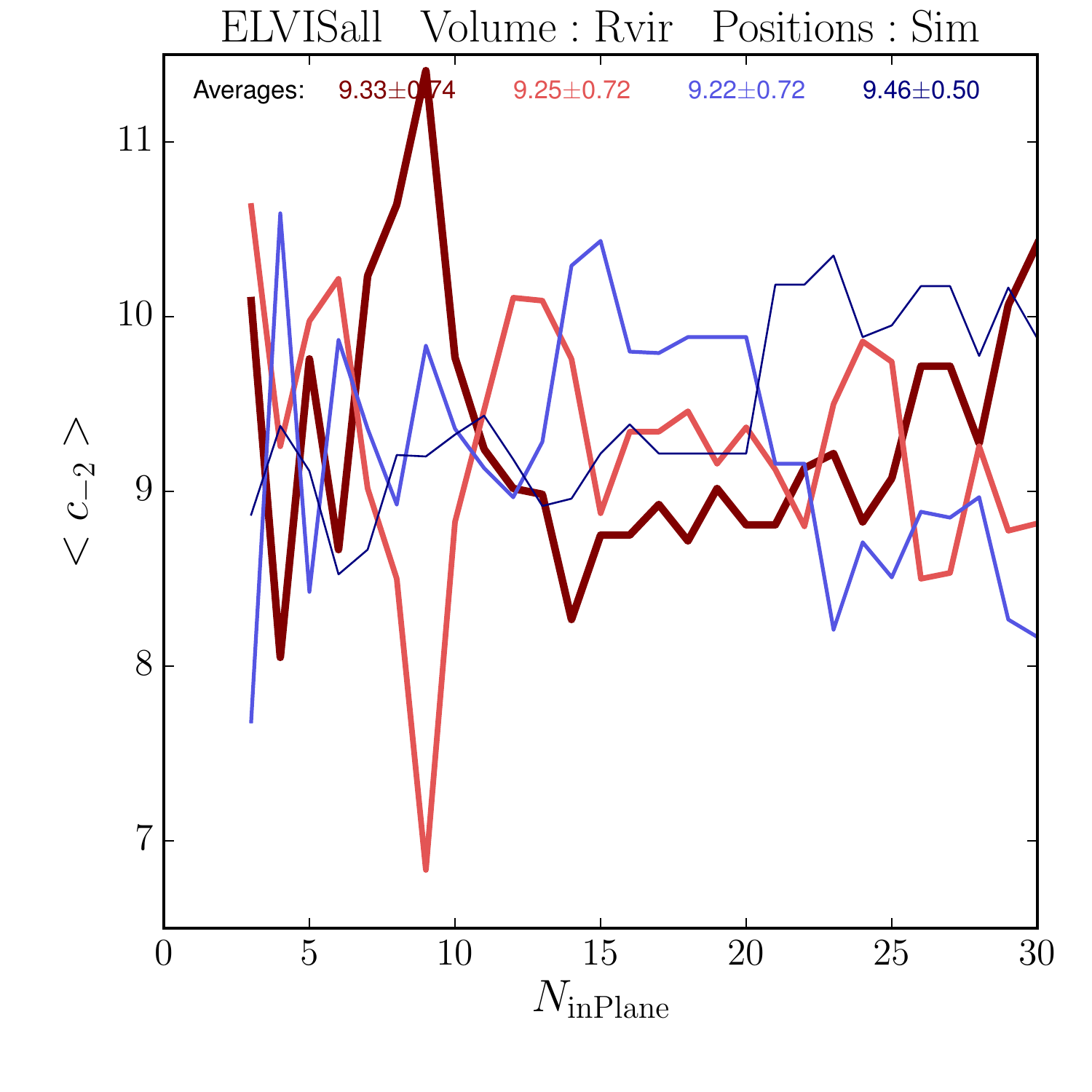}
   \includegraphics[width=44mm]{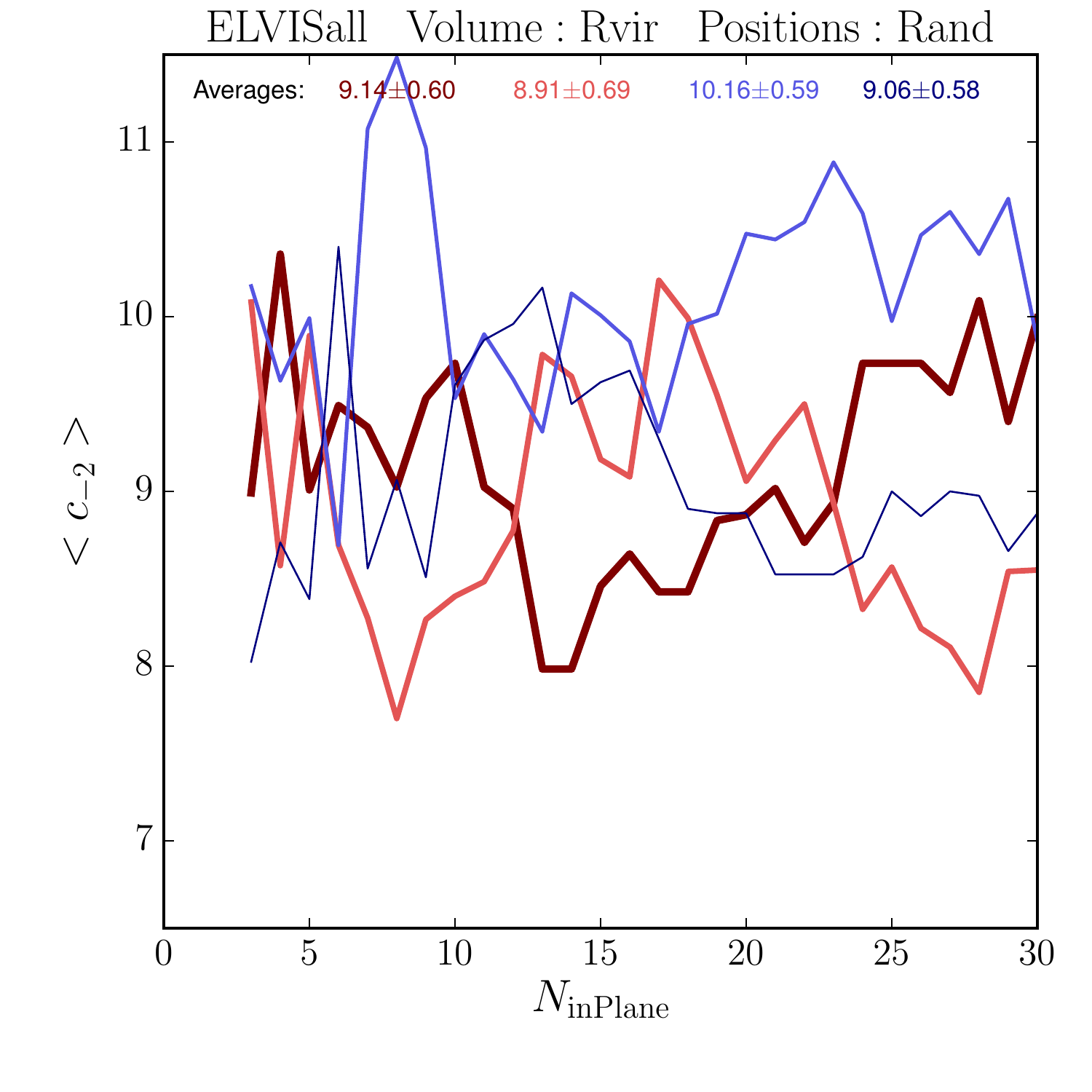}
   \includegraphics[width=44mm]{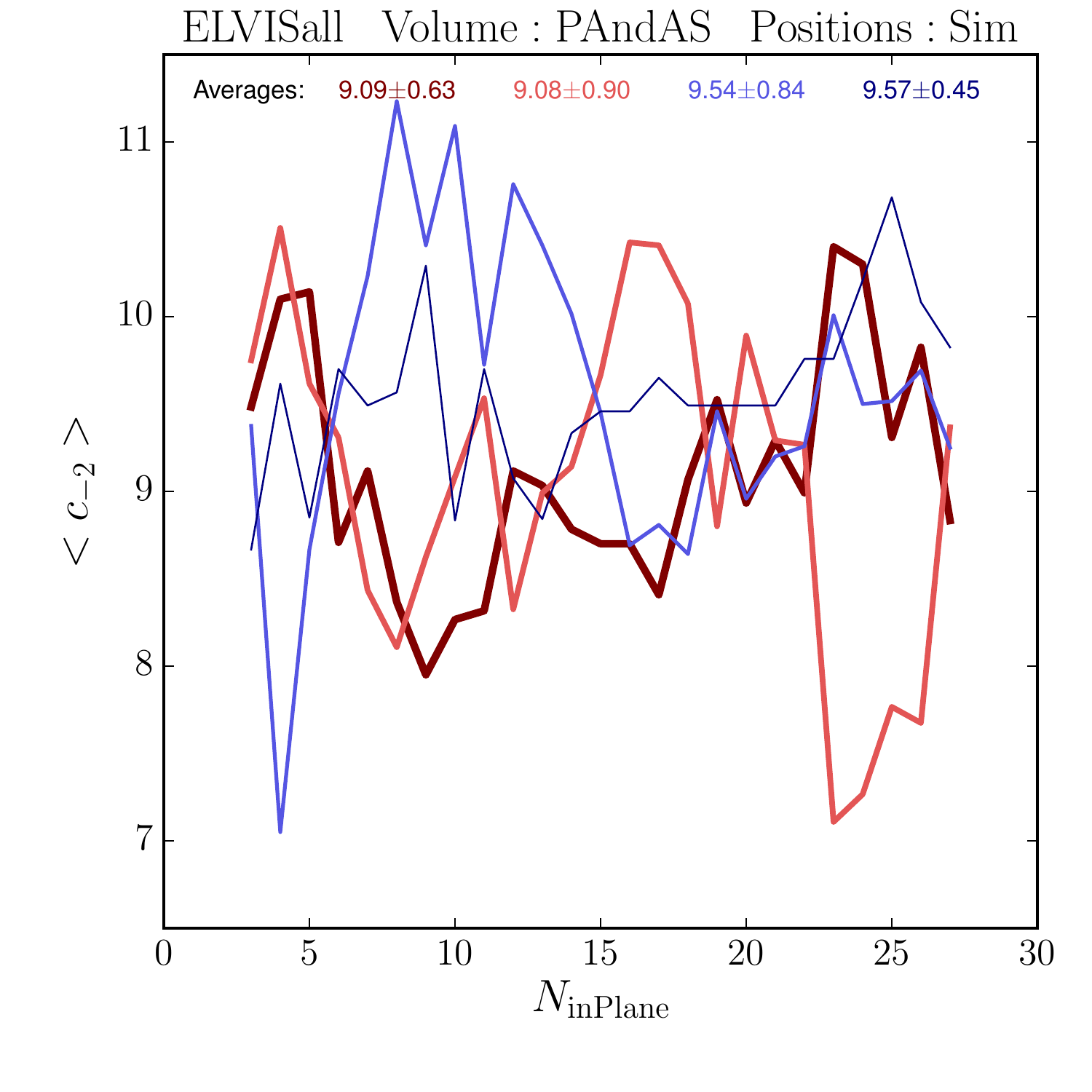}
   \includegraphics[width=44mm]{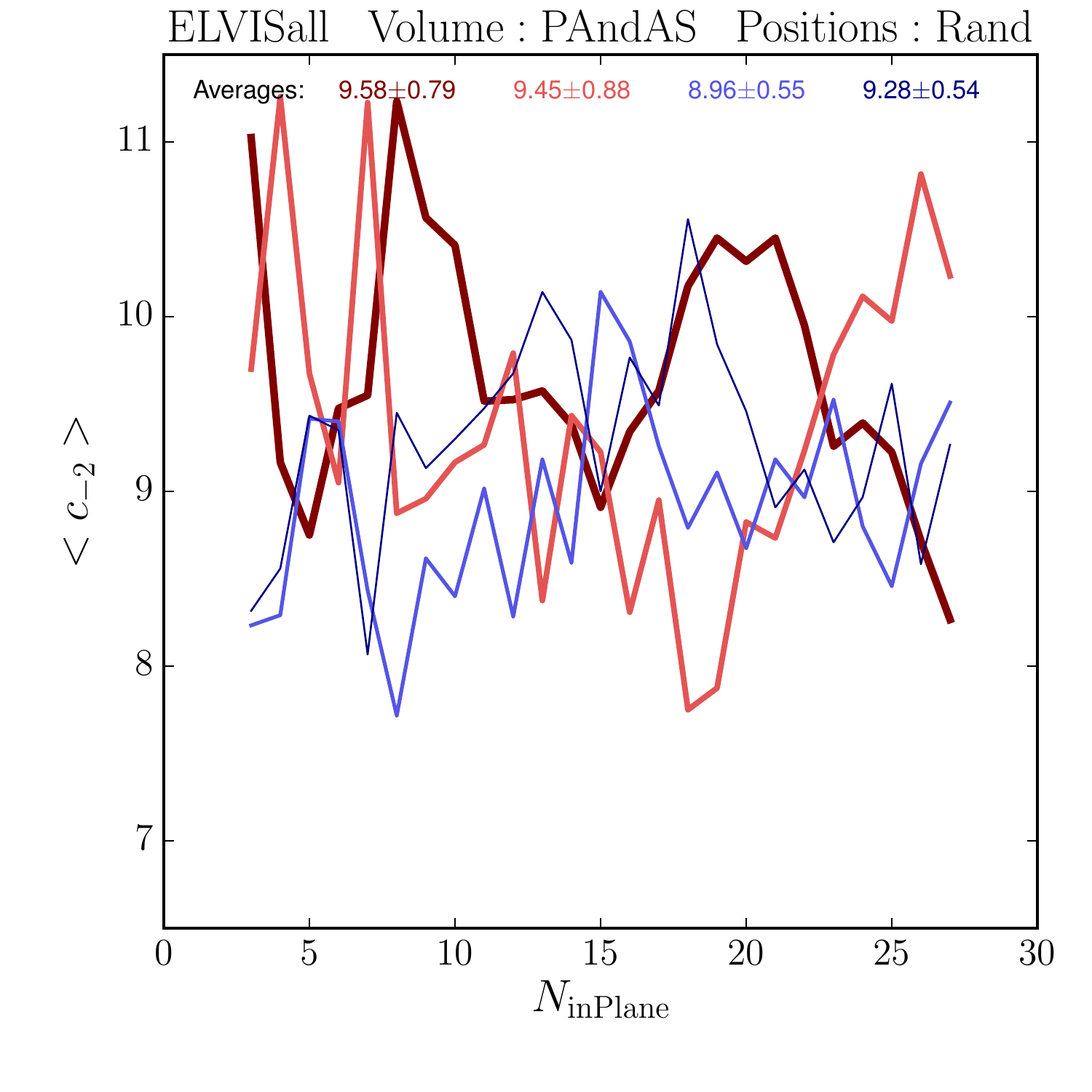}
   \includegraphics[width=44mm]{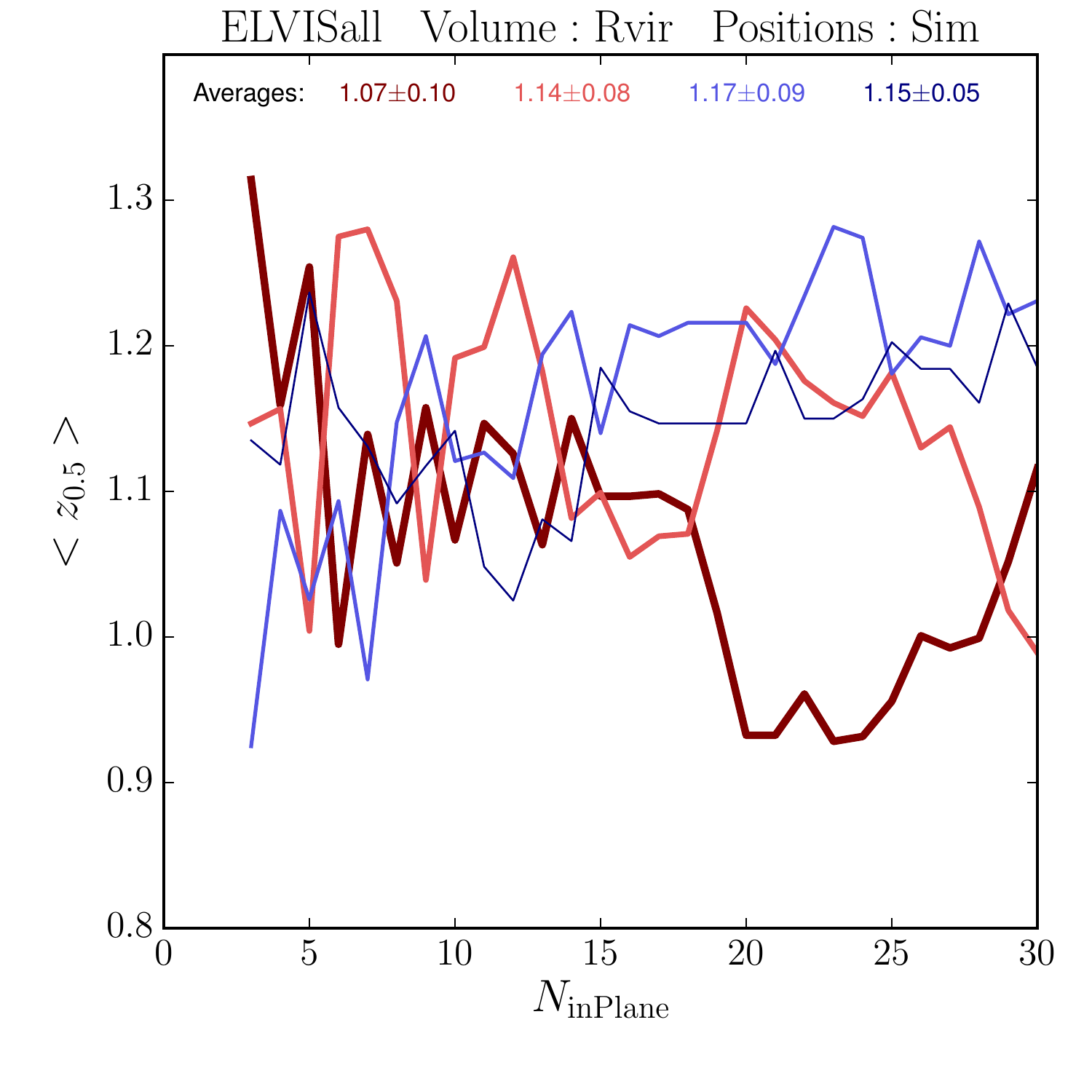}
   \includegraphics[width=44mm]{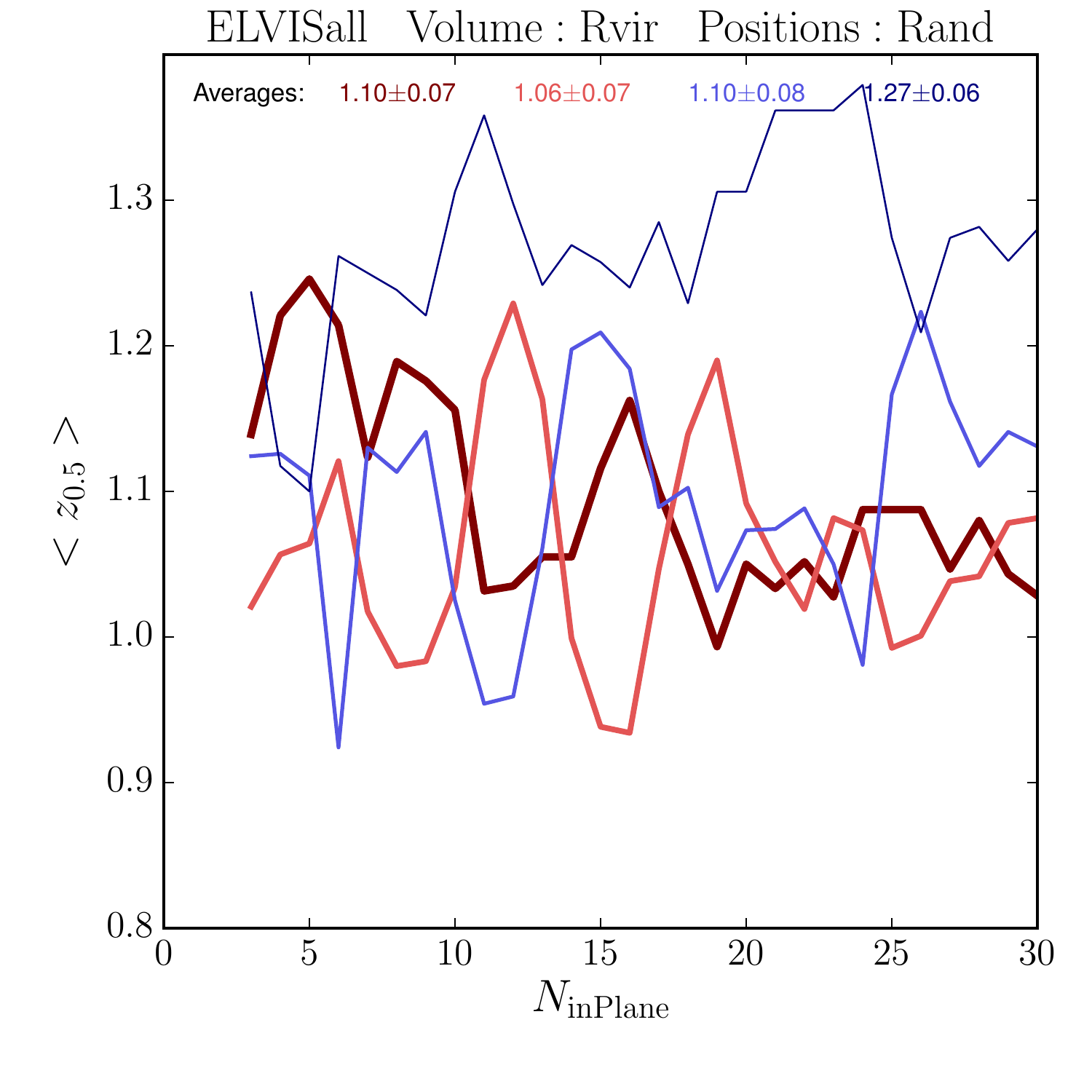}
   \includegraphics[width=44mm]{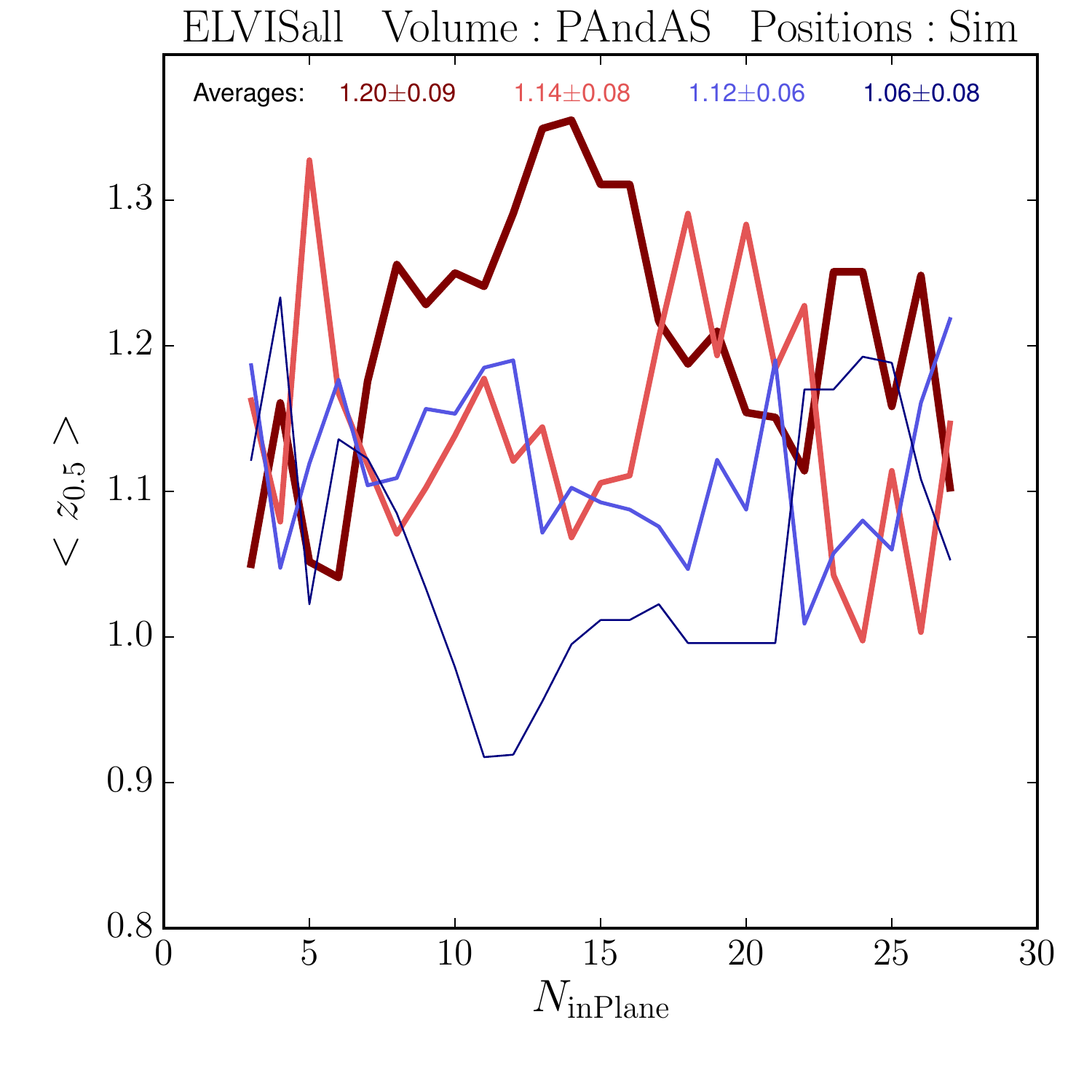}
   \includegraphics[width=44mm]{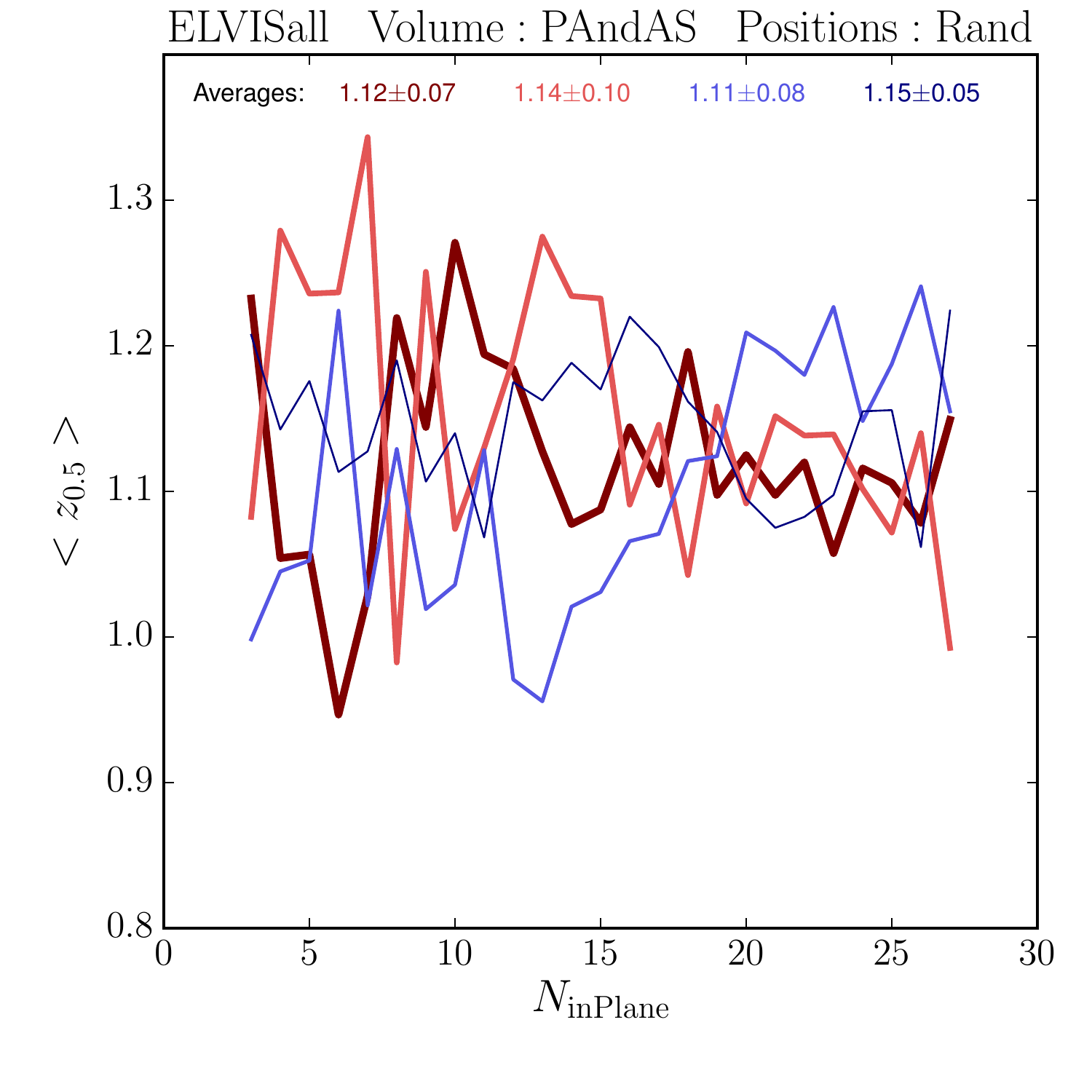}
   \includegraphics[width=44mm]{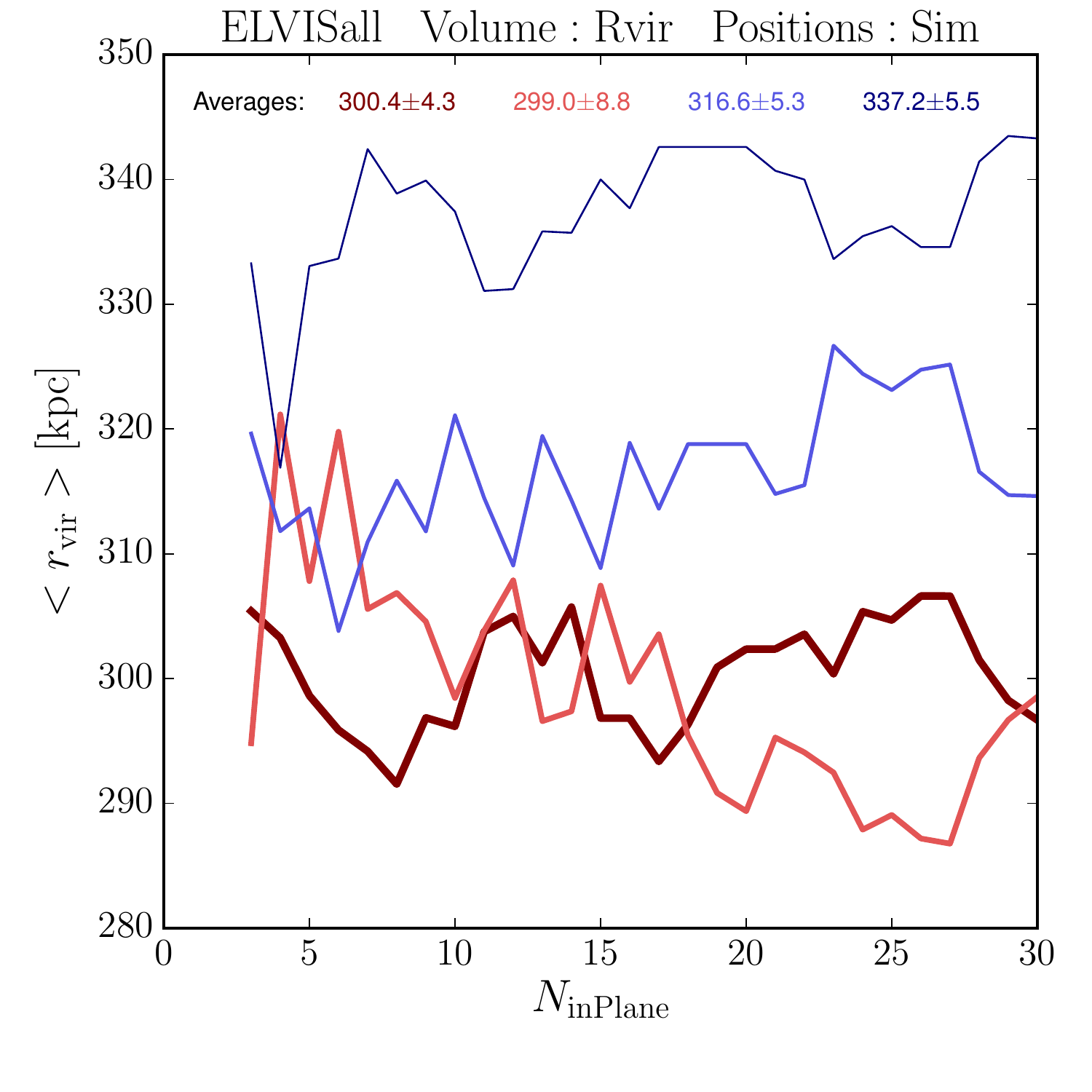}
   \includegraphics[width=44mm]{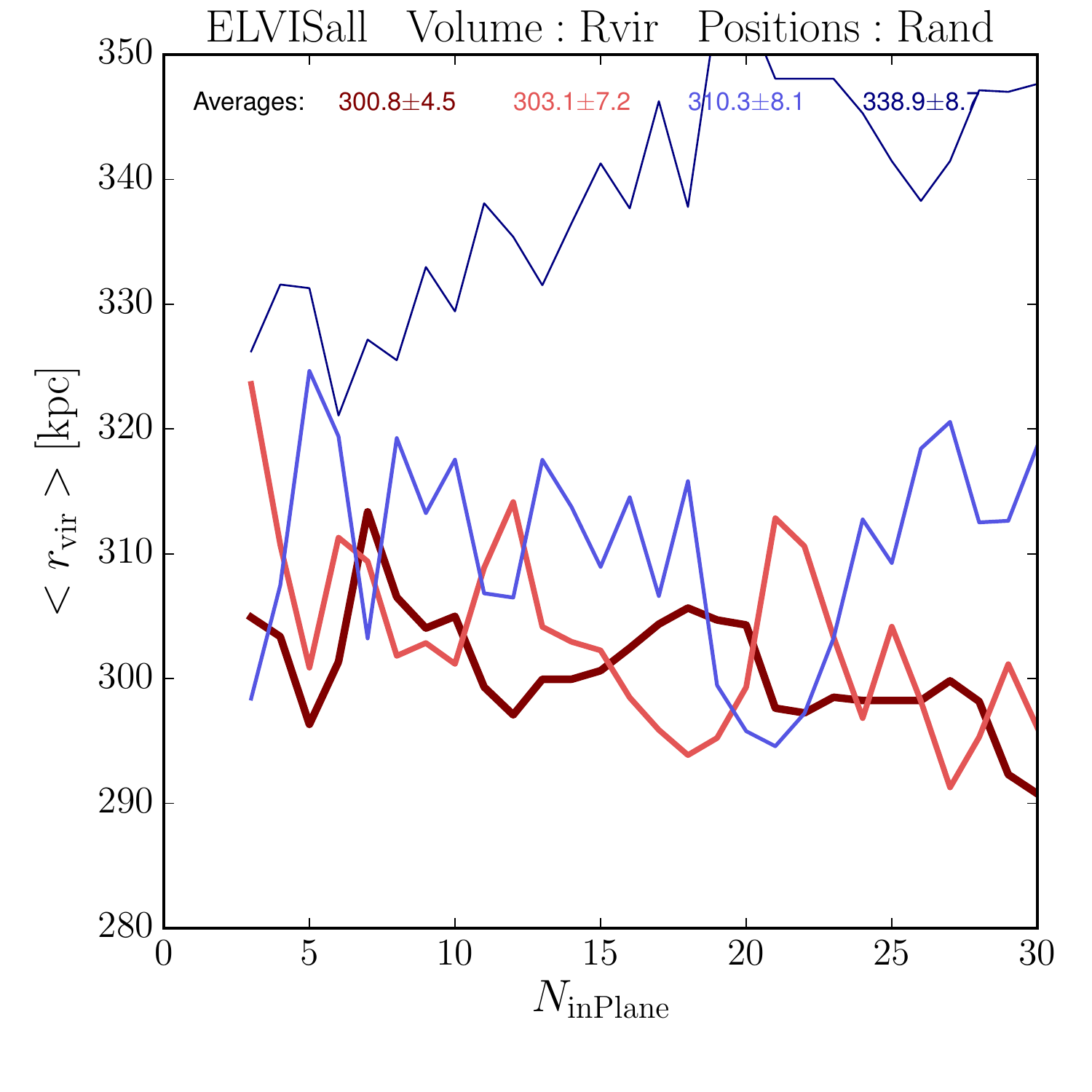}
   \includegraphics[width=44mm]{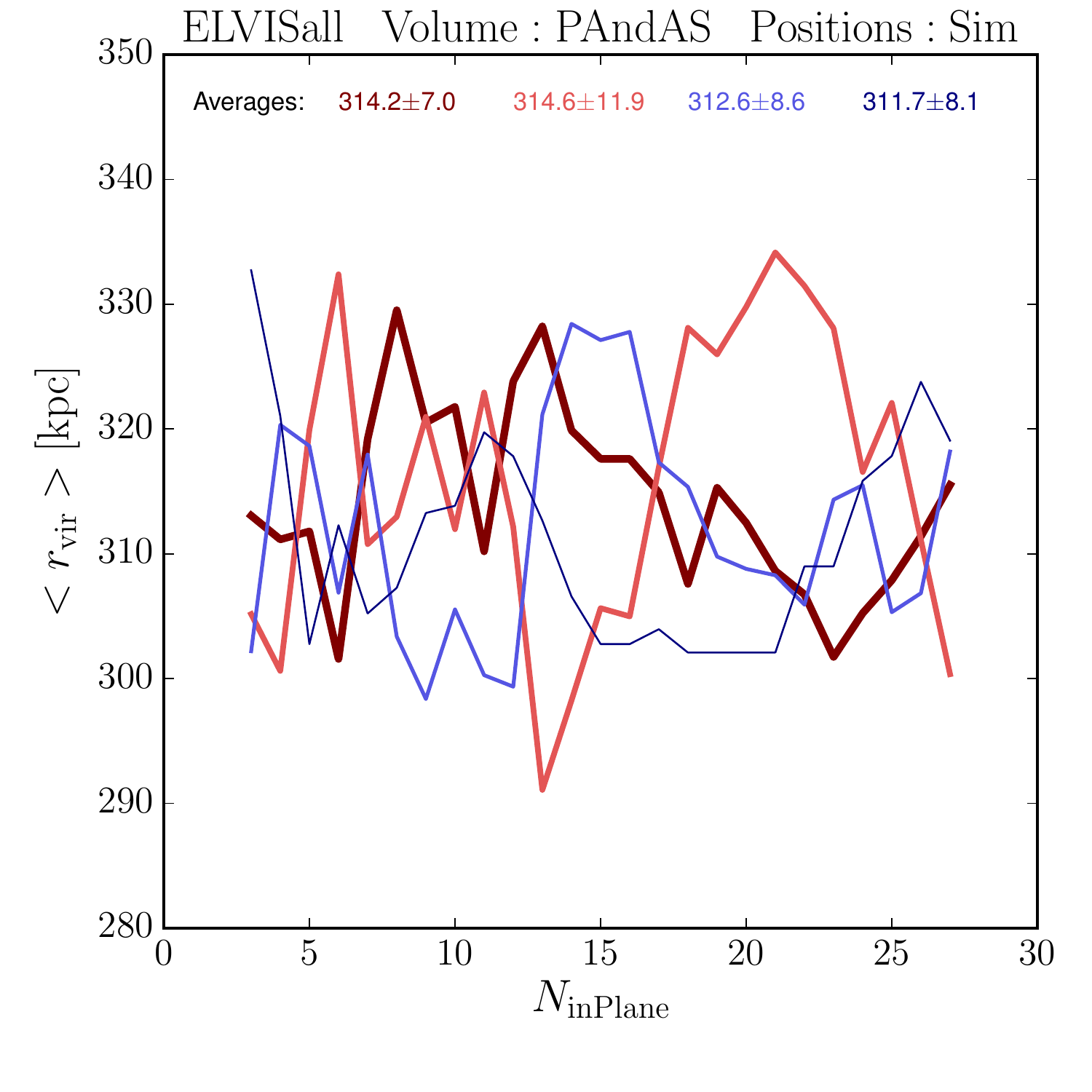}
   \includegraphics[width=44mm]{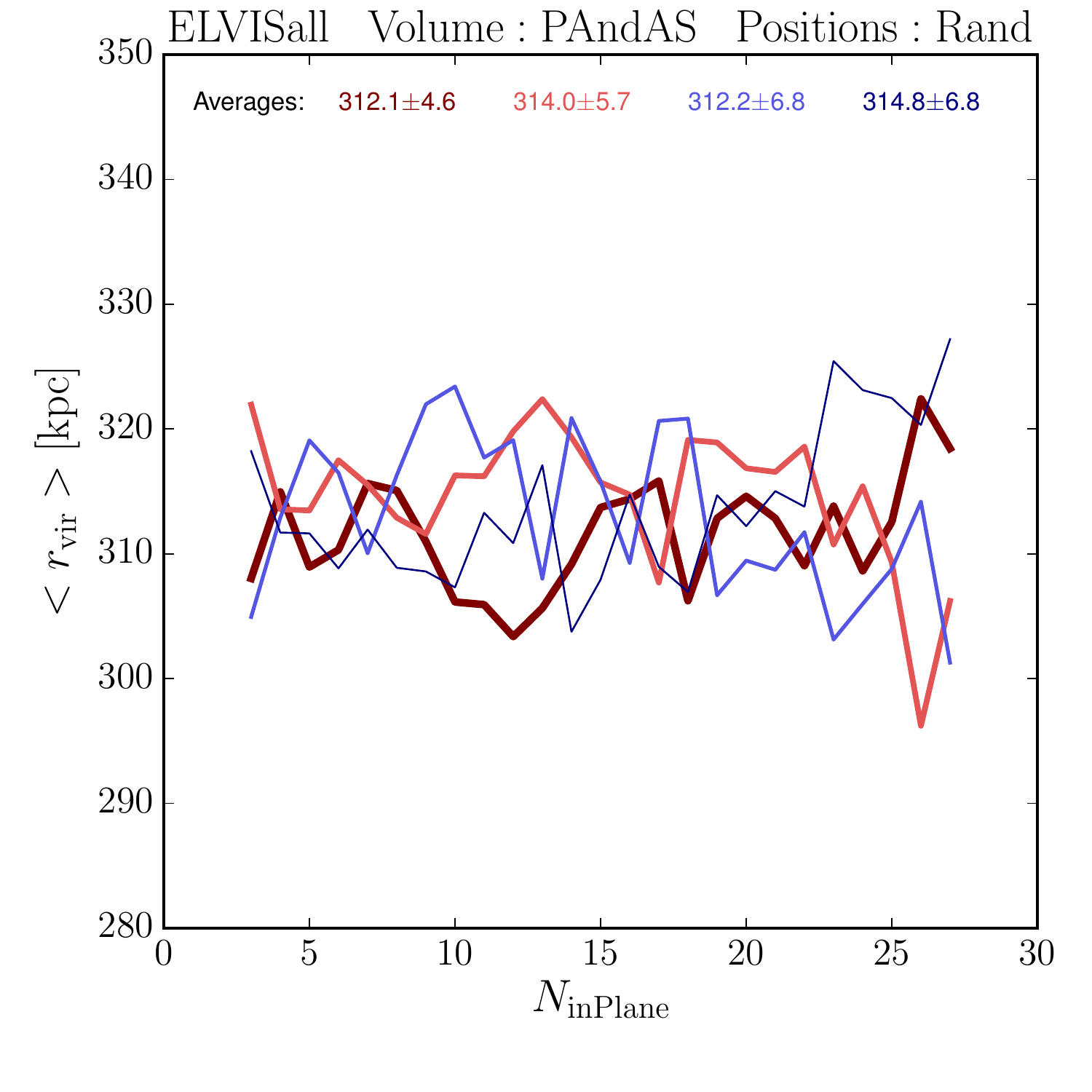}
   \includegraphics[width=44mm]{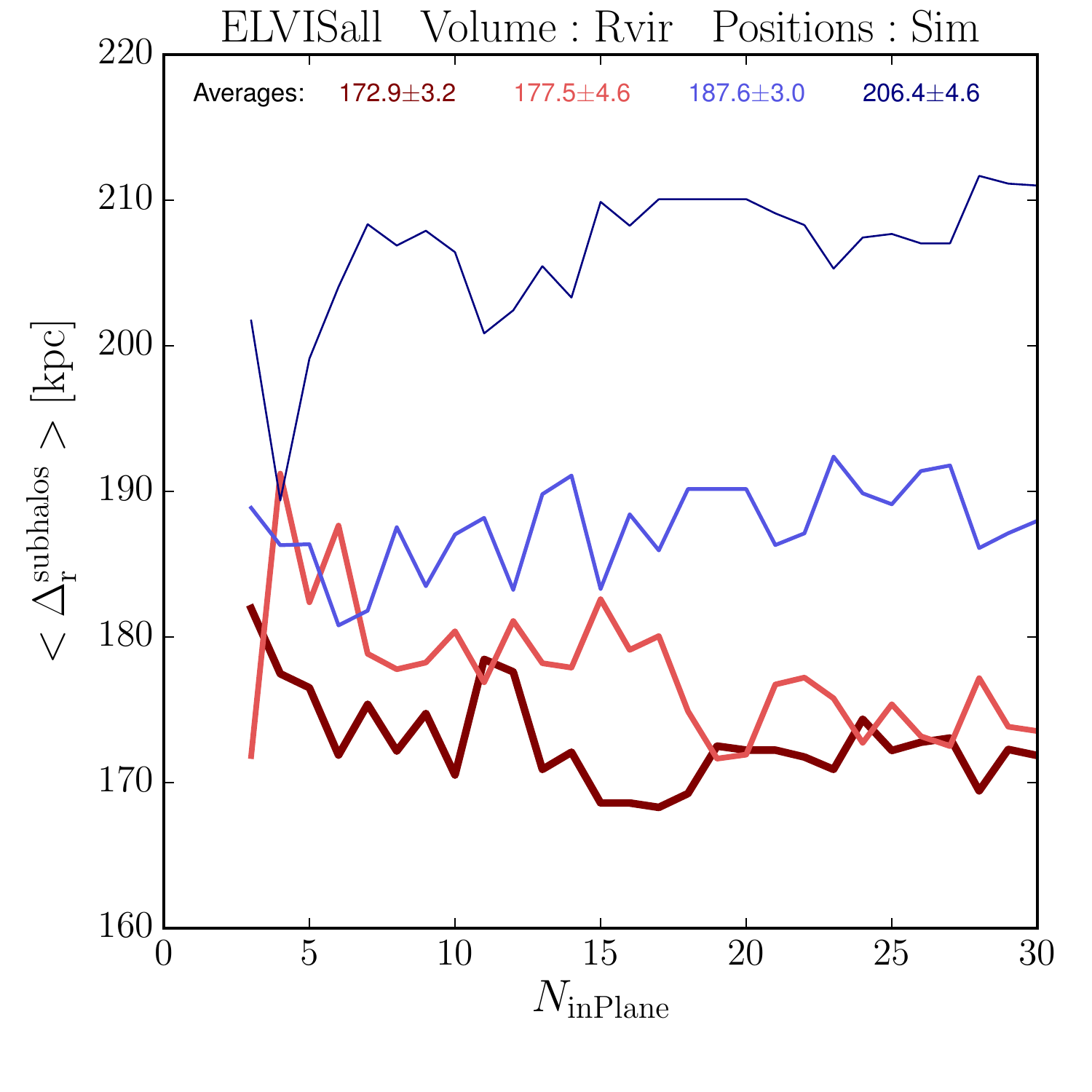}
   \includegraphics[width=44mm]{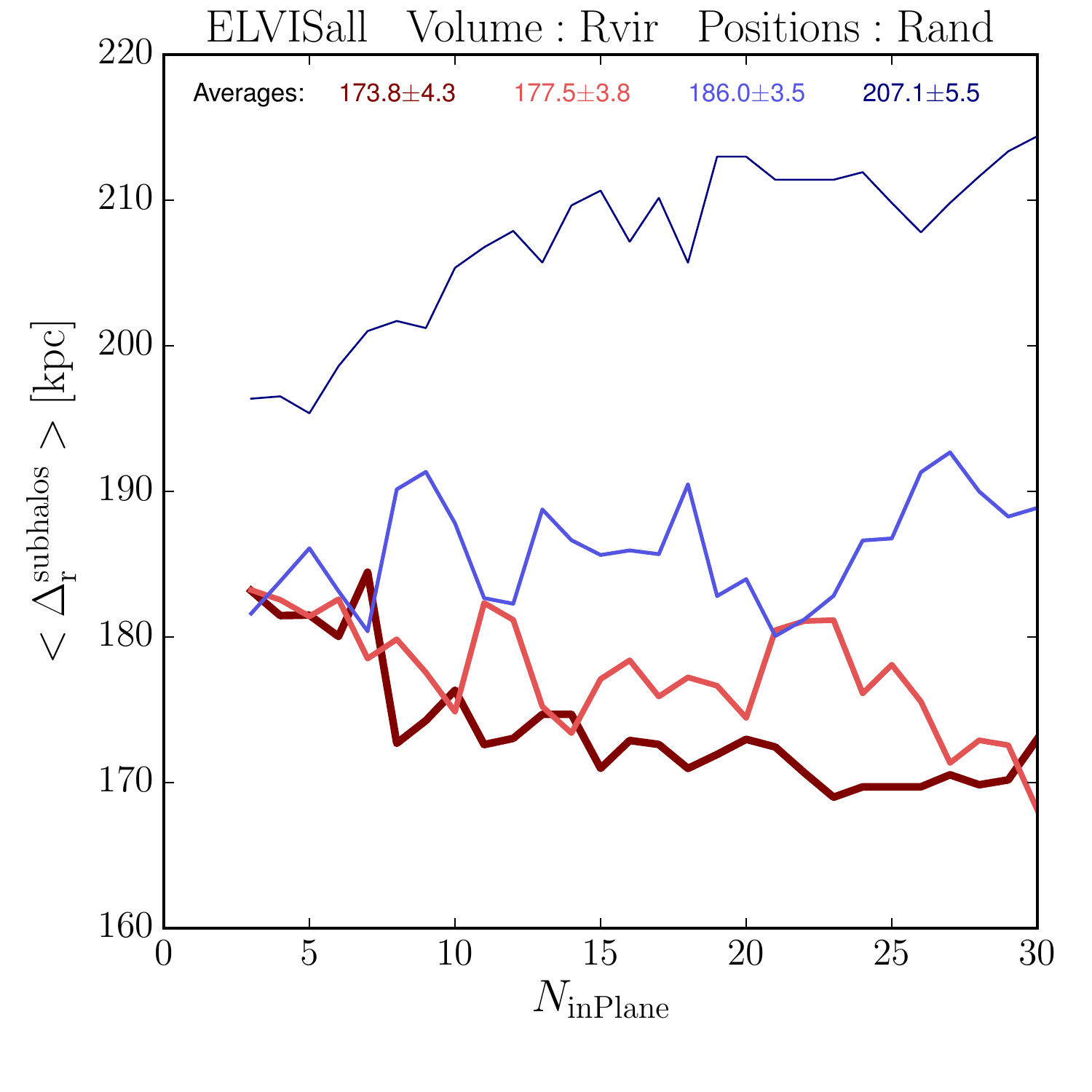}
   \includegraphics[width=44mm]{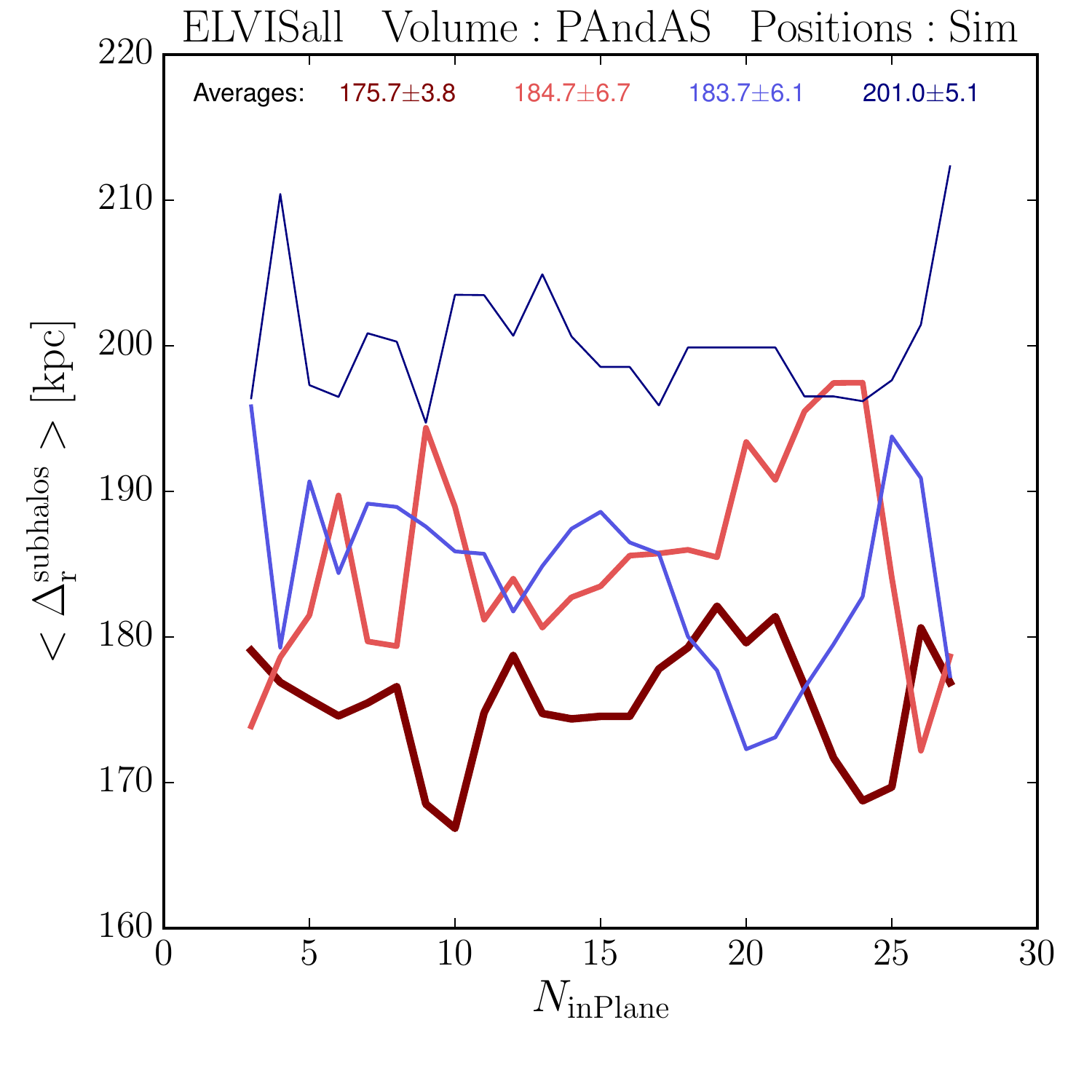}
   \includegraphics[width=44mm]{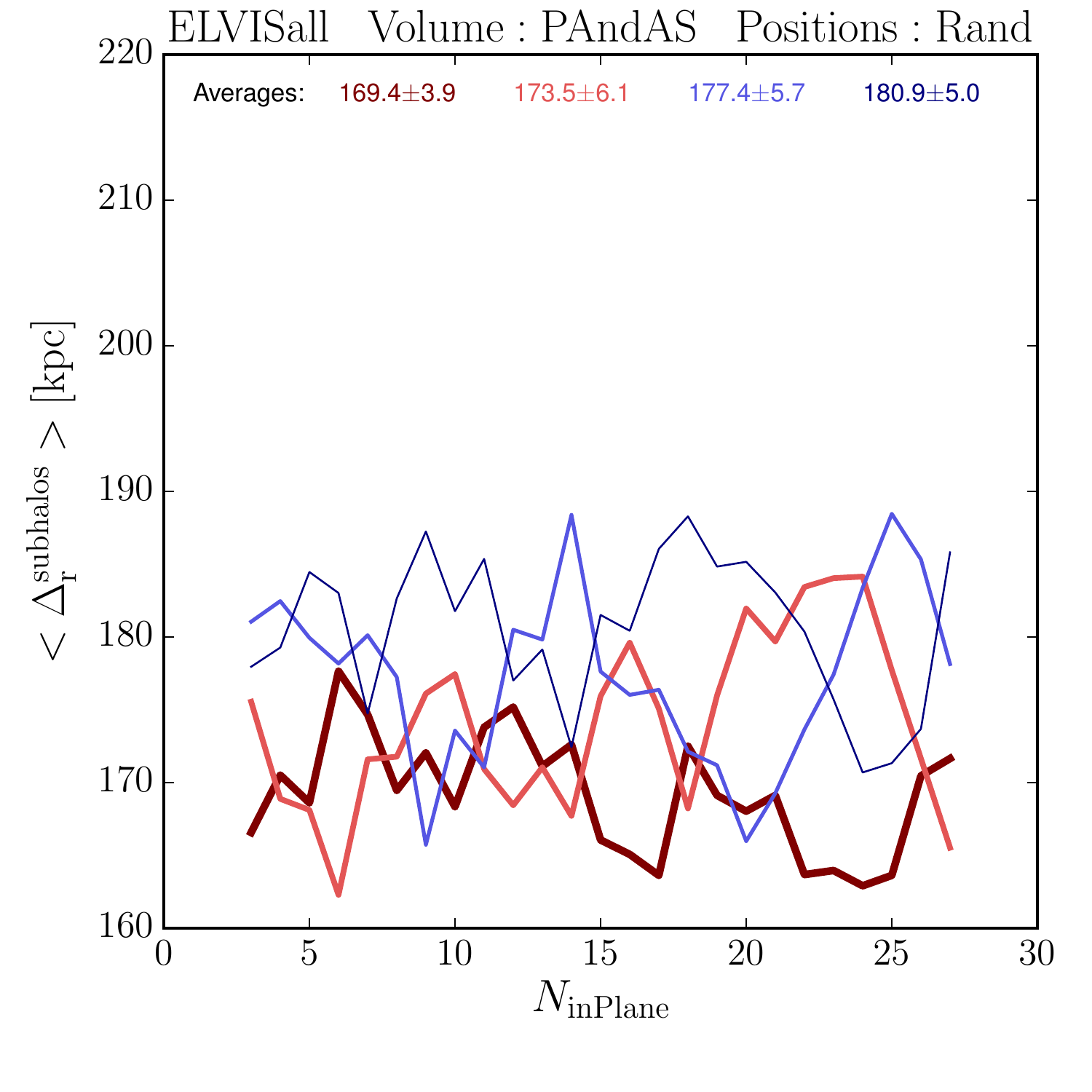}
   \caption{
Average properties of the host halos and satellite system for each number of satellites in the plane, divided into four groups of plane height each containing 12 host halos. The groups range from the 12 halos that result in the thinnest minimum plane heights (thick red line) to the 12 halos that result in the widest minimum plane heights (thin, blue line). Also given in each panel are the averages over all satellite numbers for each bin, and the rms scatter around these averages. From top to bottom, the rows give the average host halo concentration $<c_{\mathrm{-2}}>$, the average host halo formation redshift $<z_{\mathrm{0.5}}>$, the average host halo virial radius $<r_{\mathrm{vir}}>$, and the average of the rms of the radial distribution of the 30 or 27 sub-halos of each halo considered in the plane fits $<\Delta_{\mathrm{r}}^{\mathrm{subhalos}}>$. The columns correspond to the four simulated and randomized samples discussed in Sect. \ref{subsect:top30} (columns 1 and 2) and \ref{subsect:PAndAS} (columns 3 and 4).
   }
              \label{fig:halopropertiesbins}
\end{figure*}

As another test, we now invert our approach. Instead of testing whether a halo property results in more narrow satellite planes, we now test whether the most-narrow satellite planes live in halos that on average have different properties. For each number $N_\mathrm{inPlane}$\ of satellites in a plane, we separate the halos into four bins by the value of $\min \Delta_{\mathrm{rms}}$\ for these satellite planes. In other words, we split them into groups of 12 halos that range from containing the 12 thinnest satellite planes (thick red line in Fig. \ref{fig:halopropertiesbins}) to the 12 widest (thin blue line). This is done for each number of satellites in the plane $N_\mathrm{inPlane}$, such that the composition of the bins changes for different satellite numbers. We then calculate the average halo concentration, formation redshift, virial radius and sub-halo rms distance for each bin and each number of satellites. The results are shown in Figure \ref{fig:halopropertiesbins}. The average of the mean properties over all numbers of satellites in the plane (i.e. flattening the x-axis), as well as the scatter around this average, are given at the top of each panel for each bin in the corresponding color.

It is clear from the highly fluctuating and crossing curves that higher concentrations (first row) and formation redshifts (second row) of the host halos do not result in those halos having the thinnest satellite planes. The scatter in the averages over all numbers of satellites in the planes is as large or larger than the difference between them, too. If anything, the halos resulting in the most narrow planes seem to be forming slightly later (smaller $z_{\mathrm{0.5}}$) than the others for $N_\mathrm{inPlane} \geq 15$, which is the opposite of the trend suggested by \citet{Buck2015a}. However, the differences in the averages between the four bins are consistent with the scatter in each bin and therefore not significant, and the same trend is seen for the randomized satellite positions.

There is a tendency that halos that contain more narrow satellite planes (for a given number of satellites in the plane) have smaller virial radii if the satellites are selected from within the whole virial radius (third row). Averaged over all satellite numbers the mean virial radii in the four bins are consistent between the simulated sample and the one with randomized angular positions. This again evidences that the driving variable is the radial extent to which the sub-halo system is considered. Consequently, in the case of the fixed-volume PAndAS-like selection, this dependency on $r_{\mathrm{vir}}$\ is not as pronounced any more. The averages are consistent with being identical.

There is a strong tendency that sub-halo systems that are more radially concentrated result in more narrow planes (lowest row) if selected from the whole virial volume. This is again less clear in the PAndAS samples. However, this effect is again also present for the samples with random sub-halo positions (columns 2), which demonstrates that it is merely an effect of considering only the absolute thickness of the sub-halo systems. It does not stem from an increase in the the spatial and kinematic coherence of sub-halo systems in these types of halos, which is completely eradicated by choosing random angular positions of the sub-halos.

\subsection{Tests of Correlation}
\label{subsect:correlationtests}

\begin{figure*}
   \centering
   \includegraphics[width=88mm]{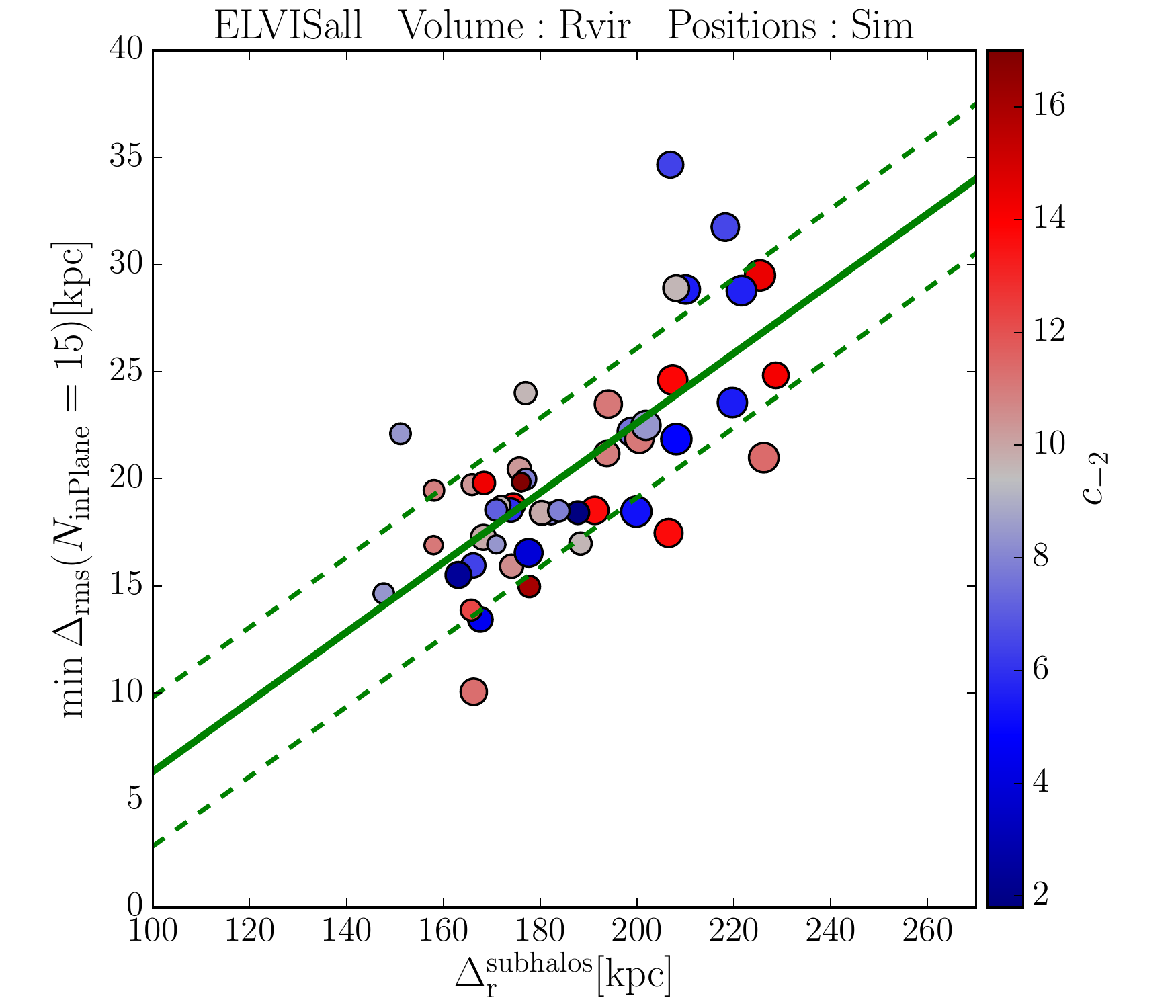}
   \includegraphics[width=88mm]{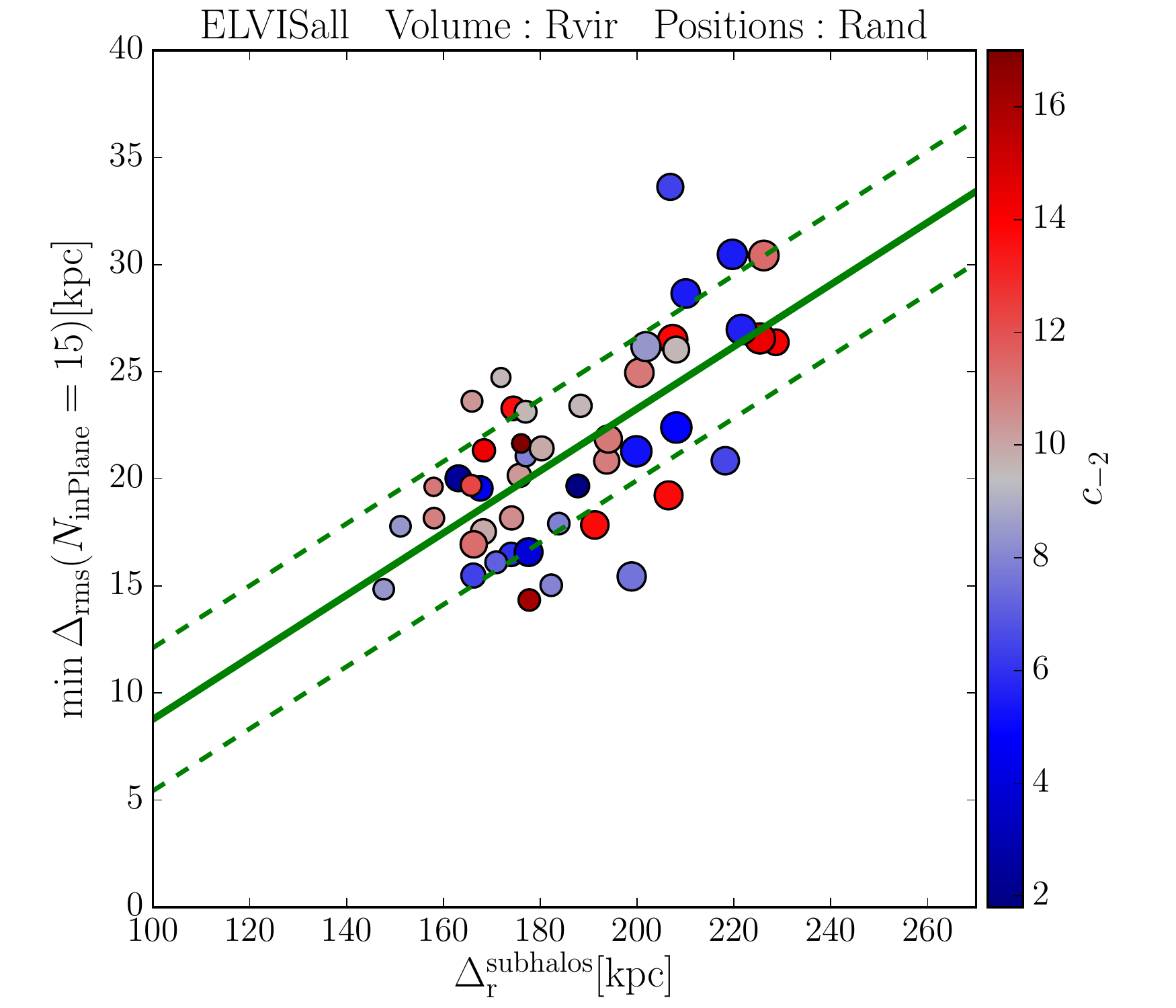}
   \includegraphics[width=88mm]{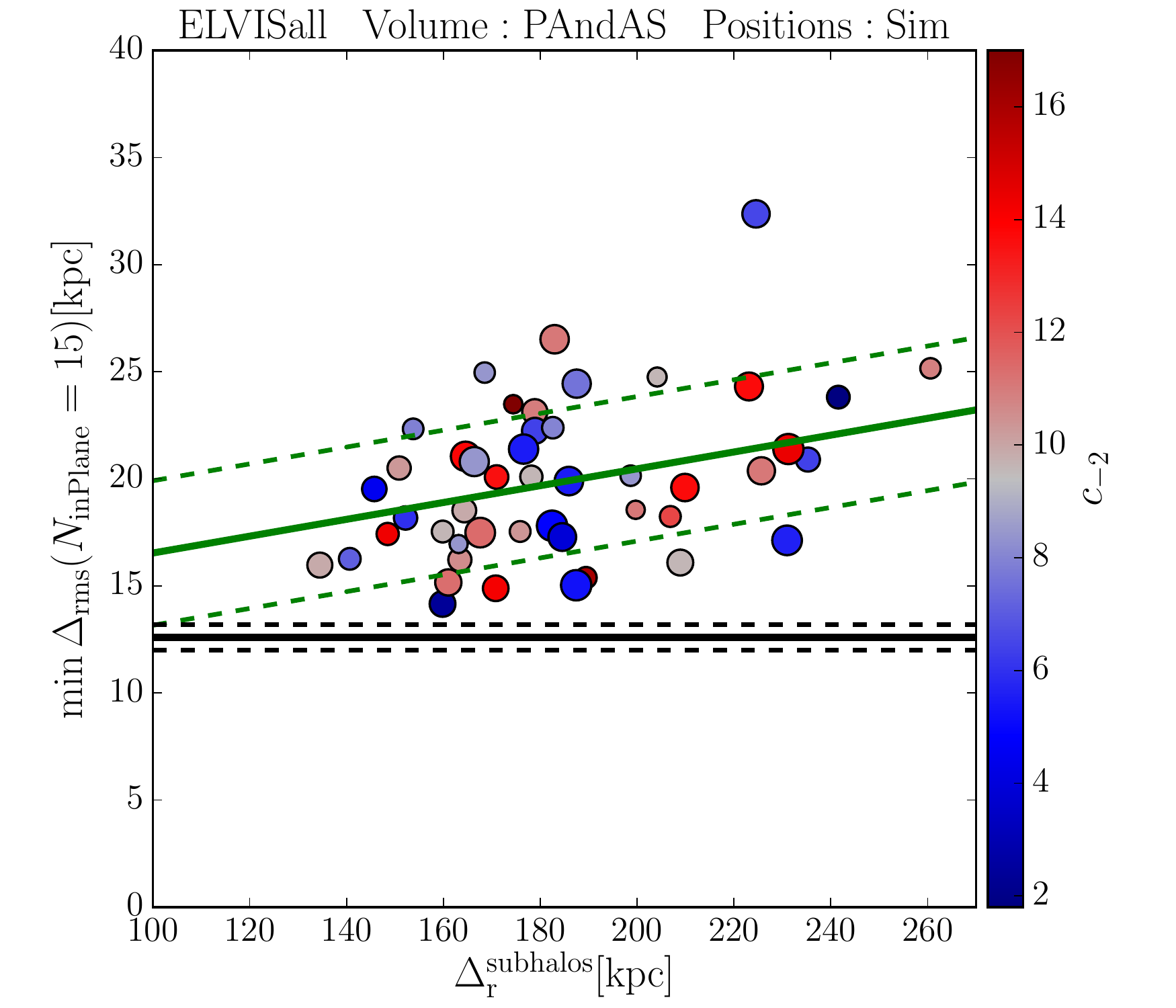}
   \includegraphics[width=88mm]{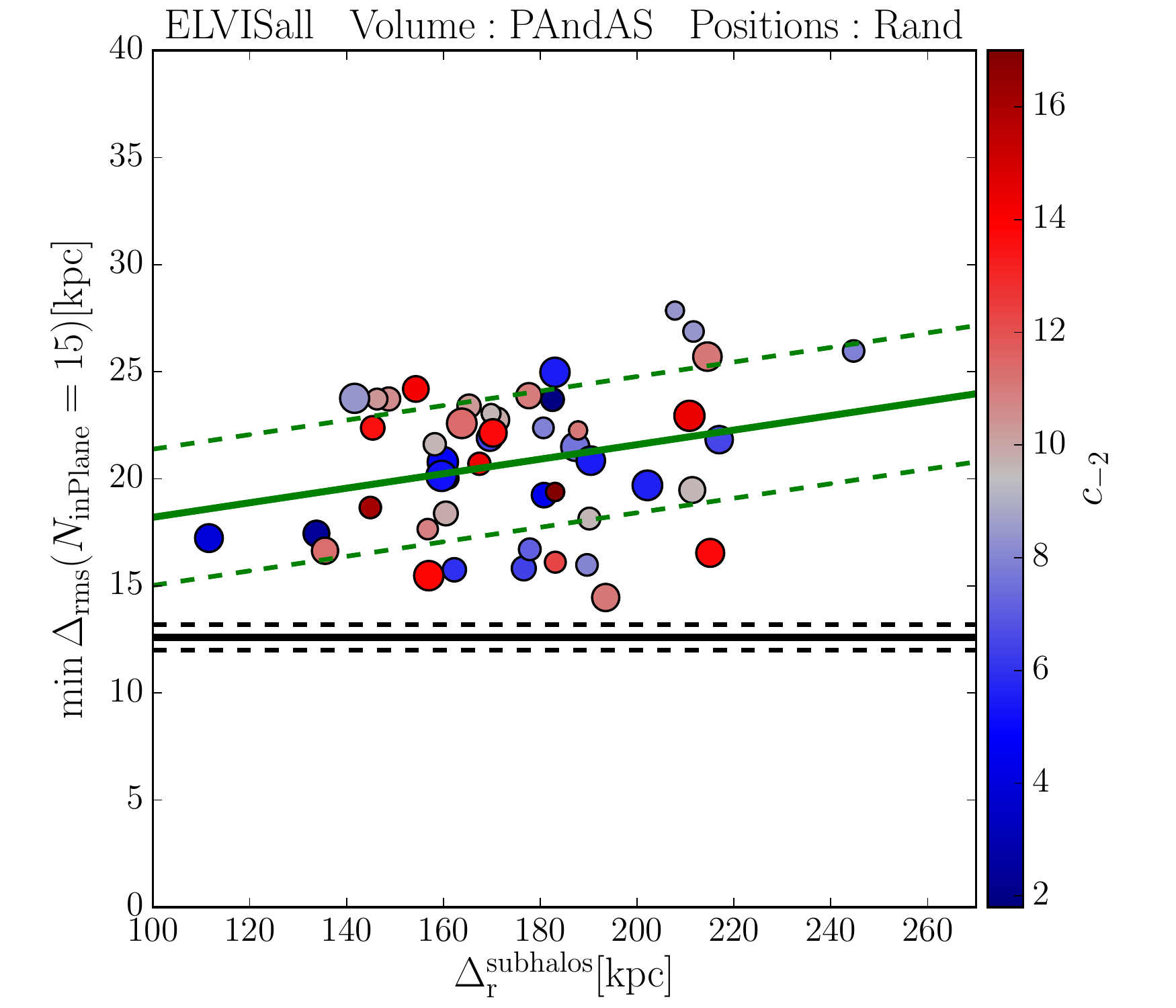}
   \caption{
Rms heights of planes consisting of 15 satellites, versus rms radii of the sub-halo systems. Each point represents one host halo's satellite system. They are color coded for host halo concentration, with larger symbols corresponding to larger virial radii. The four panels are for the same sub-halo samples as those in Fig. \ref{fig:planeheights}. The black solid and dashed lines in the lower panels indicate the plane height of the observed GPoA of $\Delta_{\mathrm{rms}} = 12.6 \pm 0.6$. 
The thick, green line is a linear fit to all datapoints, the standard deviation of the resiuduals around this fit is illustrated by the dashed lines. If sub-halos are selected from the full virial volume (upper panels) then there is a clear correlation with the overall radial distribution of the sub-halos, which is present irrespective of whether the simulated sub-halo positions are considered (left panels), or the sub-halo positions are randomized (right panels).  However, if sub-halos are selected from a more realistic volume (lower panels) then no such correlation is found.
If the selection volume is linked to the virial radius, there is also a correlation of $r_\mathrm{vir}$\ (symbol sizes) with $\Delta_\mathrm{r}^\mathrm{subhalos}$.
   }
              \label{fig:rmsradiuscorrelation}
\end{figure*}

In addition to these qualitative comparisons, we now apply more formal, quantitative tests of correlation, searching for dependencies between the various halo properties, as well as between these and the satellite plane heights. We use both the Pearson correlation coefficient $\rho$\ as well as the Spearman rank correlation coefficient $r_{\mathrm{s}}$, and report the corresponding $p$-values, which give the probability of obtaining a more extreme correlation coefficient by chance. We assume that we can reject the hypothesis of no correlation if the $p$-value is below 0.05 (or $\log p < -1.3$).

For the top-30 sub-halo sample selected from the whole virial volume we find no evidence for a correlation between 
the halo concentration $c_{\mathrm{-2}}$\ and $\Delta_{\mathrm{r}}^{\mathrm{subhalos}}$\ ($\rho = 0.01$, $p = 0.93$; $r_{\mathrm{s}} = -0.01$, $p = 0.95$), or
the virial radius $r_{\mathrm{vir}}$\ and halo formation redshift $z_{\mathrm{0.5}}$\ ($\rho = 0.09$, $p = 0.52$; $r_{\mathrm{s}} = 0.09$, $p = 0.55$).
There are hints for a weak correlation between the formation redshift $z_{\mathrm{0.5}}$\ and $\Delta_{\mathrm{r}}^{\mathrm{subhalos}}$, but the two tests are not quite significant ($\rho = 0.28$, $p = 0.06$; $r_{\mathrm{s}} = 0.20$, $p = 0.16$).
We do find 
a weak, marginally significant correlation between the halo concentration $c_{\mathrm{-2}}$\ and the halo formation redshift as measured via $z_{\mathrm{0.5}}$\ ($\rho = 0.39$, $p = 0.01$; $r_{\mathrm{s}} = 0.30$, $p = 0.04$), 
as well as a strong and highly significant correlation between $r_{\mathrm{vir}}$\ and $\Delta_{\mathrm{r}}^{\mathrm{subhalos}}$\ ($\rho = 0.77$, $p = 1.1 \times 10^{-10}$; $r_{\mathrm{s}} = 0.74$, $p = 1.4 \times 10^{-9}$).

As expected from the discussions in Sect. \ref{subsect:spatial}, our tests reveal no highly significant correlation between the formation redshift or the halo concentration and the satellite plane heights for the different numbers of satellites per plane (or averaged over them). The corresponding $p$-values of these tests mostly remain well above 0.1 or $\log p > -1$\ (see Tables \ref{tab:correlations_top30_sim} to \ref{tab:correlations_PAndAS_rand})\footnote{If the results for the different numbers of satellites per plane $N_\mathrm{inPlane}$\ were independent (which they are not), we would expect to find $p$\ values below 0.1 in about 10\,per cent of the cases in completely random data.}. For the sample of 30 sub-halos selected from the whole virial volume, we find a correlation with $r_{\mathrm{vir}}$, which is present and of similar strength for both the simulated as well as the randomized sample. Unsurprisingly, such a correlation with $r_{\mathrm{vir}}$\ is not found for the PAndAS-like selection volume. While the top-30 sub-halo samples show a strong correlation between the satellite plane heights and the rms radius of the sub-halo samples $\Delta_{\mathrm{r}}^{\mathrm{subhalos}}$, only a weak correlation like this exists in the PAndAS-like samples. This quantitatively demonstrated the importance of accurately accounting for the selection volume of the observed M31 satellite galaxies when comparing to sub-halos in simulations.

\subsection{Satellite system extent vs. plane height}
\label{subsect:rmsthicknessrelation}

Fig. \ref{fig:rmsradiuscorrelation} shows the correlation between the satellite plane height for planes consisting of 15 satellites, and the rms radius of the sub-halo samples, $\Delta_{\mathrm{r}}^{\mathrm{subhalos}}$. The latter is a good predictor of the height of the most narrow plane if the satellites are selected from the whole virial volume. This is independent of whether the actual sub-halo positions of the simulations or randomized positions are used. A linear fit of the form 
$$
\min \Delta_\mathrm{rms} = a \times \Delta_{\mathrm{r}}^{\mathrm{subhalos}} + b
$$
demonstrates the similarity. The fit parameters are $a = 0.163 \pm 0.001$\ and $b = -9.95 \pm 20.93$\,kpc for the simulated sample, and $a = 0.145 \pm 0.001$\ and $b = -5.73 \pm 19.21$\,kpc for the randomized one. Even the residuals around these best-fit lines have almost identical scatter (3.48 and 3.34\, kpc, respectively). This shows how little additional structure is present in the simulated sub-halo systems compared to sub-halo positions randomly chosen from an isotropic distribution. Knowing the radial distribution of a sub-halo system thus allows to predict the height of it's most narrow satellite plane.
However, this is not as pronounced if the sub-halo positions are constrained by the PAndAS-like selection volume, neither for the simulated sub-halos positions ($a = 0.0393 \pm 0.0002$, $b = 12.61 \pm 8.57$\,kpc, scatter of 3.37\,kpc), nor for the randomized ones ($a = 0.0339 \pm 0.0003$, $b = 14.81 \pm 10.45$\,kpc, scatter of 3.18\,kpc).

\section{Kinematic Analysis}
\label{subsect:kinematics}

\begin{figure*}
   \centering
   \includegraphics[width=88mm]{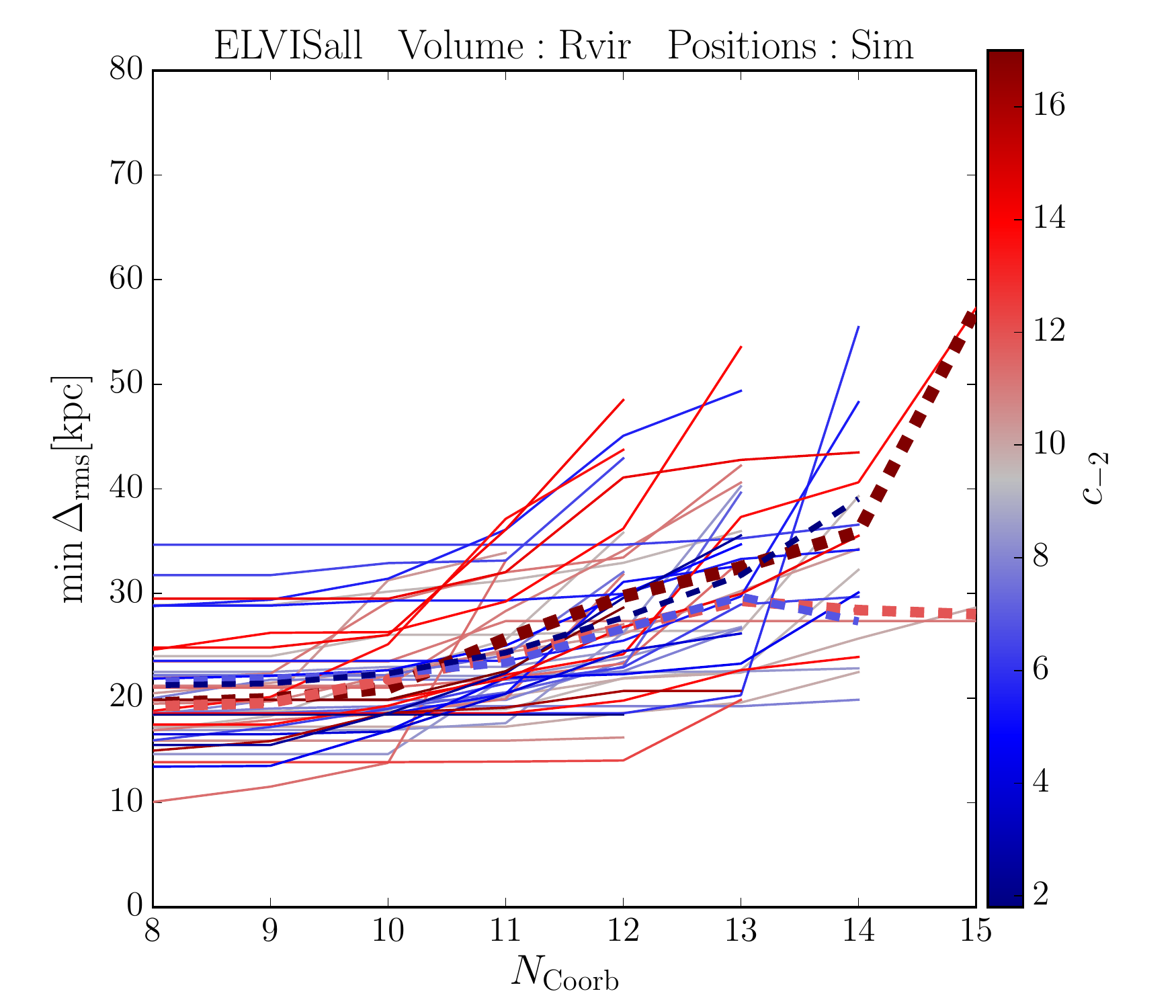}
   \includegraphics[width=88mm]{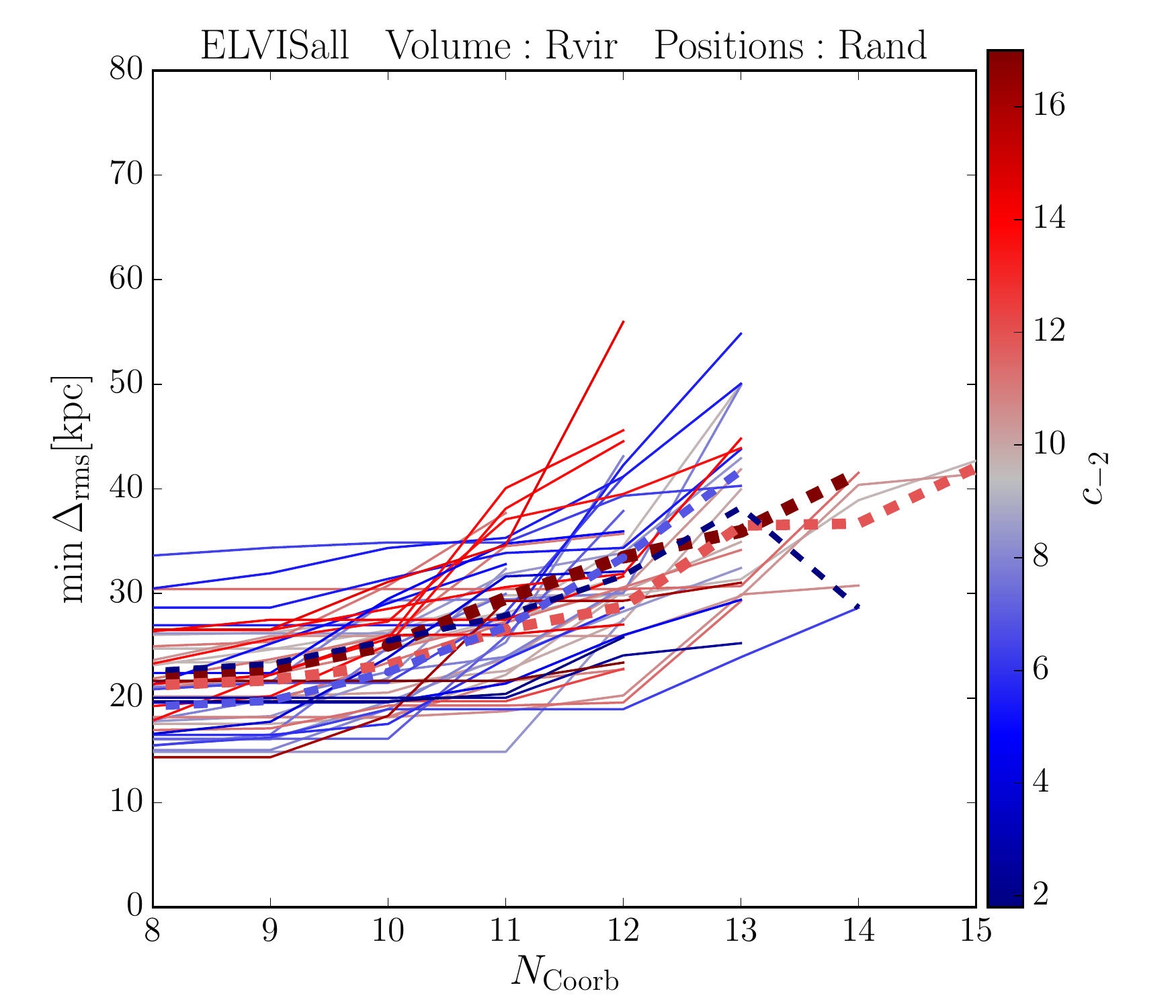}
   \includegraphics[width=88mm]{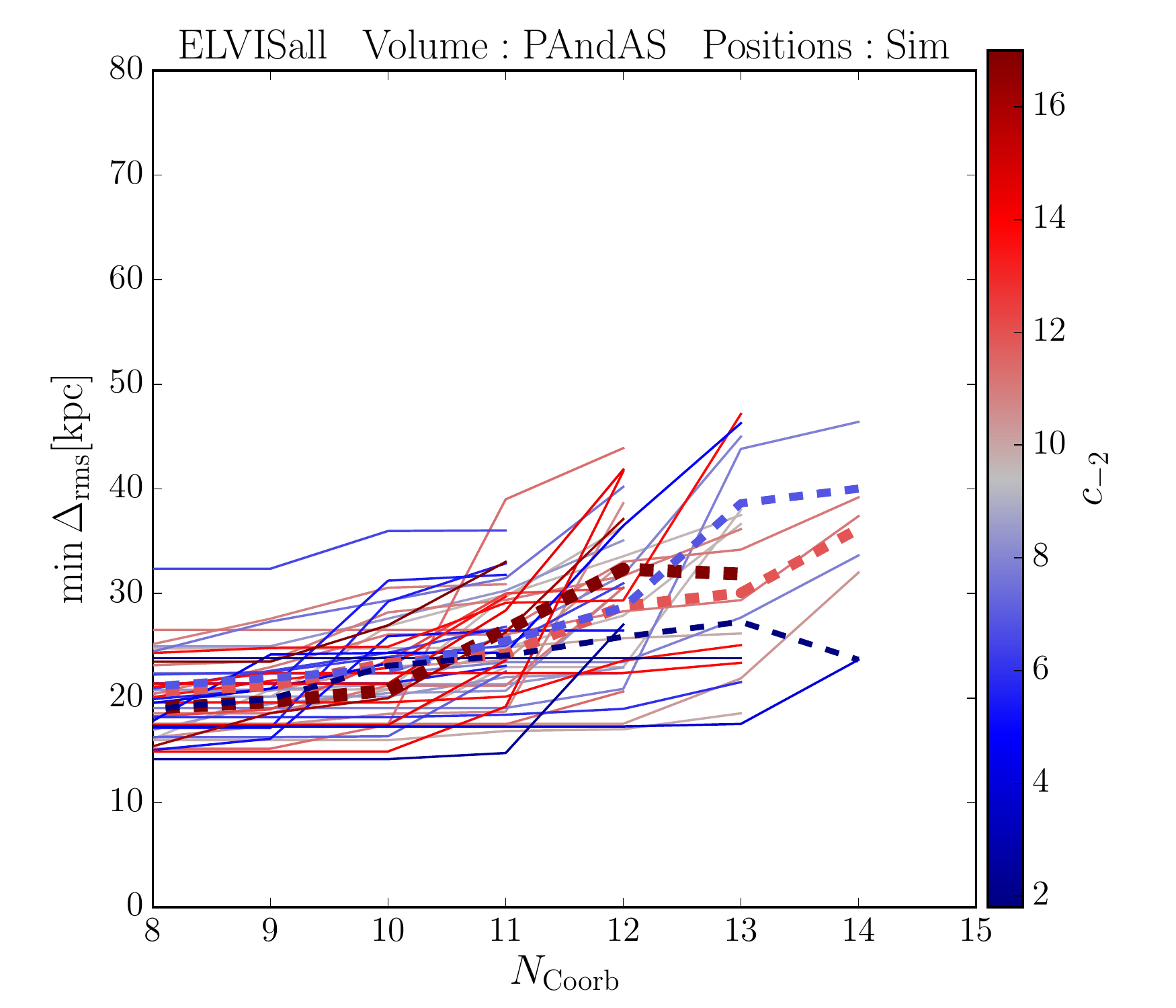}
   \includegraphics[width=88mm]{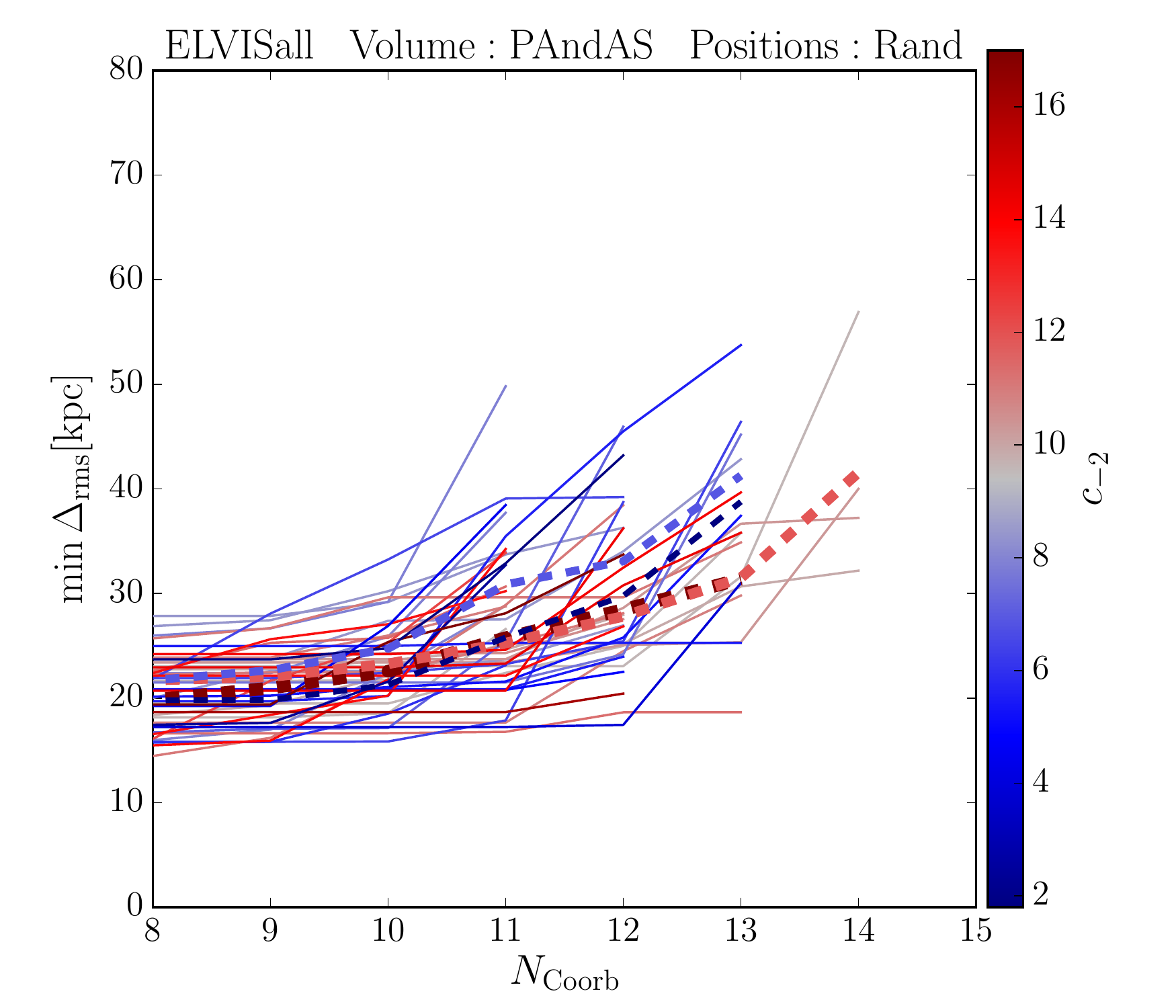}
   \caption{
Plane height versus number of co-orbiting satellites for planes consisting of 15 satellites, color coded for host halo concentration. The four panels are for the same sub-halo samples as those in Fig. \ref{fig:planeheights}. The lines stop once no plane with a given number of co-orbiting satellites $N_{\mathrm{Coorb}}$\ is present in the sample. The thick, dashed lines give the average plane heights for combinations of 12 sub-halo systems ranging from the 12 most-concentrated (red) to the 12 least concentrated (blue). As expected, the randomized satellite systems result in less narrow and less kinematically coherent satellite planes. None of the 48 ELVIS halos can reproduce the observed flattening and kinematic coherence. Furthermore, there is no indication that the more concentrated halos result in more narrow and more kinematically coherent satellite planes. Rather, the sub-halo systems of the most concentrated host halos appear to be less thin and coherent (the average of the 12 most-concentrated halos is above those of the less concentrated halos).
%Plots with color-coding by other halo properties than $c_{\mathrm{-2}}$ ($z_{\mathrm{0.5}}$, $r_{\mathrm{vir}}$, and $\Delta_{\mathrm{r}}^{\mathrm{subhalos}}$, for a total of 4 figures with 4 panels each) are available as a figure set in the online journal.
   }
              \label{fig:planeheightscorot}
\end{figure*}

While we find that there is no correlation between the concentration of a host halo and the minimum heights of sub-halo planes, there remains the possibility that a stronger \textit{kinematic} coherence is present for sub-halos in hosts of higher concentration. We investigate this possibility for planes consisting of a total of 15 sub-halos in Figure \ref{fig:planeheightscorot}. It shows the minimum of the height of satellite planes which contain a given number of co-orbiting satellites. For example, if a system contains a plane of 15 sub-halos of height 20\,kpc of which 10 co-orbit, and another one of height 25\,kpc of which 12 co-orbit, then the line for a given ELVIS halo remains at $\min \Delta_{\mathrm{rms}} = 20$\,kpc for $N_{\mathrm{Coorb}} \leq 10$, and then goes up to $\min \Delta_{\mathrm{rms}} = 25$\,kpc for $N_{\mathrm{Coorb}} = 11$\ and 12. Once no plane of any height is found for a given number of co-orbiting sub-halos, the corresponding line for this halo stops.

Figure \ref{fig:planeheightscorot} suggests that, as to be expected, the randomized samples (right panels) are a bit less kinematically coherent: the lines tend to end at lower $N_{\mathrm{Coorb}}$. Similarly, the planes in the PAndAS-like samples are less kinematically coherent than those selecting the top-30 sub halos within the full virial radius. A larger number of sub-halos to choose planes from increases the number of possible combinations for a given number of sub-halos per plane, which therefore increases the chance to find planes with a higher degree of velocity coherence. This demonstrates again that closely modelling observational selection effects is indispensable for a meaningful comparison of simulations with the observed situation. None of the sub-halo systems can reproduce the height and number of co-orbiting satellites of the observed GPoA. This allows us to put a lower limit on the frequency of such planes in cosmological simulations of $\lesssim 2$\,per cent, which is consistent with the low frequencies of 0.04 to 0.17\,per cent found in earlier studies using larger samples of host halos \citep{Ibata2014,Pawlowski2014}.

If there is an increase in the kinematic coherence for sub-halos systems of more concentrated hosts, the corresponding lines in this plot should turn upwards later than those of the less concentrated halos. This is not the case, as can be seen from the thicker dashed lines indicating the average\footnote{The assignment of halos to the four bins remains fixed, but the averaging takes into account that at higher $N_{\mathrm{Coorb}}$\ some hosts do not contribute any more because the do not contain sufficiently kinematically coherent planes.} plane heights for four bins of host halo concentration ranging from the 12 highest concentration hosts (red) to the 12 lowest-concentration hosts (blue). The lines follow a common trend and overlap within the scatter of their respective host satellite system populations.

\section{Testing for Effects of Environment and Central Host Galaxy}
\label{sect:newtests}

\begin{figure*}
   \centering
   \includegraphics[width=88mm]{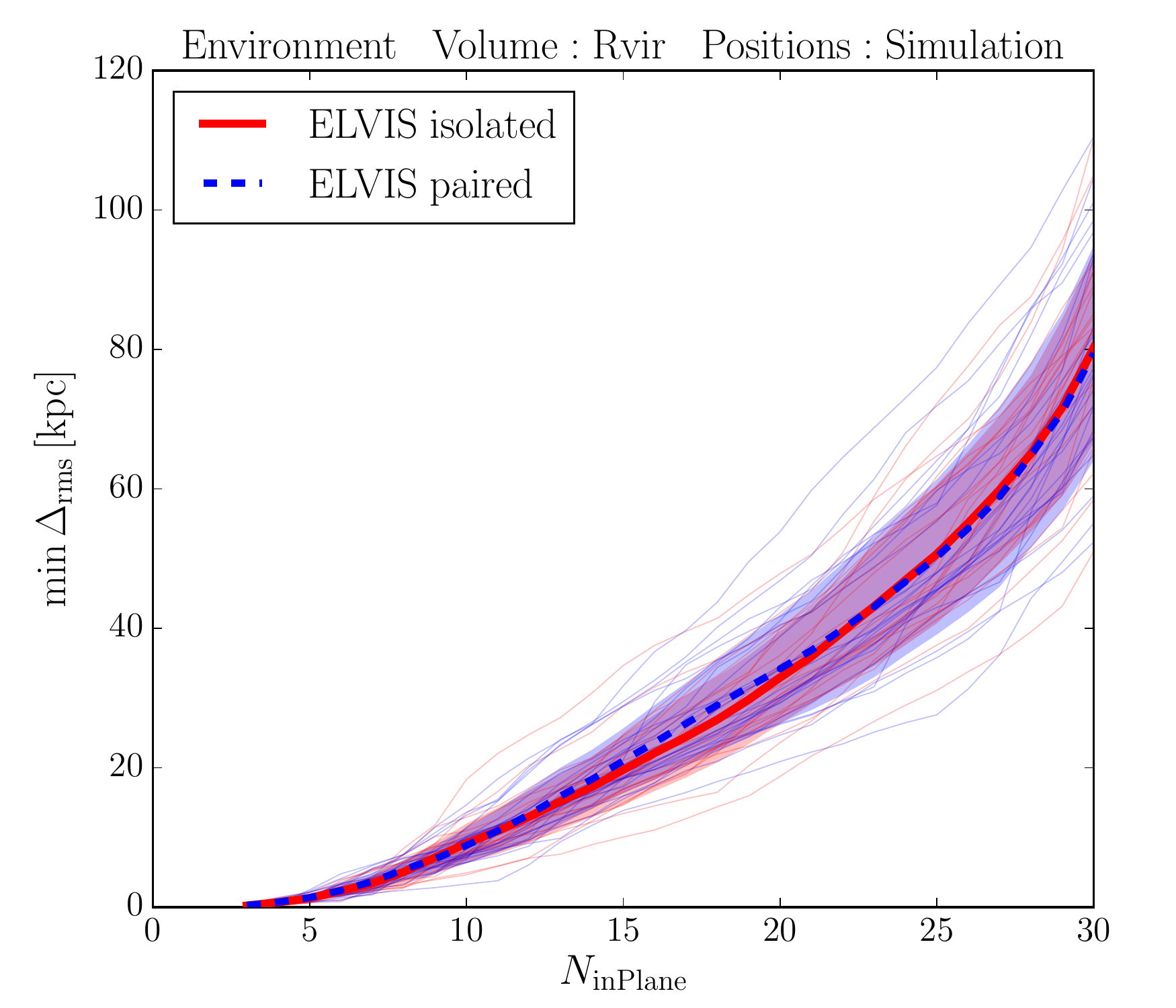}
   \includegraphics[width=88mm]{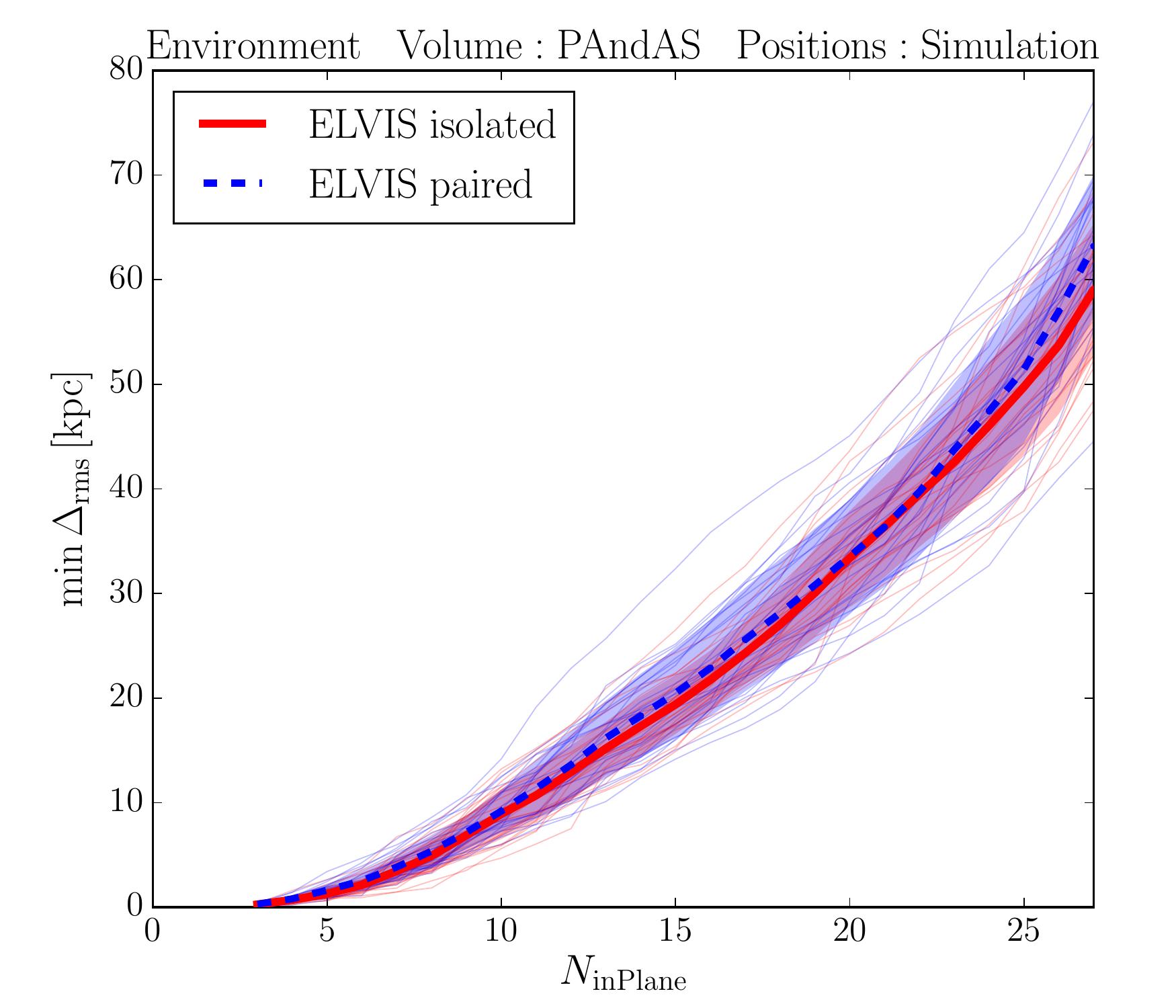}
   \includegraphics[width=88mm]{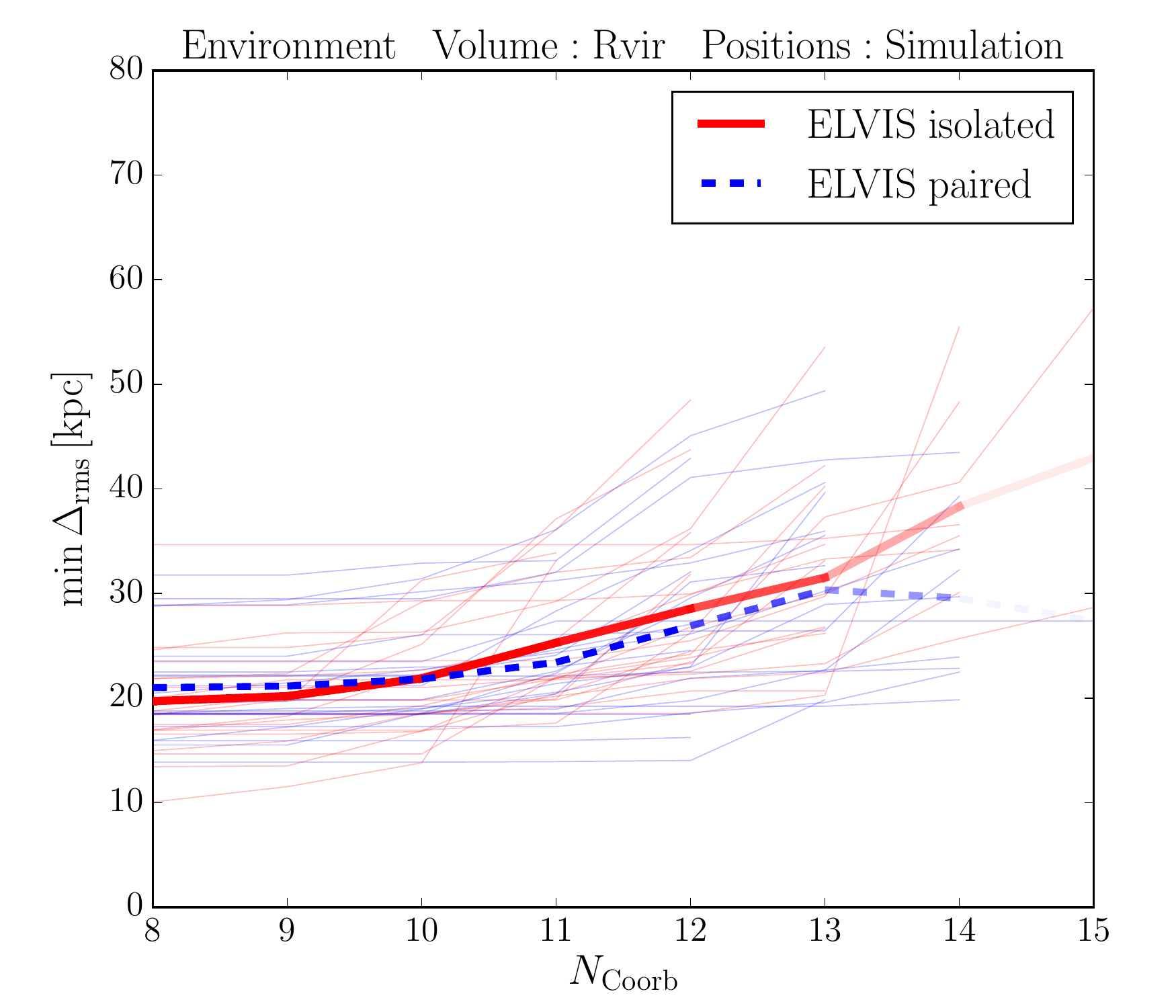}
   \includegraphics[width=88mm]{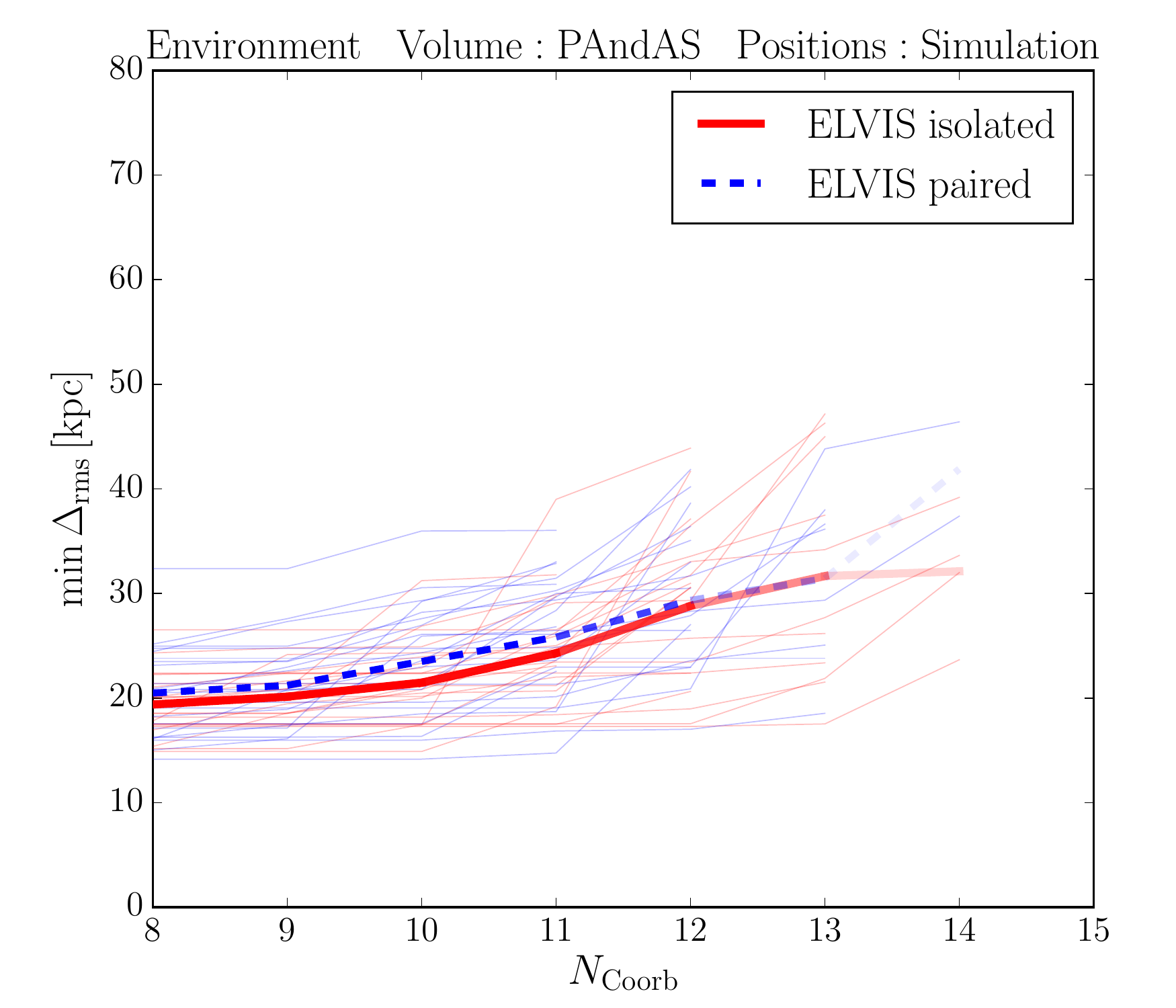}
   \caption{
    Comparison of plane heights (top panels, similar to Fig. \ref{fig:planeheights}) and kinematic coherence (bottom panels, similar to Fig. \ref{fig:planeheightscorot}) for the two ELVIS simulation sub samples of isolated (red) and paird (blue) hosts. Sub-halos are selected from the virial volume (left panels) or the PAndAS volume (right panels). The isolated and paired samples do not show significant differences, neither in their averages (thick lines) or the 50 \% scatter (shaded regions). This indicates that environment does not have an effect on the properties of planes of satellite galaxies, specifically that being part of a Local-Group like pair does not result in more narrow or more kinematically coherent satellite planes.
   }
              \label{fig:environment}
\end{figure*}

\begin{figure*}
   \centering
   \includegraphics[width=88mm]{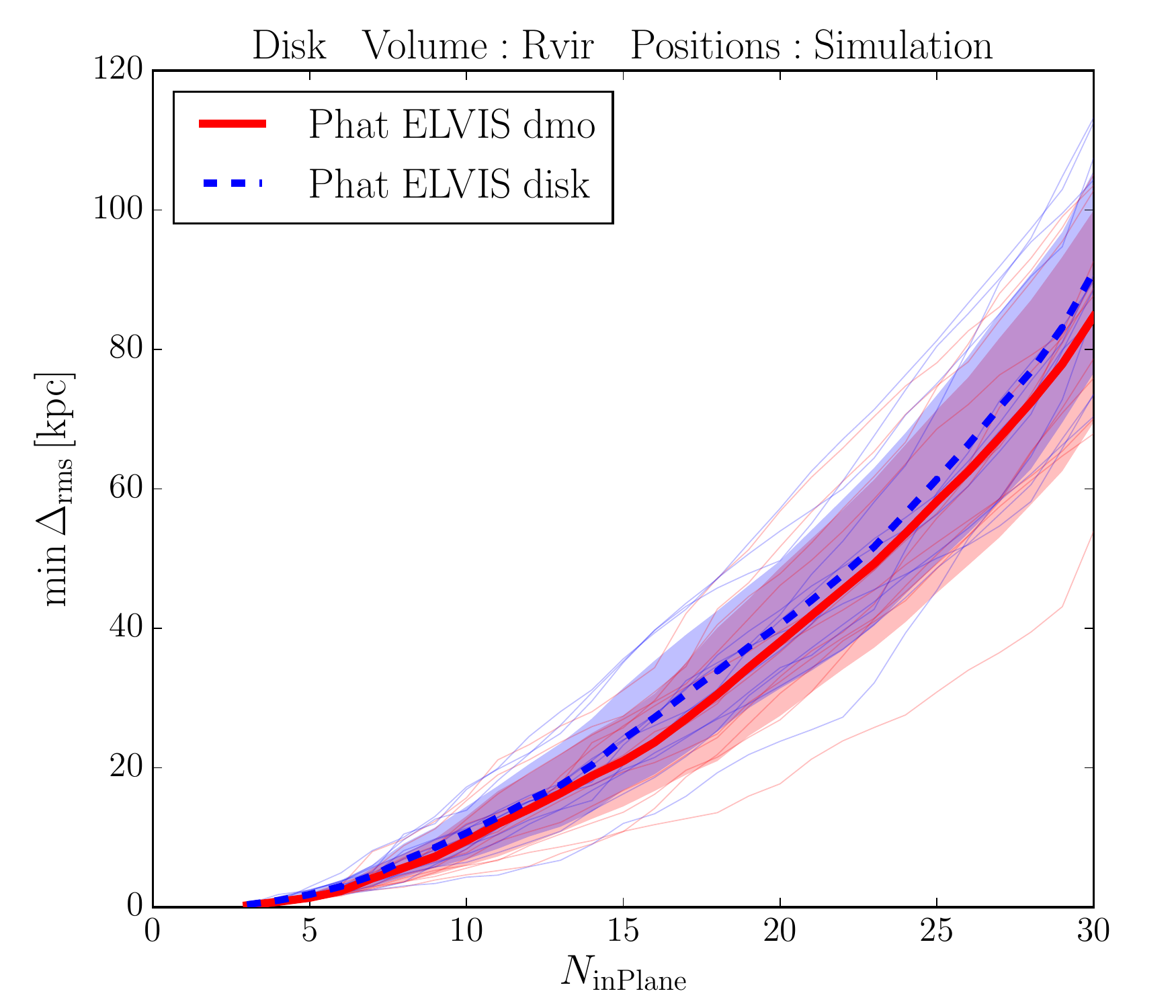}
   \includegraphics[width=88mm]{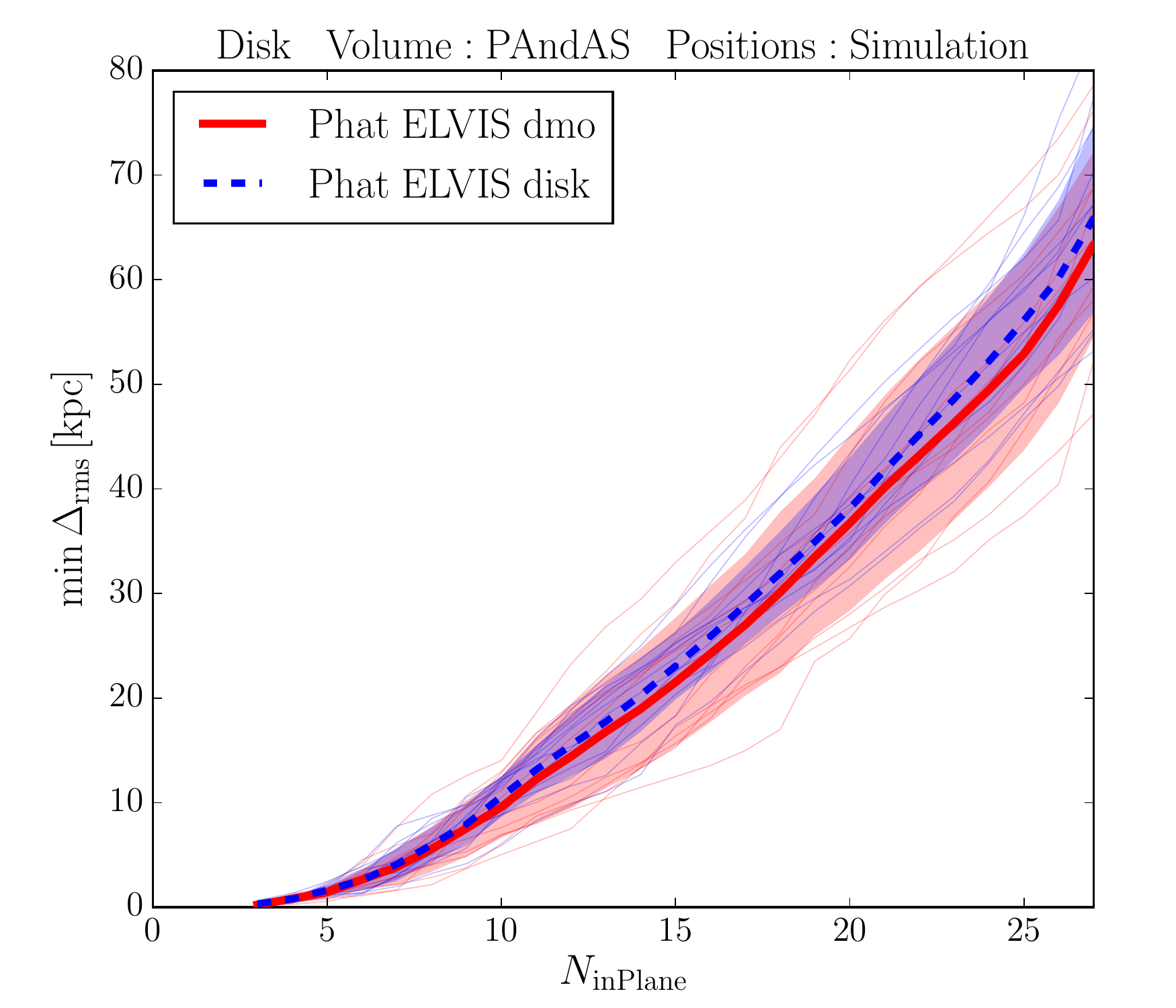}
   \includegraphics[width=88mm]{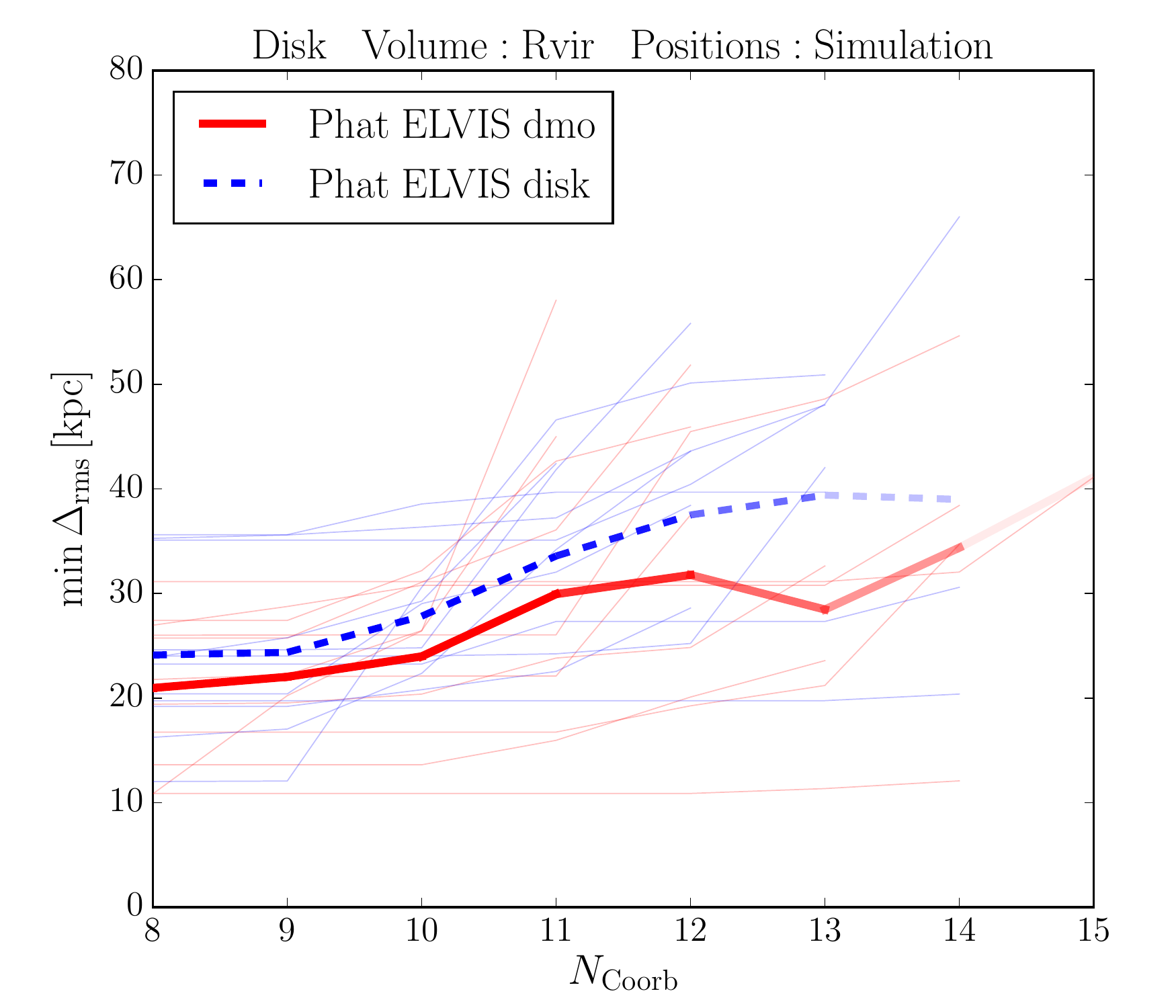}
   \includegraphics[width=88mm]{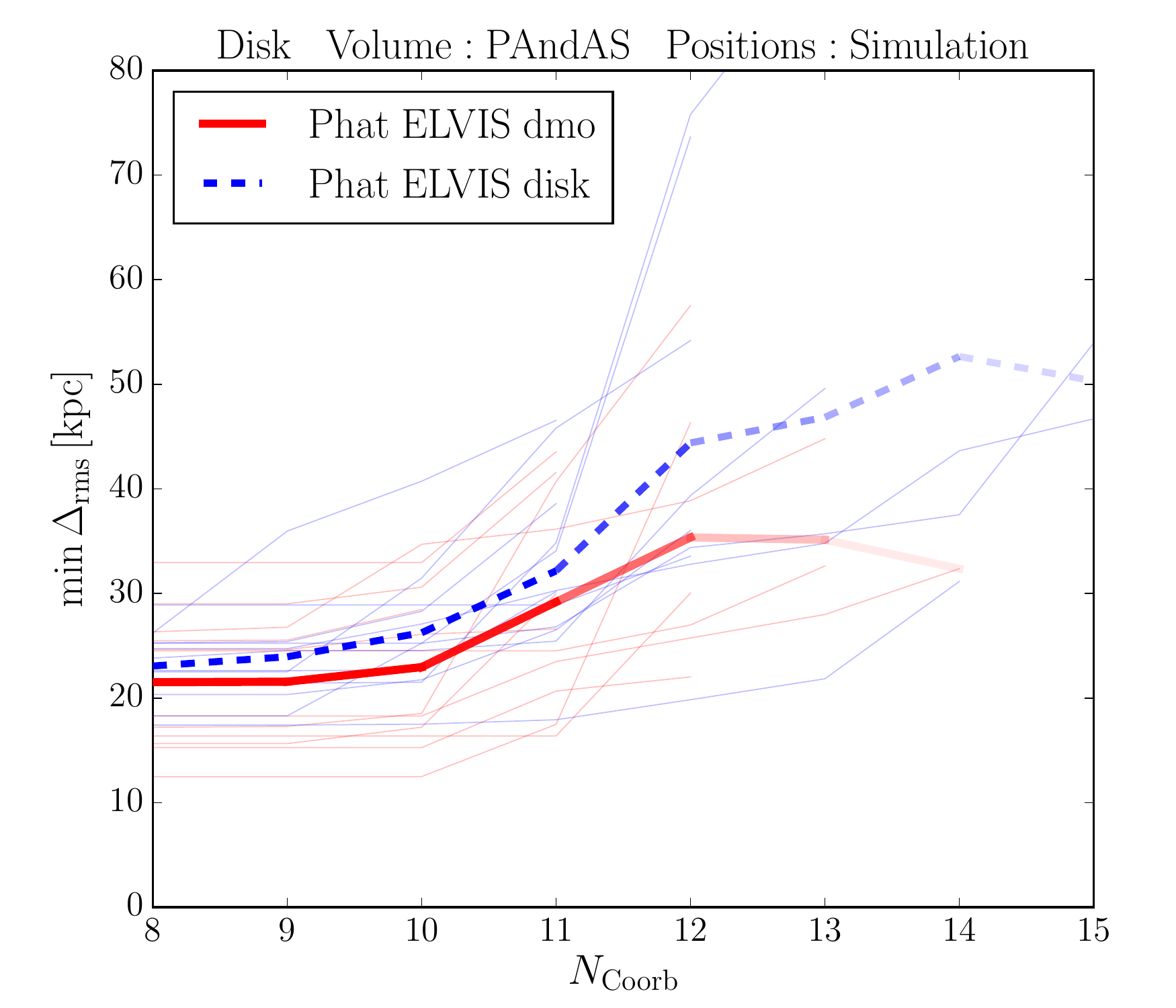}
   \caption{ 
  Plots similar to Fig. \ref{fig:environment}, but comparing the dark-matter-only (dmo, red) runs and the runs with an embedded disk galaxy potential (disk, blue) of the Phat ELVIS suite of simulations. The inclusion of a central disk galaxy potential does not result in more narrow or more kinematically coherent planes of satellite galaxies, but rather in a wider average plane height.}
              \label{fig:disk}
\end{figure*}

\subsection{Isolated vs. Paired Hosts}

We now split up the ELVIS sample into a sub-sample containing the 24 isolated hosts ({\it ELVIS isolated}), and a sub-sample containing 24 hosts that are part of 12 pairs of host galaxies  ({\it ELVIS paired}). The distribution of plane heights for these two sets are shown in the upper panels of Fig. \ref{fig:environment}, similar to Fig. \ref{fig:planeheights}, but now color-coded according to sub-sample membership. Comparing the sub-halos systems of ELVIS isolated (red) and ELVIS paired (blue) hosts, we find no significant difference in both the average (thick lines) or the scatter (shaded regions) of the plane height.
The lower panels in Fig. \ref{fig:environment} show the kinematic coherence (as in Fig. \ref{fig:planeheightscorot}). Again, the two samples of simulated sub-halo systems do not display substantial differences. We thus conclude that belonging to a host that is part of a pair configuration does not strongly influence the presence and properties of planes of satellite galaxies in $\Lambda$CDM simulations. This is in agreement with \citet{PawlowskiMcGaugh2014b}, who have looked for analogs of the Vast Polar Structure of the Milky Way in the ELVIS simulations and did not find a correlation in their incidence or properties with whether the host is in an isolated or paired configuration.

As is the case for the whole sample, neither of the two subsamples shows a correlation between the plane heights and halo concentration or formation redshift.  There is thus no indication that being part of a paired host halo configuration, nor that being an isolated host, results in a correlation with these halo properties. As expected from the full sample, if sub-halos are selected from the virial volume a weak correlation with the virial radius and a strong correlation with the extent of the subhalo distribution is again present. However, owing to the reduced sample sizes the respective significances are lower than for the full sample.

\subsection{DMO vs. Central Galaxy Potential}
 
It has been found that the addition of a central galaxy potential to a cosmological simulation, whether self-consistently via a hydrodynamical simulation or using an analytic potential, results in enhanced tidal disruption of sub-halos compared to a dark-matter-only simulation. This in turn changes the radial distribution and orbital properties of the sub-halo system \citep{GarrisonKimmel2017,Sawala2017}. The inner regions of host halos become depleted in sub-halos, and the orbits of surviving sub-halo satellites are less radial on average than in equivalent dark-matter-only runs. To investigate whether the presence of a central galaxy potential and its effect on the sub-halo distribution influence the properties and coherence of planes of satellite galaxies, we now compare the resulting plane heights and kinematic coherence of the two Phat ELVIS simulations samples by \citet{Kelley2018}: the 12 dark-matter-only runs that are equivalent to the ELVIS simulations ({\it Phat ELVIS dmo}) and the 12 runs started from the same initial conditions but with an added central galaxy potential ({\it Phat ELVIS disk}). In the bottom panel of Fig. \ref{fig:radialdistr}, the expected change in the radial distribution of sub-halos is apparent: satellite sub-halos in the Phat ELVIS disk runs selected from the PAndAS footprint have a radial distribution that is in better agreement with the observed distribution of satellite galaxies in M31 than the original ELVIS runs that did not include a central galaxy potential.

Fig. \ref{fig:disk} compiles the results of this comparison. The average plane heights (thick lines) are comparable for the dmo (red) and disk (blue) runs, though the former are slightly more narrow. The typical scatter (shaded regions) overlap well and show similar width, though there appears to be more scatter in the dmo runs if sub-halos are selected from the PAndAS footprint (right panels). For sub-halos selected from the PAndAS footprint (right panels), we find $\min \Delta_{\mathrm{rms}}$\ to be on average 7 per cent larger for the simulations that contain a disk potential compared to their disk-free analogs (compared to 11 per cent if selected from $r_\mathrm{vir}$). This is in line with the general trend of a correlation between the $\min \Delta_{\mathrm{rms}}$\ and the radial extent of a sub-halo system measured by $\Delta_{\mathrm{r}}^{\mathrm{subhalos}}$\ discussed in Sect. \ref{subsect:rmsthicknessrelation}. The median $\Delta_{\mathrm{r}}^{\mathrm{subhalos}}$\ is 188\,kpc for the dmo and 214\,kpc for the disk runs (or 184\,kpc and 191\,kpc, respectively, if the sub-halos are selected from $r_\mathrm{vir}$). Using the corresponding linear fits in \ref{subsect:rmsthicknessrelation}, this translates to an expected increases in the width of the sub-halo distributions of 5 per cent (6 per cent), which largely accounts for the found increase in plane widths. This further supports the notion that the major driver behind the width of the most narrow planes of satellites found in a given distribution is the distribution's radial extent. 
The lower panels of Fig. \ref{fig:disk} further show that the inclusion of a central galaxy potential does not result in a more pronounced kinematic correlation in the most narrow satellite planes. The main difference is again the tendency for the disk-free simulations to have slightly smaller $\min \Delta_{\mathrm{rms}}$\ for any number of co-orbiting sub-halo satellites. Fundamentally, this finding demonstrates that the addition of baryonic physics to simulations does not appear to offer a solution to the planes of satellite galaxies problem.

Using the concentration parameters of the Phat ELVIS host halos, we have also confirmed that no significant correlations with host halo concentration of formation time are present, neither for the dmo nor for the disk samples. This is in line with our previously discussed findings for the ELVIS simulations.

\section{Conclusion}
\label{sect:conclusion}

We have used the 48 host halos of the ELVIS simulations to test whether  the host halo concentration, formation time, or status as an isolated or paired halo has an effect on the existence of narrow, kinematically coherent planes of satellites. We are unable to reproduce the result of \citet{Buck2015a} that the flattening of sub-halo systems correlates with the host halo formation time. Similarly, we do not find evidence of a correlation between the host halo concentration and the height of the thinnest sub-halo planes. Only the overall radial extent of the considered sub-halo distribution, as constrained by $r_{\mathrm{vir}}$\ or measured by $\Delta_{\mathrm{r}}^{\mathrm{subhalos}}$, shows signs of a correlation. However, this correlation is present to a similar degree if the sub-halo position are randomized, indicating that it is mainly an effect of a smaller total radial extent of lower mass halos, and not of a stronger spatial or kinematic coherence among the sub-halos in certain host dark matter halos.

We showed that it is important to select halos in a similar way as the observed satellite distribution in order to avoid biases, by constructing sub-halo selections which follow the PAndAS survey volume and number of satellites. These reveal that none of the 48 ELVIS halos hosts a plane of satellites that is as narrow and as kinematically coherent as the observed GPoA, in line with earlier findings indicating that such structures are elusive in current $\Lambda$CDM simulations \citep{Ibata2014,Pawlowski2014}. In line with the results of \citet{PawlowskiMcGaugh2014b} for the VPOS of the Milky Way, we also do not find a difference between the satellite plane properties of isolated versus paired host halos.

The main effect of baryonic physics that is relevant for the spatial and orbital properties of a satellite galaxy system is the formation of a central host galaxy. Using the Phat ELVIS simulation suite \citep{Kelley2018}, we have investigated whether the presence of such a central galaxy potential results in more narrow or more kinematically coherent planes of satellite galaxies compared to dark-matter-only analogs. We found this not to be the case. On the contrary, since the added potential  of a central host galaxy results in enhanced tidal disruption, which in turn depletes the inner regions of the host halo of satellite galaxies, it results in a more extended satellite distributions. We find that more radially extended systems tend to contain less narrow planes of satellite galaxies. Consequently, while other small-scale issues of cosmology can potentially be addressed by adding baryonic physics to cosmological simulations, it appears that the dominant baryonic effect makes the planes of satellite galaxies problem worse.

We are thus left with the unsettling realisation that the incidence of satellite planes as extreme as observed remains an unsolved problem for the $\Lambda$CDM model of cosmology, and that this problem is not easily addressed by invoking baryonic effects, an exceptionally high concentration or early formation time of the host halo, or the host-pair environment of the MW and M31 halos.

%__________________________________________________________________

\acknowledgments
M. S. P. acknowledges that support for this work was provided by NASA through Hubble Fellowship grant \#HST-HF2-51379.001-A awarded by the Space Telescope Science Institute, which is operated by the Association of Universities  for  Research  in  Astronomy,  Inc.,  for NASA,  under  contract  NAS5-26555.
TK and JSB were  supported by NSF AST-1518291, HST-AR-14282, and HST-AR-13888.

%\clearpage

\begin{table*}
\small
  \begin{center}
  \caption{\label{tab:correlations_top30_sim}Correlations for simulated top-30 sub-halos within $r_{\mathrm{vir}}$.}
  \begin{tabular}{rrrrrrrrrrrrrrrrr}
  \tableline\tableline
    & \multicolumn{4}{c}{$\min \Delta_{\mathrm{rms}}$ vs. $c_{\mathrm{-2}}$} & \multicolumn{4}{c}{$\min \Delta_{\mathrm{rms}}$ vs. $z_{\mathrm{0.5}}$} & \multicolumn{4}{c}{$\min \Delta_{\mathrm{rms}}$ vs. $r_{\mathrm{vir}}$} & \multicolumn{4}{c}{$\min \Delta_{\mathrm{rms}}$ vs. $\Delta_{\mathrm{r}}^{\mathrm{subhalos}}$} \\
    & \multicolumn{2}{c}{Pearson} & \multicolumn{2}{c}{Spearman} & \multicolumn{2}{c}{Pearson} & \multicolumn{2}{c}{Spearman} & \multicolumn{2}{c}{Pearson} & \multicolumn{2}{c}{Spearman} & \multicolumn{2}{c}{Pearson} & \multicolumn{2}{c}{Spearman} \\
  $N_\mathrm{inPlane}$ & $\rho$ & $\log p$ & $r_{\mathrm{s}}$ & $\log p$ & $\rho$ & $\log p$ & $r_{\mathrm{s}}$ & $\log p$ & $\rho$ & $\log p$ & $r_{\mathrm{s}}$ & $\log p$ & $\rho$ & $\log p$ & $r_{\mathrm{s}}$ & $\log p$ \\
  \tableline
3 & -0.17 & -0.6 & -0.17 & -0.6 &-0.15 & -0.5 & -0.18 & -0.7 &0.38 & -2.1 & 0.41 & -2.4 &0.39 & -2.3 & 0.42 & -2.5 \\
4 & 0.24 & -1.0 & 0.20 & -0.7 &0.05 & -0.1 & -0.03 & -0.1 &0.22 & -0.9 & 0.18 & -0.6 &0.27 & -1.2 & 0.24 & -1.0 \\
5 & -0.03 & -0.1 & -0.06 & -0.2 &0.03 & -0.1 & 0.02 & -0.1 &0.40 & -2.4 & 0.38 & -2.1 &0.45 & -2.9 & 0.38 & -2.1 \\
6 & -0.10 & -0.3 & -0.06 & -0.2 &-0.04 & -0.1 & 0.00 & -0.0 &0.39 & -2.2 & 0.32 & -1.6 &0.54 & -4.1 & 0.49 & -3.4 \\
7 & -0.14 & -0.5 & -0.15 & -0.5 &-0.08 & -0.2 & -0.09 & -0.3 &0.52 & -3.8 & 0.50 & -3.5 &0.58 & -4.9 & 0.56 & -4.4 \\
8 & -0.15 & -0.5 & -0.13 & -0.4 &-0.04 & -0.1 & 0.01 & -0.0 &0.58 & -4.8 & 0.57 & -4.5 &0.65 & -6.3 & 0.62 & -5.5 \\
9 & -0.10 & -0.3 & -0.14 & -0.5 &-0.06 & -0.2 & 0.02 & -0.0 &0.53 & -3.9 & 0.50 & -3.5 &0.63 & -5.7 & 0.58 & -4.7 \\
10 & -0.11 & -0.3 & -0.03 & -0.1 &-0.11 & -0.3 & 0.01 & -0.0 &0.47 & -3.1 & 0.47 & -3.1 &0.61 & -5.4 & 0.61 & -5.4 \\
11 & -0.11 & -0.3 & -0.04 & -0.1 &-0.13 & -0.4 & -0.10 & -0.3 &0.42 & -2.5 & 0.37 & -2.0 &0.54 & -4.1 & 0.47 & -3.1 \\
12 & -0.12 & -0.4 & -0.07 & -0.2 &-0.13 & -0.4 & -0.13 & -0.4 &0.41 & -2.4 & 0.37 & -2.0 &0.56 & -4.5 & 0.51 & -3.6 \\
13 & -0.08 & -0.2 & -0.04 & -0.1 &-0.03 & -0.1 & -0.02 & -0.1 &0.42 & -2.5 & 0.41 & -2.4 &0.65 & -6.3 & 0.63 & -5.7 \\
14 & -0.06 & -0.2 & 0.01 & -0.0 &-0.04 & -0.1 & -0.05 & -0.1 &0.42 & -2.5 & 0.43 & -2.6 &0.66 & -6.5 & 0.66 & -6.4 \\
15 & -0.05 & -0.1 & 0.03 & -0.1 &0.00 & -0.0 & 0.01 & -0.0 &0.44 & -2.7 & 0.43 & -2.6 &0.70 & -7.5 & 0.68 & -6.9 \\
16 & -0.03 & -0.1 & 0.03 & -0.1 &0.01 & -0.0 & 0.04 & -0.1 &0.47 & -3.1 & 0.49 & -3.4 &0.72 & -7.9 & 0.71 & -7.9 \\
17 & -0.03 & -0.1 & -0.00 & -0.0 &0.04 & -0.1 & 0.05 & -0.1 &0.51 & -3.7 & 0.52 & -3.9 &0.73 & -8.3 & 0.72 & -8.2 \\
18 & 0.00 & -0.0 & 0.01 & -0.0 &0.10 & -0.3 & 0.09 & -0.3 &0.53 & -3.9 & 0.53 & -4.0 &0.74 & -8.9 & 0.75 & -9.1 \\
19 & -0.01 & -0.0 & 0.00 & -0.0 &0.11 & -0.3 & 0.14 & -0.5 &0.52 & -3.8 & 0.51 & -3.6 &0.74 & -8.6 & 0.72 & -8.1 \\
20 & -0.00 & -0.0 & 0.01 & -0.0 &0.13 & -0.4 & 0.16 & -0.6 &0.53 & -3.9 & 0.51 & -3.7 &0.74 & -8.6 & 0.72 & -8.2 \\
21 & -0.00 & -0.0 & 0.04 & -0.1 &0.12 & -0.4 & 0.15 & -0.5 &0.51 & -3.7 & 0.49 & -3.4 &0.72 & -8.1 & 0.71 & -7.9 \\
22 & -0.00 & -0.0 & 0.01 & -0.0 &0.13 & -0.4 & 0.16 & -0.5 &0.51 & -3.7 & 0.48 & -3.2 &0.71 & -7.8 & 0.68 & -7.0 \\
23 & -0.00 & -0.0 & -0.01 & -0.0 &0.16 & -0.5 & 0.16 & -0.6 &0.50 & -3.6 & 0.46 & -3.0 &0.71 & -7.8 & 0.66 & -6.5 \\
24 & 0.00 & -0.0 & -0.00 & -0.0 &0.17 & -0.6 & 0.16 & -0.6 &0.49 & -3.4 & 0.44 & -2.8 &0.71 & -7.8 & 0.64 & -6.1 \\
25 & -0.01 & -0.0 & -0.03 & -0.1 &0.17 & -0.6 & 0.13 & -0.4 &0.48 & -3.3 & 0.45 & -2.8 &0.72 & -8.0 & 0.68 & -6.9 \\
26 & -0.01 & -0.0 & -0.05 & -0.1 &0.17 & -0.6 & 0.13 & -0.4 &0.46 & -3.0 & 0.46 & -2.9 &0.70 & -7.5 & 0.69 & -7.2 \\
27 & -0.01 & -0.0 & -0.03 & -0.1 &0.16 & -0.6 & 0.14 & -0.5 &0.45 & -2.9 & 0.48 & -3.2 &0.70 & -7.4 & 0.69 & -7.3 \\
28 & -0.00 & -0.0 & -0.02 & -0.1 &0.15 & -0.5 & 0.14 & -0.4 &0.45 & -2.9 & 0.48 & -3.3 &0.70 & -7.4 & 0.70 & -7.6 \\
29 & -0.01 & -0.0 & -0.02 & -0.0 &0.17 & -0.6 & 0.14 & -0.5 &0.44 & -2.7 & 0.47 & -3.2 &0.69 & -7.1 & 0.70 & -7.5 \\
30 & -0.04 & -0.1 & -0.06 & -0.2 &0.14 & -0.4 & 0.11 & -0.3 &0.45 & -2.9 & 0.47 & -3.1 &0.68 & -7.0 & 0.69 & -7.3 \\
  \tableline
Average & -0.04 & -0.2 & -0.03 & -0.2 &0.04 & -0.3 & 0.05 & -0.3 &0.46 & -3.1 & 0.45 & -3.0 &0.64 & -6.4 & 0.62 & -6.0 \\
  \tableline
\end{tabular}
 \end{center}
 \tablecomments{Correlation coefficients and logarithms of the corresponding $p$-values for Pearson ($\rho$) and Spearman ($r_{\mathrm{s}}$) tests of correlations between the minimum plane heights $\min \Delta_{\mathrm{rms}}$ for different numbers of satellites in a plane $N_\mathrm{inPlane}$\ versus various halo parameters: halo concentration $c_{\mathrm{-2}}$, formation redshift $z_{\mathrm{0.5}}$, virial radius $r_{\mathrm{vir}}$, and rms radius of the sub-halo distribution $\Delta_{\mathrm{r}}^{\mathrm{subhalos}}$.}
\end{table*}

\begin{table*}
\small
  \begin{center}
  \caption{\label{tab:correlations_top30_rand}Correlations for randomized top-30 sub-halos within $r_{\mathrm{vir}}$.}
  \begin{tabular}{rrrrrrrrrrrrrrrrr}
  \tableline\tableline
    & \multicolumn{4}{c}{$\min \Delta_{\mathrm{rms}}$ vs. $c_{\mathrm{-2}}$} & \multicolumn{4}{c}{$\min \Delta_{\mathrm{rms}}$ vs. $z_{\mathrm{0.5}}$} & \multicolumn{4}{c}{$\min \Delta_{\mathrm{rms}}$ vs. $r_{\mathrm{vir}}$} & \multicolumn{4}{c}{$\min \Delta_{\mathrm{rms}}$ vs. $\Delta_{\mathrm{r}}^{\mathrm{subhalos}}$} \\
    & \multicolumn{2}{c}{Pearson} & \multicolumn{2}{c}{Spearman} & \multicolumn{2}{c}{Pearson} & \multicolumn{2}{c}{Spearman} & \multicolumn{2}{c}{Pearson} & \multicolumn{2}{c}{Spearman} & \multicolumn{2}{c}{Pearson} & \multicolumn{2}{c}{Spearman} \\
  $N_\mathrm{inPlane}$ & $\rho$ & $\log p$ & $r_{\mathrm{s}}$ & $\log p$ & $\rho$ & $\log p$ & $r_{\mathrm{s}}$ & $\log p$ & $\rho$ & $\log p$ & $r_{\mathrm{s}}$ & $\log p$ & $\rho$ & $\log p$ & $r_{\mathrm{s}}$ & $\log p$ \\
  \tableline
3 & -0.06 & -0.2 & -0.04 & -0.1 &0.14 & -0.5 & 0.14 & -0.5 &0.10 & -0.3 & 0.11 & -0.3 &0.21 & -0.8 & 0.21 & -0.8 \\
4 & -0.12 & -0.4 & -0.12 & -0.4 &0.07 & -0.2 & 0.07 & -0.2 &0.33 & -1.7 & 0.31 & -1.5 &0.31 & -1.5 & 0.34 & -1.7 \\
5 & -0.14 & -0.5 & -0.09 & -0.3 &-0.14 & -0.5 & -0.07 & -0.2 &0.45 & -2.8 & 0.48 & -3.2 &0.29 & -1.4 & 0.28 & -1.2 \\
6 & 0.03 & -0.1 & 0.05 & -0.1 &0.03 & -0.1 & -0.04 & -0.1 &0.22 & -0.9 & 0.23 & -0.9 &0.37 & -2.0 & 0.31 & -1.5 \\
7 & 0.01 & -0.0 & 0.03 & -0.1 &0.15 & -0.5 & 0.11 & -0.4 &0.20 & -0.7 & 0.16 & -0.6 &0.30 & -1.4 & 0.27 & -1.2 \\
8 & 0.11 & -0.3 & 0.10 & -0.3 &0.07 & -0.2 & 0.05 & -0.1 &0.24 & -1.0 & 0.27 & -1.2 &0.48 & -3.2 & 0.49 & -3.4 \\
9 & -0.03 & -0.1 & 0.00 & -0.0 &0.04 & -0.1 & 0.02 & -0.1 &0.34 & -1.8 & 0.34 & -1.8 &0.50 & -3.6 & 0.48 & -3.2 \\
10 & -0.02 & -0.1 & 0.00 & -0.0 &0.13 & -0.4 & 0.11 & -0.3 &0.33 & -1.6 & 0.33 & -1.6 &0.52 & -3.8 & 0.49 & -3.4 \\
11 & 0.03 & -0.1 & 0.09 & -0.3 &0.19 & -0.7 & 0.20 & -0.8 &0.42 & -2.5 & 0.43 & -2.6 &0.58 & -4.8 & 0.55 & -4.2 \\
12 & 0.01 & -0.0 & 0.09 & -0.3 &0.10 & -0.3 & 0.14 & -0.4 &0.45 & -2.9 & 0.45 & -2.9 &0.61 & -5.3 & 0.58 & -4.7 \\
13 & 0.07 & -0.2 & 0.12 & -0.4 &0.10 & -0.3 & 0.13 & -0.4 &0.42 & -2.5 & 0.43 & -2.6 &0.60 & -5.2 & 0.57 & -4.6 \\
14 & 0.05 & -0.1 & 0.12 & -0.4 &0.10 & -0.3 & 0.18 & -0.7 &0.43 & -2.7 & 0.45 & -2.8 &0.62 & -5.6 & 0.58 & -4.9 \\
15 & 0.04 & -0.1 & 0.11 & -0.3 &0.11 & -0.3 & 0.15 & -0.5 &0.47 & -3.1 & 0.45 & -2.8 &0.67 & -6.8 & 0.61 & -5.4 \\
16 & 0.02 & -0.0 & 0.09 & -0.3 &0.14 & -0.5 & 0.16 & -0.6 &0.49 & -3.3 & 0.44 & -2.7 &0.67 & -6.7 & 0.58 & -4.8 \\
17 & -0.03 & -0.1 & 0.05 & -0.1 &0.15 & -0.5 & 0.14 & -0.5 &0.52 & -3.9 & 0.46 & -2.9 &0.68 & -6.9 & 0.61 & -5.3 \\
18 & -0.05 & -0.1 & 0.02 & -0.0 &0.16 & -0.5 & 0.16 & -0.6 &0.53 & -3.9 & 0.45 & -2.9 &0.69 & -7.3 & 0.62 & -5.6 \\
19 & -0.05 & -0.1 & 0.02 & -0.1 &0.15 & -0.5 & 0.19 & -0.7 &0.52 & -3.8 & 0.46 & -3.0 &0.68 & -7.0 & 0.62 & -5.6 \\
20 & -0.04 & -0.1 & 0.00 & -0.0 &0.17 & -0.6 & 0.19 & -0.7 &0.50 & -3.6 & 0.44 & -2.8 &0.67 & -6.7 & 0.62 & -5.5 \\
21 & -0.01 & -0.0 & 0.01 & -0.0 &0.21 & -0.8 & 0.23 & -0.9 &0.51 & -3.7 & 0.46 & -3.0 &0.69 & -7.3 & 0.64 & -5.9 \\
22 & -0.01 & -0.0 & 0.02 & -0.0 &0.21 & -0.8 & 0.23 & -0.9 &0.52 & -3.8 & 0.49 & -3.4 &0.69 & -7.2 & 0.64 & -6.0 \\
23 & -0.01 & -0.0 & -0.01 & -0.0 &0.24 & -1.0 & 0.24 & -1.0 &0.53 & -3.9 & 0.48 & -3.2 &0.71 & -7.9 & 0.66 & -6.5 \\
24 & -0.02 & -0.0 & -0.02 & -0.0 &0.24 & -1.0 & 0.22 & -0.9 &0.53 & -3.9 & 0.50 & -3.5 &0.73 & -8.4 & 0.69 & -7.2 \\
25 & -0.02 & -0.1 & -0.02 & -0.0 &0.23 & -1.0 & 0.20 & -0.8 &0.53 & -3.9 & 0.50 & -3.5 &0.74 & -8.6 & 0.69 & -7.2 \\
26 & -0.05 & -0.1 & -0.07 & -0.2 &0.23 & -1.0 & 0.19 & -0.7 &0.53 & -4.0 & 0.51 & -3.6 &0.74 & -8.9 & 0.69 & -7.2 \\
27 & -0.03 & -0.1 & -0.07 & -0.2 &0.27 & -1.2 & 0.21 & -0.8 &0.54 & -4.1 & 0.55 & -4.2 &0.77 & -9.8 & 0.72 & -8.2 \\
28 & -0.03 & -0.1 & -0.07 & -0.2 &0.29 & -1.3 & 0.22 & -0.9 &0.58 & -4.8 & 0.58 & -4.7 &0.79 & -10.7 & 0.73 & -8.3 \\
29 & 0.01 & -0.0 & -0.02 & -0.0 &0.33 & -1.7 & 0.23 & -0.9 &0.61 & -5.4 & 0.62 & -5.6 &0.81 & -11.6 & 0.74 & -8.8 \\
30 & -0.01 & -0.0 & -0.07 & -0.2 &0.35 & -1.8 & 0.25 & -1.0 &0.66 & -6.6 & 0.68 & -7.0 &0.83 & -12.5 & 0.76 & -9.4 \\
  \tableline
Average & -0.01 & -0.1 & 0.01 & -0.2 &0.16 & -0.7 & 0.15 & -0.6 &0.45 & -3.1 & 0.43 & -2.9 &0.61 & -6.2 & 0.56 & -5.1 \\
  \tableline
\end{tabular}
 \end{center}
 \tablecomments{Correlation coefficients and logarithms of the corresponding $p$-values for Pearson ($\rho$) and Spearman ($r_{\mathrm{s}}$) tests of correlations between the minimum plane heights $\min \Delta_{\mathrm{rms}}$ for different numbers of satellites in a plane $N_\mathrm{inPlane}$\ versus various halo parameters: halo concentration $c_{\mathrm{-2}}$, formation redshift $z_{\mathrm{0.5}}$, virial radius $r_{\mathrm{vir}}$, and rms radius of the sub-halo distribution $\Delta_{\mathrm{r}}^{\mathrm{subhalos}}$.}
\end{table*}

\begin{table*}
\small
  \begin{center}
  \caption{\label{tab:correlations_PAndAS_sim}Correlations for simulated top-27 sub-halos within PAndAS volume.}
  \begin{tabular}{rrrrrrrrrrrrrrrrr}
  \tableline\tableline
    & \multicolumn{4}{c}{$\min \Delta_{\mathrm{rms}}$ vs. $c_{\mathrm{-2}}$} & \multicolumn{4}{c}{$\min \Delta_{\mathrm{rms}}$ vs. $z_{\mathrm{0.5}}$} & \multicolumn{4}{c}{$\min \Delta_{\mathrm{rms}}$ vs. $r_{\mathrm{vir}}$} & \multicolumn{4}{c}{$\min \Delta_{\mathrm{rms}}$ vs. $\Delta_{\mathrm{r}}^{\mathrm{subhalos}}$} \\
    & \multicolumn{2}{c}{Pearson} & \multicolumn{2}{c}{Spearman} & \multicolumn{2}{c}{Pearson} & \multicolumn{2}{c}{Spearman} & \multicolumn{2}{c}{Pearson} & \multicolumn{2}{c}{Spearman} & \multicolumn{2}{c}{Pearson} & \multicolumn{2}{c}{Spearman} \\
  $N_\mathrm{inPlane}$ & $\rho$ & $\log p$ & $r_{\mathrm{s}}$ & $\log p$ & $\rho$ & $\log p$ & $r_{\mathrm{s}}$ & $\log p$ & $\rho$ & $\log p$ & $r_{\mathrm{s}}$ & $\log p$ & $\rho$ & $\log p$ & $r_{\mathrm{s}}$ & $\log p$ \\
  \tableline
3 & -0.06 & -0.2 & -0.11 & -0.3 &0.03 & -0.1 & 0.08 & -0.2 &0.25 & -1.0 & 0.20 & -0.8 &0.27 & -1.2 & 0.26 & -1.1 \\
4 & -0.09 & -0.3 & -0.16 & -0.6 &0.08 & -0.2 & 0.07 & -0.2 &0.14 & -0.5 & 0.12 & -0.4 &0.36 & -2.0 & 0.37 & -2.0 \\
5 & -0.09 & -0.3 & -0.14 & -0.5 &-0.08 & -0.2 & -0.03 & -0.1 &0.01 & -0.0 & -0.05 & -0.1 &0.33 & -1.6 & 0.33 & -1.7 \\
6 & 0.13 & -0.4 & 0.11 & -0.4 &0.01 & -0.0 & 0.05 & -0.1 &0.03 & -0.1 & -0.02 & -0.1 &0.32 & -1.6 & 0.32 & -1.6 \\
7 & 0.14 & -0.5 & 0.18 & -0.7 &-0.03 & -0.1 & -0.03 & -0.1 &-0.04 & -0.1 & -0.13 & -0.4 &0.38 & -2.1 & 0.39 & -2.3 \\
8 & 0.20 & -0.8 & 0.25 & -1.1 &-0.12 & -0.4 & -0.11 & -0.3 &-0.13 & -0.4 & -0.24 & -1.0 &0.34 & -1.8 & 0.31 & -1.5 \\
9 & 0.18 & -0.6 & 0.23 & -1.0 &-0.24 & -1.0 & -0.25 & -1.1 &-0.08 & -0.2 & -0.16 & -0.6 &0.32 & -1.6 & 0.28 & -1.3 \\
10 & 0.06 & -0.2 & 0.14 & -0.5 &-0.28 & -1.3 & -0.27 & -1.2 &0.01 & -0.0 & -0.08 & -0.2 &0.35 & -1.8 & 0.37 & -2.0 \\
11 & 0.06 & -0.2 & 0.12 & -0.4 &-0.31 & -1.5 & -0.32 & -1.6 &0.06 & -0.2 & -0.00 & -0.0 &0.31 & -1.5 & 0.39 & -2.2 \\
12 & 0.00 & -0.0 & 0.04 & -0.1 &-0.34 & -1.7 & -0.36 & -1.9 &0.03 & -0.1 & -0.02 & -0.1 &0.30 & -1.4 & 0.32 & -1.6 \\
13 & 0.04 & -0.1 & 0.05 & -0.1 &-0.33 & -1.6 & -0.38 & -2.1 &-0.04 & -0.1 & -0.09 & -0.3 &0.35 & -1.8 & 0.36 & -1.9 \\
14 & 0.04 & -0.1 & 0.06 & -0.2 &-0.32 & -1.6 & -0.33 & -1.6 &-0.01 & -0.0 & -0.07 & -0.2 &0.33 & -1.7 & 0.35 & -1.8 \\
15 & -0.00 & -0.0 & 0.02 & -0.0 &-0.29 & -1.3 & -0.30 & -1.4 &-0.01 & -0.0 & -0.07 & -0.2 &0.35 & -1.8 & 0.37 & -2.0 \\
16 & 0.03 & -0.1 & 0.05 & -0.1 &-0.26 & -1.2 & -0.29 & -1.4 &-0.02 & -0.1 & -0.06 & -0.2 &0.31 & -1.5 & 0.34 & -1.7 \\
17 & 0.05 & -0.1 & 0.05 & -0.1 &-0.26 & -1.1 & -0.27 & -1.2 &-0.08 & -0.2 & -0.12 & -0.4 &0.29 & -1.3 & 0.31 & -1.5 \\
18 & 0.03 & -0.1 & 0.01 & -0.0 &-0.25 & -1.1 & -0.29 & -1.3 &-0.08 & -0.2 & -0.12 & -0.4 &0.27 & -1.2 & 0.29 & -1.4 \\
19 & 0.03 & -0.1 & -0.01 & -0.0 &-0.24 & -1.0 & -0.26 & -1.1 &-0.07 & -0.2 & -0.12 & -0.4 &0.24 & -1.0 & 0.24 & -1.0 \\
20 & 0.05 & -0.1 & 0.03 & -0.1 &-0.16 & -0.6 & -0.24 & -1.0 &-0.02 & -0.0 & -0.09 & -0.3 &0.18 & -0.7 & 0.20 & -0.7 \\
21 & 0.05 & -0.1 & 0.03 & -0.1 &-0.10 & -0.3 & -0.17 & -0.6 &0.00 & -0.0 & -0.09 & -0.3 &0.16 & -0.5 & 0.18 & -0.6 \\
22 & 0.02 & -0.1 & 0.00 & -0.0 &-0.05 & -0.1 & -0.09 & -0.3 &0.04 & -0.1 & -0.05 & -0.1 &0.15 & -0.5 & 0.14 & -0.5 \\
23 & 0.01 & -0.0 & 0.05 & -0.1 &-0.06 & -0.2 & -0.11 & -0.3 &0.08 & -0.2 & 0.01 & -0.0 &0.30 & -1.4 & 0.23 & -0.9 \\
24 & 0.06 & -0.2 & 0.08 & -0.2 &-0.07 & -0.2 & -0.09 & -0.3 &0.10 & -0.3 & 0.07 & -0.2 &0.32 & -1.5 & 0.26 & -1.1 \\
25 & 0.10 & -0.3 & 0.12 & -0.4 &-0.08 & -0.2 & -0.10 & -0.3 &0.09 & -0.3 & 0.06 & -0.2 &0.33 & -1.7 & 0.29 & -1.3 \\
26 & 0.09 & -0.3 & 0.11 & -0.3 &-0.06 & -0.2 & -0.10 & -0.3 &0.10 & -0.3 & 0.09 & -0.3 &0.36 & -1.9 & 0.31 & -1.5 \\
27 & 0.09 & -0.3 & 0.10 & -0.3 &-0.04 & -0.1 & -0.06 & -0.2 &0.09 & -0.3 & 0.11 & -0.3 &0.44 & -2.7 & 0.37 & -2.0 \\
  \tableline
Average & 0.05 & -0.2 & 0.06 & -0.3 &-0.15 & -0.7 & -0.17 & -0.8 &0.02 & -0.2 & -0.04 & -0.3 &0.31 & -1.5 & 0.30 & -1.5 \\
  \tableline
\end{tabular}
 \end{center}
 \tablecomments{Correlation coefficients and logarithms of the corresponding $p$-values for Pearson ($\rho$) and Spearman ($r_{\mathrm{s}}$) tests of correlations between the minimum plane heights $\min \Delta_{\mathrm{rms}}$ for different numbers of satellites in a plane $N_\mathrm{inPlane}$\ versus various halo parameters: halo concentration $c_{\mathrm{-2}}$, formation redshift $z_{\mathrm{0.5}}$, virial radius $r_{\mathrm{vir}}$, and rms radius of the sub-halo distribution $\Delta_{\mathrm{r}}^{\mathrm{subhalos}}$.}
\end{table*}

\begin{table*}
\small
  \begin{center}
  \caption{\label{tab:correlations_PAndAS_rand}Correlations for randomized top-27 sub-halos within PAndAS volume.}
  \begin{tabular}{rrrrrrrrrrrrrrrrr}
  \tableline\tableline
    & \multicolumn{4}{c}{$\min \Delta_{\mathrm{rms}}$ vs. $c_{\mathrm{-2}}$} & \multicolumn{4}{c}{$\min \Delta_{\mathrm{rms}}$ vs. $z_{\mathrm{0.5}}$} & \multicolumn{4}{c}{$\min \Delta_{\mathrm{rms}}$ vs. $r_{\mathrm{vir}}$} & \multicolumn{4}{c}{$\min \Delta_{\mathrm{rms}}$ vs. $\Delta_{\mathrm{r}}^{\mathrm{subhalos}}$} \\
    & \multicolumn{2}{c}{Pearson} & \multicolumn{2}{c}{Spearman} & \multicolumn{2}{c}{Pearson} & \multicolumn{2}{c}{Spearman} & \multicolumn{2}{c}{Pearson} & \multicolumn{2}{c}{Spearman} & \multicolumn{2}{c}{Pearson} & \multicolumn{2}{c}{Spearman} \\
  $N_\mathrm{inPlane}$ & $\rho$ & $\log p$ & $r_{\mathrm{s}}$ & $\log p$ & $\rho$ & $\log p$ & $r_{\mathrm{s}}$ & $\log p$ & $\rho$ & $\log p$ & $r_{\mathrm{s}}$ & $\log p$ & $\rho$ & $\log p$ & $r_{\mathrm{s}}$ & $\log p$ \\
  \tableline
3 & -0.25 & -1.1 & -0.33 & -1.6 &-0.07 & -0.2 & -0.01 & -0.0 &0.11 & -0.3 & 0.02 & -0.1 &0.29 & -1.3 & 0.18 & -0.7 \\
4 & -0.18 & -0.7 & -0.21 & -0.8 &0.03 & -0.1 & 0.01 & -0.0 &0.01 & -0.0 & -0.03 & -0.1 &0.29 & -1.4 & 0.23 & -0.9 \\
5 & 0.06 & -0.2 & 0.05 & -0.1 &0.07 & -0.2 & 0.08 & -0.2 &0.06 & -0.2 & 0.04 & -0.1 &0.31 & -1.5 & 0.24 & -1.0 \\
6 & -0.03 & -0.1 & -0.02 & -0.1 &0.10 & -0.3 & 0.14 & -0.5 &-0.10 & -0.3 & -0.05 & -0.1 &0.13 & -0.4 & 0.09 & -0.3 \\
7 & -0.16 & -0.6 & -0.16 & -0.6 &-0.05 & -0.1 & -0.03 & -0.1 &-0.05 & -0.1 & -0.07 & -0.2 &0.06 & -0.2 & 0.03 & -0.1 \\
8 & -0.15 & -0.5 & -0.21 & -0.8 &-0.01 & -0.0 & -0.05 & -0.1 &-0.08 & -0.2 & -0.06 & -0.2 &0.20 & -0.7 & 0.14 & -0.5 \\
9 & -0.15 & -0.5 & -0.16 & -0.6 &-0.13 & -0.4 & -0.15 & -0.5 &-0.00 & -0.0 & -0.03 & -0.1 &0.16 & -0.6 & 0.15 & -0.5 \\
10 & -0.13 & -0.4 & -0.11 & -0.3 &-0.11 & -0.4 & -0.14 & -0.5 &0.07 & -0.2 & 0.02 & -0.1 &0.13 & -0.4 & 0.12 & -0.4 \\
11 & -0.09 & -0.3 & -0.05 & -0.1 &-0.11 & -0.3 & -0.11 & -0.4 &0.11 & -0.3 & 0.06 & -0.2 &0.13 & -0.4 & 0.09 & -0.3 \\
12 & -0.03 & -0.1 & -0.01 & -0.0 &-0.09 & -0.3 & -0.12 & -0.4 &0.06 & -0.1 & 0.02 & -0.1 &0.15 & -0.5 & 0.11 & -0.4 \\
13 & 0.02 & -0.0 & 0.05 & -0.1 &-0.06 & -0.2 & -0.09 & -0.2 &-0.02 & -0.0 & 0.00 & -0.0 &0.17 & -0.6 & 0.09 & -0.3 \\
14 & 0.00 & -0.0 & 0.03 & -0.1 &-0.00 & -0.0 & 0.01 & -0.0 &-0.08 & -0.2 & -0.08 & -0.2 &0.21 & -0.8 & 0.10 & -0.3 \\
15 & -0.01 & -0.0 & 0.03 & -0.1 &0.00 & -0.0 & -0.02 & -0.0 &-0.07 & -0.2 & -0.07 & -0.2 &0.26 & -1.2 & 0.14 & -0.5 \\
16 & 0.03 & -0.1 & 0.06 & -0.2 &0.05 & -0.1 & 0.04 & -0.1 &-0.08 & -0.2 & -0.08 & -0.2 &0.26 & -1.1 & 0.15 & -0.5 \\
17 & 0.01 & -0.0 & 0.02 & -0.0 &0.07 & -0.2 & 0.07 & -0.2 &-0.07 & -0.2 & -0.06 & -0.2 &0.23 & -0.9 & 0.17 & -0.6 \\
18 & 0.03 & -0.1 & 0.00 & -0.0 &0.07 & -0.2 & 0.01 & -0.0 &-0.05 & -0.1 & -0.04 & -0.1 &0.20 & -0.8 & 0.15 & -0.5 \\
19 & 0.01 & -0.0 & -0.00 & -0.0 &0.04 & -0.1 & 0.02 & -0.0 &-0.03 & -0.1 & -0.04 & -0.1 &0.21 & -0.8 & 0.16 & -0.5 \\
20 & 0.00 & -0.0 & -0.02 & -0.1 &0.02 & -0.1 & -0.01 & -0.0 &-0.01 & -0.0 & -0.02 & -0.1 &0.15 & -0.5 & 0.14 & -0.5 \\
21 & -0.02 & -0.0 & -0.04 & -0.1 &0.00 & -0.0 & -0.03 & -0.1 &0.02 & -0.1 & 0.04 & -0.1 &0.15 & -0.5 & 0.13 & -0.4 \\
22 & 0.00 & -0.0 & -0.01 & -0.0 &-0.02 & -0.0 & -0.02 & -0.1 &0.04 & -0.1 & 0.05 & -0.1 &0.14 & -0.5 & 0.13 & -0.4 \\
23 & 0.01 & -0.0 & -0.00 & -0.0 &-0.00 & -0.0 & -0.01 & -0.0 &0.07 & -0.2 & 0.08 & -0.2 &0.16 & -0.5 & 0.11 & -0.3 \\
24 & -0.00 & -0.0 & -0.02 & -0.0 &-0.01 & -0.0 & 0.01 & -0.0 &0.08 & -0.2 & 0.10 & -0.3 &0.19 & -0.7 & 0.13 & -0.4 \\
25 & -0.00 & -0.0 & -0.04 & -0.1 &0.01 & -0.0 & 0.02 & -0.0 &0.09 & -0.3 & 0.10 & -0.3 &0.21 & -0.8 & 0.16 & -0.6 \\
26 & 0.04 & -0.1 & -0.00 & -0.0 &0.03 & -0.1 & 0.01 & -0.0 &0.05 & -0.1 & 0.06 & -0.2 &0.23 & -1.0 & 0.17 & -0.6 \\
27 & 0.13 & -0.4 & 0.07 & -0.2 &0.08 & -0.2 & 0.08 & -0.2 &-0.00 & -0.0 & 0.05 & -0.1 &0.25 & -1.1 & 0.24 & -1.0 \\
  \tableline
Average & -0.03 & -0.2 & -0.04 & -0.2 &-0.00 & -0.1 & -0.01 & -0.2 &0.00 & -0.2 & 0.00 & -0.1 &0.20 & -0.8 & 0.14 & -0.5 \\
  \tableline
\end{tabular}
 \end{center}
 \tablecomments{Correlation coefficients and logarithms of the corresponding $p$-values for Pearson ($\rho$) and Spearman ($r_{\mathrm{s}}$) tests of correlations between the minimum plane heights $\min \Delta_{\mathrm{rms}}$ for different numbers of satellites in a plane $N_\mathrm{inPlane}$\ versus various halo parameters: halo concentration $c_{\mathrm{-2}}$, formation redshift $z_{\mathrm{0.5}}$, virial radius $r_{\mathrm{vir}}$, and rms radius of the sub-halo distribution $\Delta_{\mathrm{r}}^{\mathrm{subhalos}}$.}
\end{table*}

\end{document}